\documentclass[]{aastex631}
\usepackage{graphicx}
\usepackage{xcolor}
\usepackage{natbib}

\newcommand{\ha}{H$\alpha$}

\newcommand{\htwo}{H$_2$}
\newcommand{\hi}{\ion{H}{1}}
\newcommand{\nai}{\ion{Na}{1}}

\newcommand{\feii}{\ion{Fe}{2}}
\newcommand{\sii}{\ion{S}{2}}

\newcommand{\hei}{\ion{He}{1}}
\newcommand{\caii}{\ion{Ca}{2}}
\newcommand{\oi}{\ion{O}{1}}
\newcommand{\oii}{\ion{O}{2}}
\newcommand{\nii}{\ion{N}{2}}
\newcommand{\n}{[\ion{N}{1}]}

\newcommand{\av}{A$_V$}

\newcommand{\lsun}{L$_{\odot}$}
\newcommand{\msun}{M$_{\odot}$}
\newcommand{\rsun}{R$_{\odot}$}
\newcommand{\msunyr}{M$_{\odot}$\,yr$^{-1}$}
\newcommand{\macc}{$\dot{M}_{\mathrm{acc}}$}

\newcommand{\lacc}{$L_{\mathrm{acc}}$}
\newcommand{\lacci}{$L_{\mathrm{acc(i)}}$}
\newcommand{\lumi}{$L_{\mathrm{i}}$}
\newcommand{\rstar}{$R_{\mathrm{*}}$}

\newcommand{\lstar}{$L_{\mathrm{*}}$}
\newcommand{\mstar}{$M_{\mathrm{*}}$}
\newcommand{\Mbol}{$M_{\mathrm{bol}}$}
\newcommand{\teff}{$T_\mathrm{eff}$}

\newcommand{\f}{\scriptsize}

\newcommand{\lburst}{$L_{\mathrm{6000,burst}}$}
\newcommand{\lquiesc}{$L_{\mathrm{6000,quiesc}}$}
\newcommand{\laccb}{$L_{\mathrm{acc,burst}}$}
\newcommand{\laccq}{$L_{\mathrm{acc,quiesc}}$}

\shorttitle{EYs observed with LBT}
\shortauthors{Giannini et al.}

\graphicspath{{./}{figures/}}

\begin{document}

%

   \title{Spectroscopic follow-up of Gaia alerted YSO variables: the LBT view \thanks{Based on observations made with the {\it Large Binocular Telescope}  operated by the {\it Fundaci\'on Galileo Galilei} (FGG) of the {\it Istituto Nazionale di Astrofisica} (INAF) at the {\it  Observatorio del Roque de los Muchachos} (La Palma, Canary Islands, Spain), under programs A36TAC\_22, XX, XX, XX (PI: S. Antoniucci).}
   }


 \author[0000-0002-7035-8513]{Teresa Giannini}
\affiliation{INAF - Osservatorio Astronomico di Roma, Via di Frascati, 33, 00078, Monte Porzio Catone, Italy}
\email{teresa.giannini@inaf.it}

\author[0000-0002-8364-7795]{Manuele Gangi}
\affiliation{ASI, Italian Space Agency, Via del Politecnico snc, 00133, Rome, Italy}

\author[0000-0002-4283-2185]{Fernando Cruz-S\'aenz de Miera}
\affiliation{Institut de Recherche en Astrophysique et Plan\'etologie, Universit\'e de Toulouse, UT3-PS, OMP, CNRS, 9 av. du Colonel Roche, 31028 Toulouse Cedex 4, France}
\affiliation{HUN-REN Research Centre for Astronomy and Earth Sciences, Konkoly Observatory,
MTA Centre of Excellence, Konkoly Thege Miklós út 15-17., H-1121 Budapest, Hungary}

\author[0000-0002-9190-0113]{Brunella Nisini}
\affiliation{INAF - Osservatorio Astronomico di Roma, Via di Frascati, 33, 00078, Monte Porzio Catone, Italy}

\author[0000-0002-3648-433X]{M\'at\'e Szil\'agyi}
\affiliation{HUN-REN Research Centre for Astronomy and Earth Sciences, Konkoly Observatory,
MTA Centre of Excellence, Konkoly Thege Miklós út 15-17., H-1121 Budapest, Hungary}

\author[0000-0002-1892-2180]{Katia Biazzo}
\affiliation{INAF - Osservatorio Astronomico di Roma, Via di Frascati, 33, 00078, Monte Porzio Catone, Italy}

\author[0000-0001-7157-6275]{\'Agnes K\'osp\'al}
\affiliation{HUN-REN Research Centre for Astronomy and Earth Sciences, Konkoly Observatory,
MTA Centre of Excellence, Konkoly Thege Miklós út 15-17., H-1121 Budapest, Hungary}
\affiliation{ELTE E\"otv\"os Lor\'and University, Institute of Physics and Astronomy, P\'azm\'any P\'eter s\'et\'any 1A, Budapest 1117, Hungary}

\author[0000-0001-6015-646X]{P\'eter \'Abrah\'am}
\affiliation{HUN-REN Research Centre for Astronomy and Earth Sciences, Konkoly Observatory,
MTA Centre of Excellence, Konkoly Thege Miklós út 15-17., H-1121 Budapest, Hungary}
\affiliation{ELTE E\"otv\"os Lor\'and University, Institute of Physics and Astronomy, P\'azm\'any P\'eter s\'et\'any 1A, Budapest 1117, Hungary}

\author[0000-0002-0666-3847]{Simone Antoniucci}
\affiliation{INAF - Osservatorio Astronomico di Roma, Via di Frascati, 33, 00078, Monte Porzio Catone, Italy}

\author[0000-0003-1604-2064]{Roberta Carini}
\affiliation{INAF - Osservatorio Astronomico di Roma, Via di Frascati, 33, 00078, Monte Porzio Catone, Italy}
 
\author[0000-0002-5261-6216]{Eleonora Fiorellino}
\affiliation{Alma Mater Studiorum – Universitá di Bologna, Dipartimento di Fisica e Astronomia “Augusto Righi”, Via Gobetti 93/2, I-40129, Bologna, Italy}
\affiliation{INAF – Osservatorio Astronomico di Trieste, via Tiepolo 11, I-34143 Trieste}

\author[0000-0002-3351-1216]{Adriana Gargiulo}
\affiliation{INAF – Istituto di Astrofisica Spaziale e Fisica Cosmica Milano, Via A. Corti 12, 20133 Milano, Italy}

\author[0000-0002-6894-1267]{Ester Marini}
\affiliation{INAF - Osservatorio Astronomico di Roma, Via di Frascati, 33, 00078, Monte Porzio Catone, Italy}

\author[0000-0002-3632-1194]{Zs\'ofia Nagy}
\affiliation{HUN-REN Research Centre for Astronomy and Earth Sciences, Konkoly Observatory,
MTA Centre of Excellence, Konkoly Thege Miklós út 15-17., H-1121 Budapest, Hungary}

\author[0000-0002-1860-2304]{Maria Gabriela Navarro}
\affiliation{INAF - Osservatorio Astronomico di Roma, Via di Frascati, 33, 00078, Monte Porzio Catone, Italy}

\author[0000-0001-8332-4227]{Fabrizio Vitali}
\affiliation{INAF - Osservatorio Astronomico di Roma, Via di Frascati, 33, 00078, Monte Porzio Catone, Italy}



 
\begin{abstract}
We analyzed optical/near-IR Large Binocular Telescope spectra of 16 sources alerted by Gaia between 2021 and 2024 due to significant photometric variability. Half of the spectra were taken during quiescence and the rest during a burst or at intermediate brightness. 
Our analysis of their ten-year light curves and photometric/spectroscopic features provide evidence that all 16 sources are accreting Young Stellar Objects (YSOs).  
One object, Gaia23bab, is a known EXor source. Other light curves either have peaks over a stable baseline, or significant variability throughout the entire observation period, suggesting multiple contributing processes.
All spectra exhibit emission lines  from accretion columns, and over half of them show atomic forbidden lines as signatures of outflowing gas. We determined stellar parameters, accretion luminosity (\lacc\,) and mass accretion rate (\macc\,) at different brightness phases. Only two sources showed variability primarily due to extinction.
During quiescence, our sources exhibit \lacc\, and \macc\, values typical of T Tauri and Herbig Ae/Be (HAeBe) sources, supporting the hypothesis that any YSO may undergo episodic accretion. In bursts, the \lacc\,  and \macc\, of sources with photometric variations exceeding 2 mag follow a shallower relation with stellar luminosity and mass, typical of known EXor sources. 
This group includes one Class\,I, one flat-spectrum, and two Class\,II sources. Notably, the other Class\,I source, Gaia24beh, shows an \lacc\, value about ten times higher than typical EXor bursts of the same mass. In the other cases, \lacc\, and \macc\, align with variability seen in T Tauri and HAeBe sources.
\end{abstract}


\section{Introduction}\label{sec:sec1}
Variability is a defining feature of accreting Young Stellar Objects (YSOs). Around 50\% of YSOs are subject to short-term events, occurring on minute-to-day timescales with photometric oscillations of a few tenths of a magnitude (e.g. Megeath et al. 2012). This type of variability is typically associated with stellar activity such as surface spots, stellar flares, and coronal mass ejections.

Superimposed on this, longer-term variability, ranging from months to years and even centuries, is sometimes observed (Fischer et al. 2023). This long-term variability is typically induced by changes in extinction caused by inner-disk warps and clumps (e.g. Kennedy et al. 2017), or by abrupt variations in the mass accretion rate (\macc) from typical values of 10$^{-10}$-10$^{-8}$ \msunyr\, to 10$^{-6}$-10$^{-4}$ \msunyr\, (Hartmann \& Kenyon, 1996; Audard et al. 2014). Sources exhibiting this latter class of variability are known as Eruptive Young Stars (EYs).

EYs encompass sources with diverse photometric and spectroscopic features. The original classification (Herbig 1977, 1989) was based on optical observations and identified two classes: FU Ori-type objects (FUors), which brighten by 4$-$5 magnitudes for decades or longer and exhibit spectra with absorption lines, and EX Lupi-type objects (EXors), which show repetitive bursts of 1$-$3 magnitudes lasting from months to one year, with spectra displaying emission lines. 
Until about twenty years ago, only a few dozen eruptive variables were known. 
In the last decade, photometric monitoring from surveys like 
Gaia\footnote{https://www.esa.int/Science\_Exploration/Space\_Science/Gaia} (Gaia Collaboration et al. 2016), 
ZTF\footnote{https://www.ztf.caltech.edu/} (Bellm et al. 
2019), Pan-STARSS\footnote{https://outerspace.stsci.edu/display/PANSTARRS/} (Chambers et al. 2016), ASAS-SN\footnote{https:www.astronomy.ohio-state.edu/asassn/} (Jayasinghe et al. 2018),  VVV/VVVX\footnote{https://vvvsurvey.org/} (Minniti et al. 2010), and 
AllWISE/NEOWISE\footnote{https://www.nasa.gov/mission\_pages/WISE/main/index.html} (Wright et al. 2010, Mainzer et al. 2011) has significantly increased both the 
observational cadence and the covered wavelength range.
This has led to a more than threefold increase in the number of EYs, with hundreds of new candidates. This progress has facilitated the discovery of EYs at early evolutionary stages (e.g. Safron et al. 2015; Lee et al. 2021; Zakri et al. 2022) and across a wide mass range extending to high-mass protostars (e.g. Caratti o Garatti et al. 2017; Hunter et al. 2017). Aspin \& Reipurth (2003), and, more recently, Connelley \& Reipurth (2018) and Contreras-Pe\~na et al. (2025) have defined, together with FUors and EXors, three more EYs sub-classes: FUor-like, namely objects that share with FUors the same absoprtion features but where no outburst has been observed, PVM (Peculiar/V1647Ori-like/MNors) sources, which appear intermediate between FUors and EXors, and Periodic objects, showing a quasi-periodic variability.

The rising number of the discovered EYs has increased the interest of the community in the role that outbursts may have on the formation and evolution of the stellar-disk system.  Outbursts, for instance, can modify the chemistry of disks and envelopes and influence the planet formation process. The elevated temperatures they induce in the disk can temporarily shift at larger disk radii the location of various ice snowlines (Hubbard 2017). Additionally, outbursts are believed to contribute over 20\% of a star's total mass (Fischer et al. 2019, Cruz-S{\'a}enz de Miera et al. 2023, Giannini et al. 2024), and significantly impact the stellar luminosity. They also, at least partially, account for the wide spread observed in the relation between accretion and stellar luminosity (Fischer et al. 2023).

The wide variety of photometric variability described above makes unlikely that a single cause can be responsible for the observed outbursts. Indeed, from the theoretical point of view, a large number of mechanisms have  been proposed as the basis for outburst triggering, essentially falling into two main schools of  thought: 'internal' and 'external' triggering. Internal triggering mechanisms involve processes within the star-disk system, such as: thermal viscous instability (Bell \& Lin 1994; Nayakshin et al. 2024), gravitational and magnetorotational instabilities (Zhu et al. 2009; Bae et al. 2014; Kadam et al.
2020), disk fragmentation (Vorobyov \& Basu 2015), planet-disk interaction (Lodato \& Clarke 2004), and extreme evaporation of planets in hot thermally unstable protoplanetary disks (Nayakshin et al. 2023). External triggering mechanisms include: binary interactions (Bonnell \& Bastien 1992), capture of fragments of the native molecular cloud 
(Dullemond et al. 2019), and collisions of discs in stellar flybys  (Borchert et al. 2022; Dong et al. 2022).

While significant progress has been made in photometric observations and modeling, spectroscopic monitoring has not developed at a similar level (e.g. Contreras Pe{\~n}a et al. 2025; Giannini et al. 2022; Connelley \& Reipurth 2018), despite crucial for distinguishing between various types of variability and, consequently, for discovering genuine young eruptive variables. 
In the last decade, the Gaia Photometric Science Alert system\footnote{https://gsaweb.ast.cam.ac.uk/alerts/alertsindex} (Hodgkin et al. 2021) has provided alerts of significant photometric variability for about 850 YSOs or candidates. 
In depth studies of the photometric and spectroscopic characteritics of Gaia-alerted sources have led to the identification of a number of new FUors (Gaia17bpi, Hillenbrand et al. 2018; Gaia18dvy, Szegedi-Elek et al. 2020; and Gaia21elv, Nagy et al. 2023) and EXors (Gaia18dvz, Hodapp et al. 2019; Gaia20eae, Cruz-S{\'a}enz de Miera et al. 2022, Ghosh et al. 2022; Gaia19fct, Park et al. 2022; Gaia23bab, Giannini et al. 2024, Nagy et al. 2025) as well as sources belonging to new subclasses (Gaia19ajj, Hillenbrand et al. 2019; Gaia19bey, Hodapp et al. 2020;  Gaia21bty, Siwak et al. 2023; Gaia18cjb, Fiorellino et al. 2024).

This paper is aimed at examining the eruptive accretion phenomenon from a more statistical perspective, being based on Large Binocular Telescope (LBT) spectroscopic observations of a sample of 16 Gaia-alerted sources. Rather than focusing on an in-depth analysis of individual sources, our goal is to define the key photometric and spectroscopic characteristics that identify eruptive variables, specifically their light curve features and their accretion luminosity and mass accretion rate.
Our paper is organized as follows: Section \ref{sec:sec2} presents our sample of Gaia-alerted sources and the all-sky surveys used for the photometric analysis; Section \ref{sec:sec3} describes the observations and data reduction procedures; Section\,\ref{sec:sec4} presents the multi-wavelength light curves; Section\,\ref{sec:sec5} 
illustrates the methods adopted to compute the extinction in each source;  Section\,\ref{sec:sec6} describes the Spectral Energy Distribution of our sources, along with an analysis of the optical and infrared colors; Section\,\ref{sec:sec7}  presents the spectra and the determination of the stellar and accretion parameters; Section\,\ref{sec:sec8} discusses our results, and  Section\,\ref{sec:sec9} presents the conclusions. Additionally, in Appendix\,\ref{appendix:A} we give notes on the individual sources and in Appendix\,\ref{appendix:B} we present the photometric data. Finally, in Appendix\,\ref{appendix:C}  and \ref{appendix:D}, the observed spectra and the tables of the line fluxes are shown.

\section{The sample}\label{sec:sec2}

\begin{table*}
\small
\footnotesize
\center
\caption{\label{tab:tab1} The sample.}
\begin{tabular}{lcccccccccc}
\hline
\hline
ID& Source    & R.A.(J2000.0) & Decl.(J2000.0)  &  Region          &  Distance   & RUWE & Gaia               & Simbad  \\
  &           & (h:m:s) & ($^{\circ}$:$^\prime$:$^{\prime\prime}$) &  & (pc)        &       & classification$^1$ &name\\
\hline
1  & Gaia21bkw & 03:28:56.97  & +31:16:22.12   & NGC 1333            &   265$^a$        & 1.04  & YSO                & 2MASS J03285694+3116222\\
2  & Gaia22efa & 04:30:37.49  & +35:50:31.56   & Auriga-California   &   532            & 2.29  & YSO-c              & AKARI-IRC-V1 J0430375+355031  \\
3  & Gaia22bvi & 04:56:57.02  & +51:30:50.40   & L1438               &   209$^b$        & 2.16  & YSO                & V347 Aur  \\
4  & Gaia22ehn & 04:33:19.09  & +22:46:33.71   & LDN1536             &   152$^c$        & 4.82  & YSO-c              & IRAS 04303+2240  \\
5  & Gaia22dbd & 05:31:55.50  & $-$02:52:19.27 & Orion               &   345            & 1.17  & YSO                & UCAC4 436-010349  \\
6  & Gaia21arv & 05:35:05.62  & $-$05:29:22.38 & Orion               &   390            & 1.09  & YSO                & V407 Ori   \\
7  & Gaia23bri & 05:41:13.79  & +27:39:38.45   & -                   &  1586            & 1.04  & YSO-c              & 2MASS J05411378+2739385 \\
8  & Gaia21ebu & 06:31:36.86  & +04:51:04.32 & Rosette nebula        &  1383           & 1.62  & YSO                & V546 Mon   \\
9  & Gaia21aul & 18:30:06.18  & +00:42:33.30 & Serpens               &   379           & 2.74  & YSO                & IRAS 18275+0040  \\
10 & Gaia23bab & 19:04:26.68  & +04:23:57.37 & G38.3-0.9             & 900$^d$          & 1.00  & YSO-c              & 2MASS J19042667+0423575    \\
11 & Gaia23dhi & 19:35:56.88  & +16:28:23.38 & -                     &  2735            & 1.02  & YSO-c              &  ZTF J193556.88+162823.4\\
12 & Gaia24afw & 19:43:44.62  & +23:14:44.66 & -                     &  2150            & 2.51  & YSO-c              &  2MASS J19434462+2314447\\
13 & Gaia21faq & 20:41:20.62  & +39:29:32.06 & Cygnus X              &  1190            & 6.42  & YSO-c              & [KMH2014] J204120.63+392931.95 \\
14 & Gaia24beh & 20:50:50.39  & +44:50:11.44 & Cygnus                &  741             & 1.07  & YSO                &[KW97] 50-20 \\
15 & Gaia21fji & 21:03:58.14  & +50:14:40.20 & L998                  &  225/626$^c$     & 9.04  & YSO-c              & HBC727 \\
16 & Gaia21csu & 23:02:44.39  & +61:39:31.86 & LDN1218               &  834             & 1.45  & YSO                & ZTF J230247.47+613919.1\\
\hline\end{tabular}	
\footnotesize
\textbf{References.} $^a$Fiorellino et al.\,2021; $^b$Dahm \& Hillenbrand 2020;  $^c$Nagy et al., in preparation; $^d$Giannini et al. 2024.\\
\textbf{Note.} $^1$YSO-c stands for 'YSO candidate'. \\
\end{table*} 
  
\begin{table*}
\small
\center
\caption{\label{tab:tab2} Stellar parameters retrieved from the literature.}
\begin{tabular}{lcccccccccc}
\hline
\hline
Source   & Class & A$_V$  & SpT & \teff     &  \lstar    &   \lacc    &  \mstar       & \macc             &  Ref.\\
         &       & (mag)  &     &   (K) &  (\lsun)      &   (\lsun)  &  (\msun)
   & (\msunyr)        &         \\
\hline
Gaia21bkw & II     &  9.8 & M4  & 3190        &  0.45        & $<$ 5 10$^{-3}$ $^q$ &  0.25  & $<$ 1.5 10$^{-10}$ $^q$ & 1 \\
Gaia22bvi & flat   & 2-5  & M4 &  3200       &     -         &     -    &  0.35 &     -&  2\\      
Gaia21arv & flat   &  -   & K3 & 4180       &    -     &      -     &     -    &  -   &  3,4\\
Gaia23bab &  II   &  5.5$^b$-3.2$^q$ &  M1 &3630       &  0.72   & 3.7$^b$ & 0.4 & 2 10$^{-7}$$^b$  &5,6 \\

\hline
\end{tabular}	
\begin{quotation}
\textbf{References.} 1\,-\,Fiorellino et al. 2021; 2\,-\,Dahm \& Hillenbrand 2020; 3\,-\,Kounkel et al. 2019; 4\,-\,Gro{\ss}schedl et al. 2019; 5\,-\,Giannini et al. 2024; 6\,-\,Nagy et al. 2025.\\
\textbf{Note.} $^q$Value measured in quiescence;  $^b$Value measured in burst.
\end{quotation}
\end{table*}

Our initial sample was derived from the publicly available Gaia alerts catalog, specifically selecting sources classified\footnote{We considered both the Gaia 'class' and 'comment' columns in the Gaia Alert Index.} as 'YSO' or 'YSO candidate' (Hodgkin et al. 2021). While this selection increases the probability of focusing on genuine YSOs, it is important to note that a substantial number of potentially outbursting YSOs were likely excluded, as approximately 75\% of Gaia alerts currently lack a definitive classification.
From the initial set of about 850 objects, we further selected sources that showed a significant brightening  ($\sim$ 1$-$2 mag) in the $G$-band during the period between 2021 and 2024.  To ensure they are readily observable with the LBT, we further required them to be visible from the Northern hemisphere and have an average $G$ magnitude $\la$ 17.5. Approximately 60 objects met these criteria. Here we present a study of a subsample comprising 16 of these objects. Future observations are planned to extend the sample and will be presented in forthcoming papers.\\

In Table\,\ref{tab:tab1} we present our sample.  Along with the coordinates, location, and distance (Section\,\ref{sec:sec2.2}), we include the Gaia RUWE\footnote{The Gaia RUWE (re-normalised unit weight error) parameter is a powerful indicator of multiplicity. Typically, it approximates unity for well-behaved individual sources.} parameter, the classification (YSO or YSO candidate) obtained from the Gaia Alert Index, and the name provided in the Simbad\footnote{https://simbad.cds.unistra.fr/simbad/sim-fid} database. Out of the 8 YSO candidates, 3 objects, namely Gaia22ehn, Gaia23bri, and Gaia21faq, are present in the catalogue of Marton et al. (2019), which indicates for them a probability of being a YSO of 80\%, 96\%, and 86\%, respectively. \\
Two of the investigated sources, Gaia22bvi and Gaia23bab, have been previously classified as eruptive variables. Gaia22bvi (i.e. V347 Aur), is an isolated pre-main sequence star exhibiting periodic brightness variations on timescales of $\sim$\,150 days likely related to accretion instabilities (Dahm \& Hillenbrand 2020). Gaia23bab is a new EXor that has experienced three outbursts of $\sim$ 2 mag in roughly 10 years (2013, 2017, and 2023, Giannini et al. 2024, Nagy et al. 2025).  The spectrum discussed here was acquired after the most recent outburst in 2023, when the source had returned to a quiescent state.
The remaining 14 objects are presented here for the first time. Only two of these objects, Gaia21bkw and Gaia21arv, have been mentioned in two articles not focused on variability studies. Specifically, Gaia21bkw is a low-mass source in Perseus and is part of a binary system with a separation $\sim$ 2\arcsec\, (Fiorellino et al. 2021, source \#174), while Gaia21arv is included in the APOGEE-2 survey, which examined close companions of YSOs in Orion (Kounkel et al. 2019).

Table\,\ref{tab:tab2} summarizes the stellar and accretion parameters we compiled from the literature for these four sources (including their evolutionary class, visual extinction, spectral type, effective temperature, stellar and accretion luminosity, stellar mass and mass accretion rate). For the subsequent analysis, we will adopt the literature stellar parameters (\teff, \lstar, \mstar) of these sources . The extinction and accretion parameters, which are potentially variable, will be newly estimated using the LBT spectra.

Assuming all our sources are YSOs, we have determined their evolutionary stage using near- and mid-infrared photometry from publicly available surveys (Section\,\ref{sec:sec2.1}) obtained during their quiescent phases.
We calculated the spectral index $\alpha = \frac{d\log (\lambda F_\lambda)}{d\log 
(\lambda)}$ using photometry at 2.16\,$\mu$m from 2MASS\footnote{https://irsa.ipac.caltech.edu/Missions/2mass.html}  (Skrutskie et al. 2006)  or LBT-LUCI\footnote{https://scienceops.lbto.org/luci/} (Seifert et al. 2003) and either 
22\,$\mu$m (AllWISE) or 24\,$\mu$m (Spitzer-MIPS\footnote{https://www.spitzer.caltech.edu/}, Rieke et al. 2004)). Based on the classification scheme by Greene \& Lada (1996), our sample includes 3 Class I sources ($\alpha > 0.3$), 2 flat spectrum sources ($-0.3 \le \alpha \le 0.3$), and 9 Class II sources ($\alpha \le -0.3$). For two sources, Gaia21arv and Gaia21csu, we could not compute $\alpha$. This was due to the absence of 22/24 $\mu$m photometry for Gaia21arv, and because quiescent near-infrared photometry is not available for Gaia21csu. The derived values of $\alpha$ for the analyzed sources and the correspondent Class are presented in Table\,\ref{tab:tab5}.

\subsection{Public surveys photometry}\label{sec:sec2.1}
To better understand the historical variability of our targets, we have gathered data from several public photometric surveys, including optical data from Gaia, Pan-STARRS 
and ZTF and infrared data from NEOWISE. 

Pan-STARRS generally offers light curves in the $grizy$ filters from 2009 to 2014, with measurements roughly every one to two years. Since 2015, optical light curves have been more frequently sampled in the Gaia $G$ filter ($\lambda_{eff}$ = 6251.50 Å), typically with four to ten observations per year. The $gri$ filters of ZTF have provided better sampling since 2018, with data points ranging from daily to monthly intervals. The NEOWISE survey provides the 3.4 $\mu$m and 4.6 $\mu$m magnitudes since mid-2014 until 2024, typically twice per year. 
Additional photometry is provided by 2MASS in the $JHK_s$ bands, and, for  some of our sources, by 
Akari\footnote{https://www.ir.isas.jaxa.jp/AKARI/Archive/} (Murakami et al. 2007) at 9, 18, 65, 90, 140, and 160 $\mu$m, by AllWISE at 3.4, 4.6, 12, and 22 $\mu$m, and by Spitzer at 3.6, 4.5, 5.8, 8.0, and 24 $\mu$m.

\subsection{Distances}\label{sec:sec2.2}
As shown in Table\,\ref{tab:tab1}, distance estimates are available in the literature for Gaia21bkw, Gaia22bvi,  Gaia22ehn, Gaia23bab, and Gaia21fji. For the other sources, we determined the distance using the following method. The distance of each source was estimated by considering the distance of the nearby clusters taken from Hunt \& Reffert (2024). In columns 1 to 3 of Figures \ref{fig:fig1} and \ref{fig:fig2} we show\,: 1) the spatial distribution of the cluster members and targets, including proper motions represented as arrows; 2) the distance distributions alongside the photogeometric distances of the targets, with their uncertainties, and, 3) the proper motion distributions.
To further validate distances and potential cluster membership, we additionally examined the extinction distribution toward the targets (Figures \ref{fig:fig1} and \ref{fig:fig2}, column 4), using 3D dust maps from the Python package \textit{dustmaps} (Green 2018). For YSOs with distances less than 1.2 kpc, we used the dust map of Edenhofer et al. (2024), for more distant YSOs, we used the dust map \textit{Bayestar19} of Green et al. (2019). We also estimated and validated the distance by identifying the location where the derivative of the extinction vs. distance relation is largest. If multiple extinction rises are visible, we adopted the value closest to the cluster distance. In Appendix\,\ref{appendix:A} we provide a detailed discussion of the distance determination for each source.

\begin{figure}
    \centering
    \includegraphics[width=0.8\textwidth]{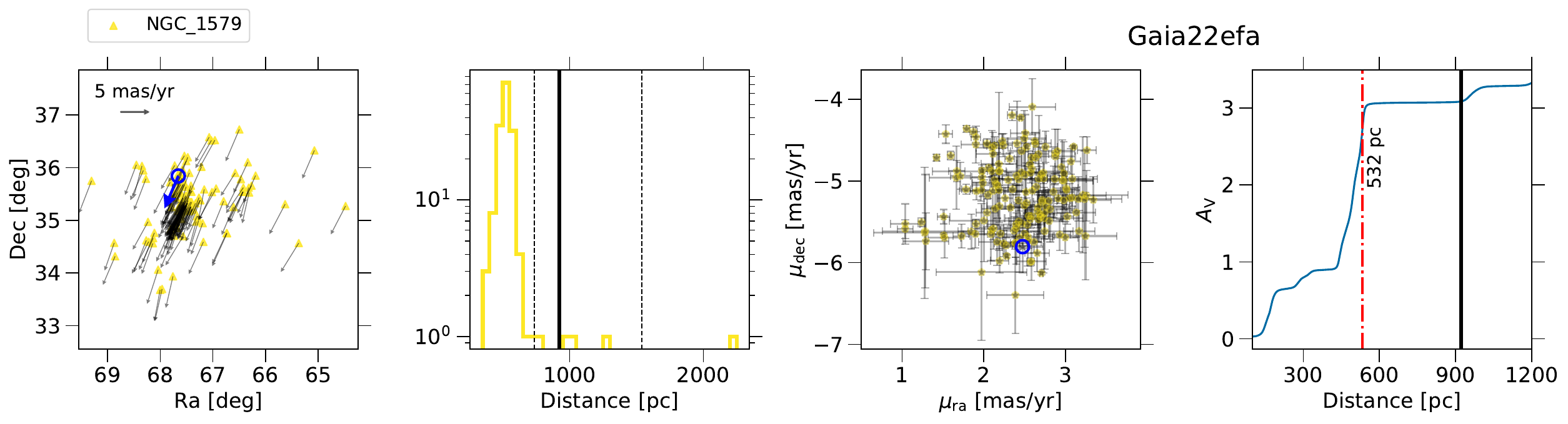}
    \includegraphics[width=0.8\textwidth]{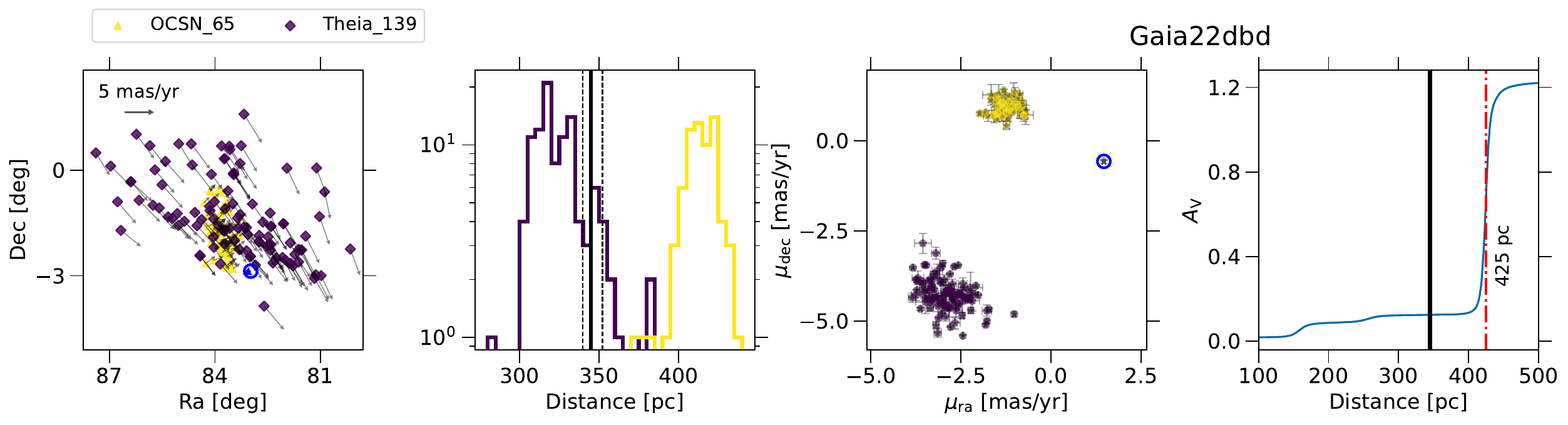}
    \includegraphics[width=0.8\textwidth]{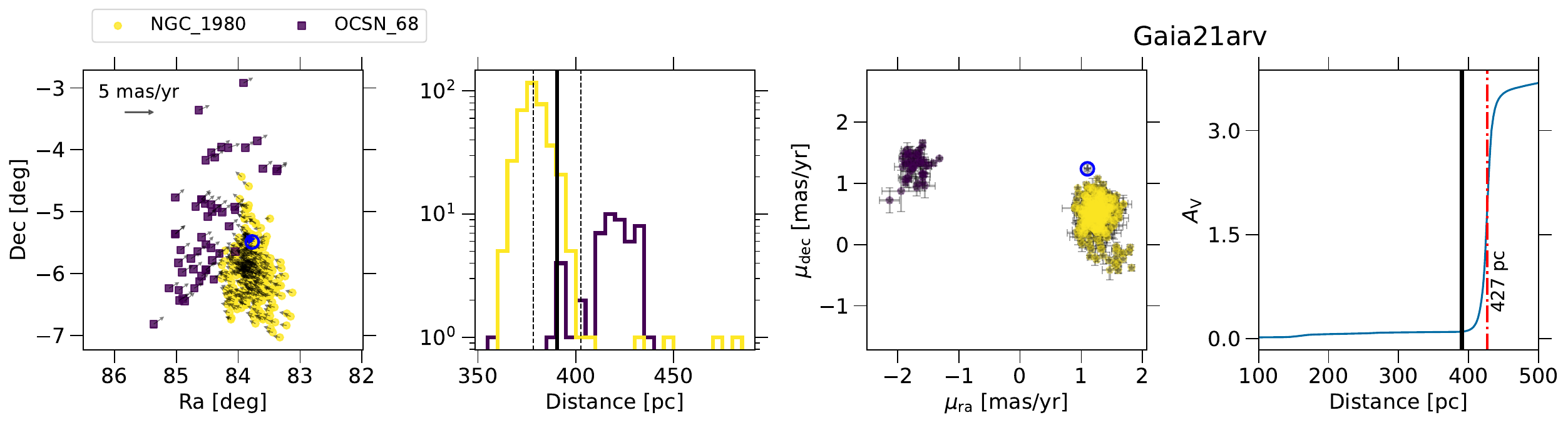}
    \includegraphics[width=0.8\textwidth]{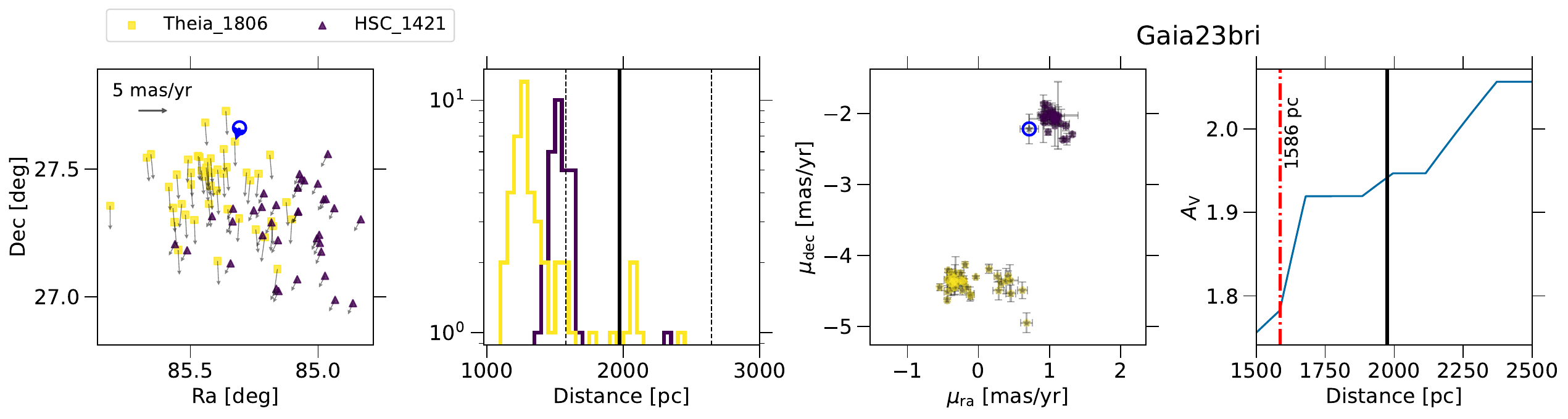}
    \includegraphics[width=0.8\textwidth]{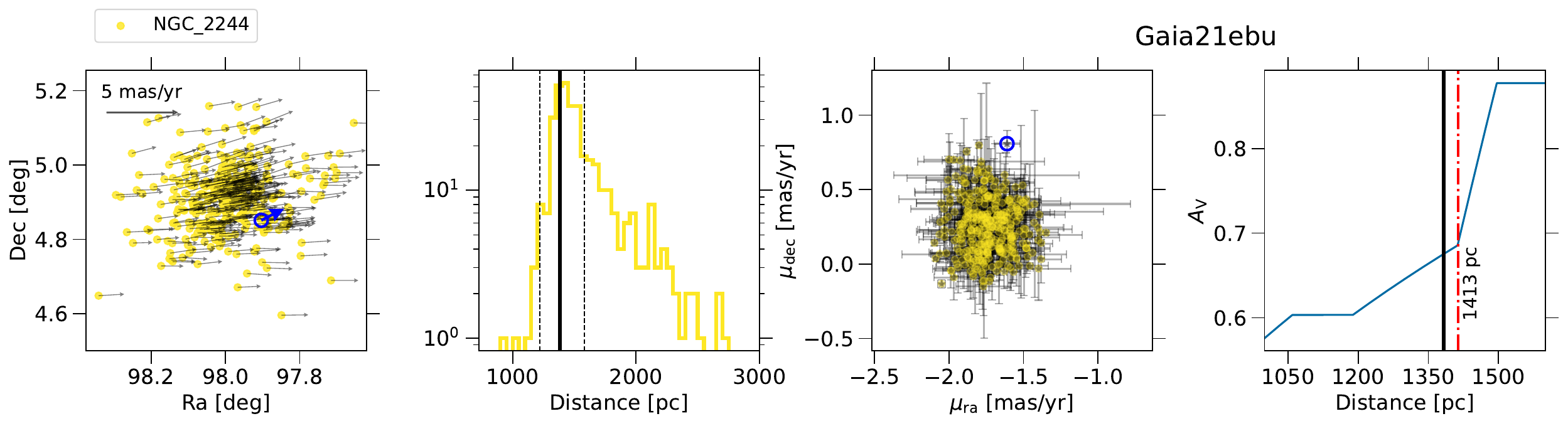}
    \caption{ \textit{Columns 1-3}: Clusters from Hunt \& Reffert (2024) close to the target YSOs. Different clusters are depicted with different colors and symbols, as indicated on top. Col.\,1: Spatial distribution of cluster members. The target YSO is shown as a blue circle. Arrows indicate proper motions. Col.\,2: Histogram of the distance distribution. The solid and dashed vertical line represents the distance and its uncertainties of the target YSO from Bailer-Jones et al. (2021). Col.\,3: Proper motion distribution with uncertainties. The target YSO is shown as a blue circle. \textit{Column 4} : Extinction distribution along the line of sight toward the target. The solid vertical line shows the photogeometric distance from Bailer-Jones et al. (2021). Dashed vertical lines mark the distance of the largest extinction increase. See Section \ref{sec:sec2.2}. for details.}
    \label{fig:fig1}
\end{figure}

\begin{figure}
    \centering
    \includegraphics[width=0.8\textwidth]{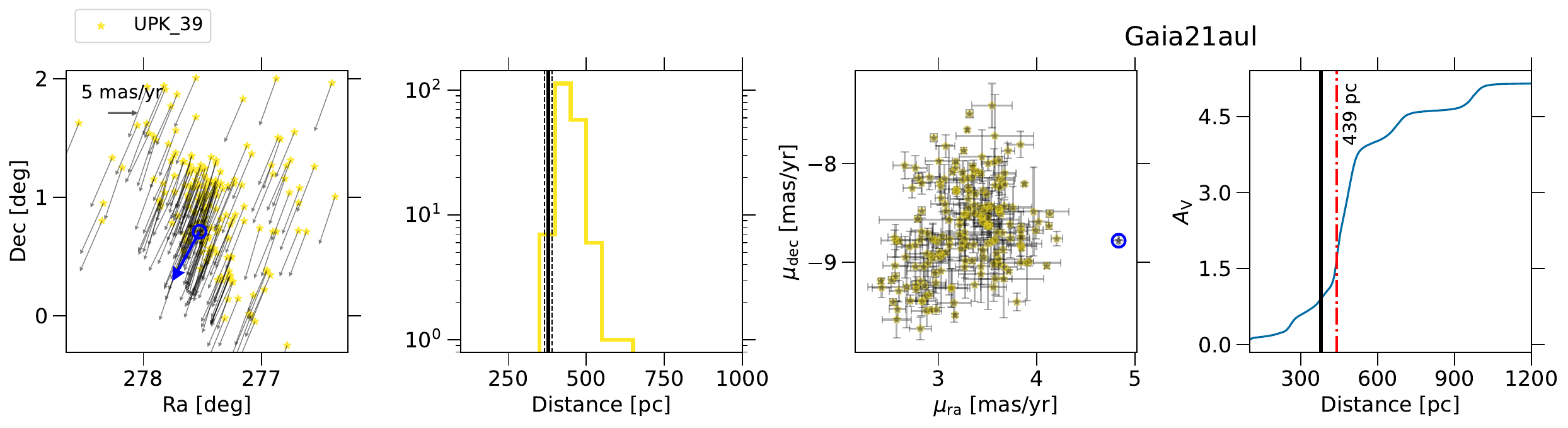}
    \includegraphics[width=0.8\textwidth]{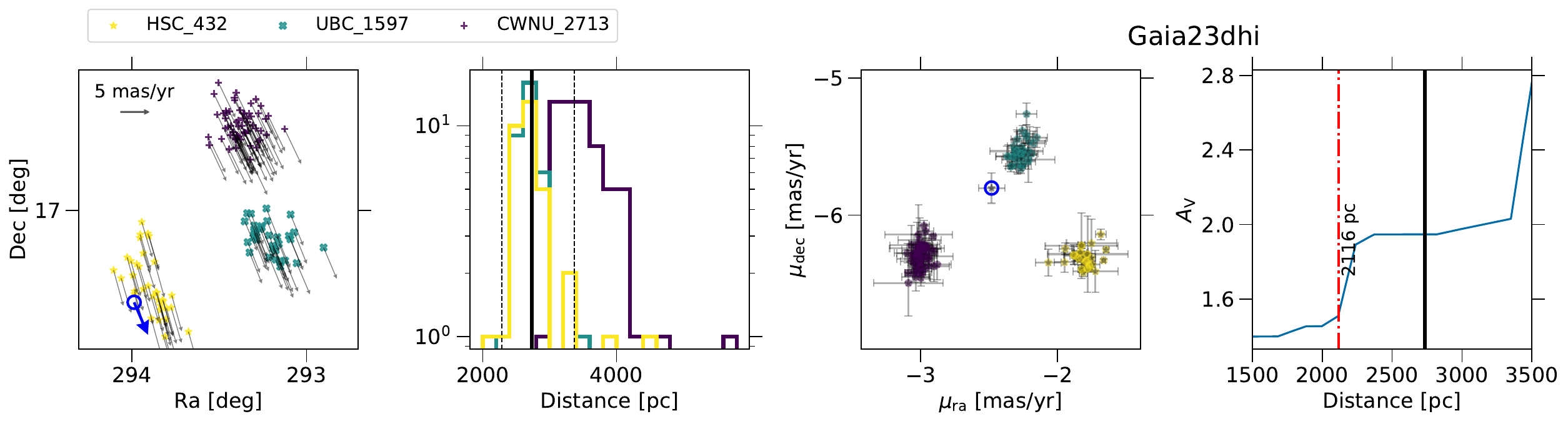}
    \includegraphics[width=0.8\textwidth]{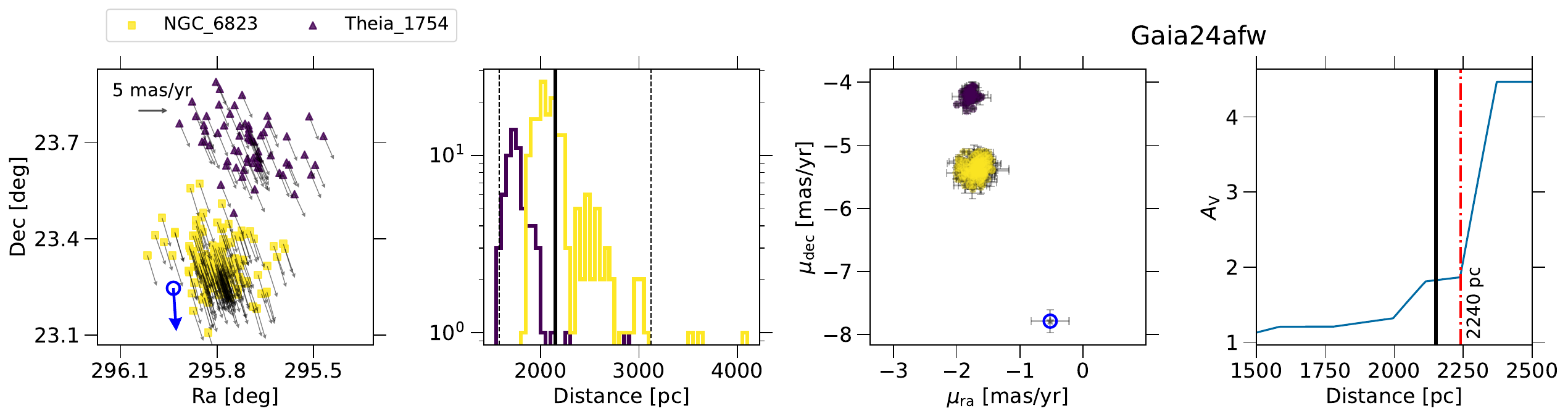}
    \includegraphics[width=0.8\textwidth]{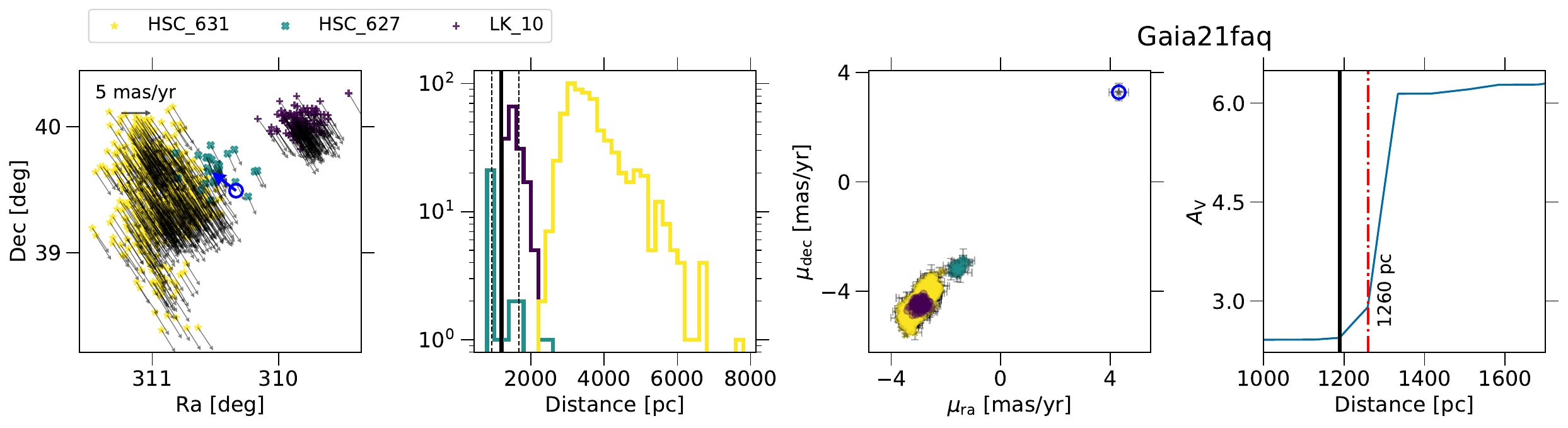}
    \includegraphics[width=0.8\textwidth]{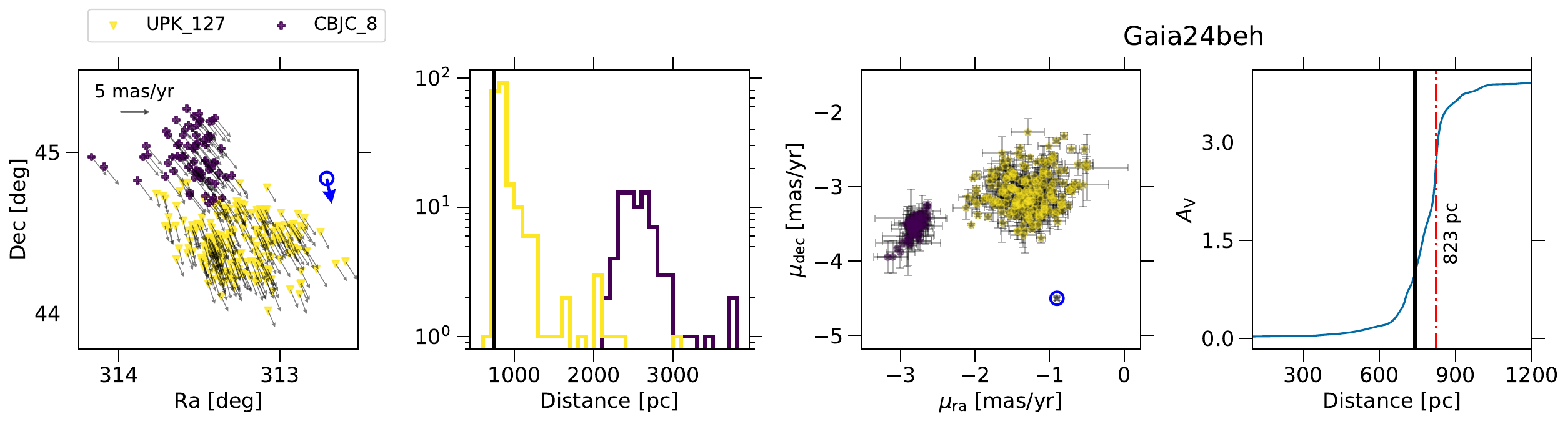}
    \includegraphics[width=0.8\textwidth]{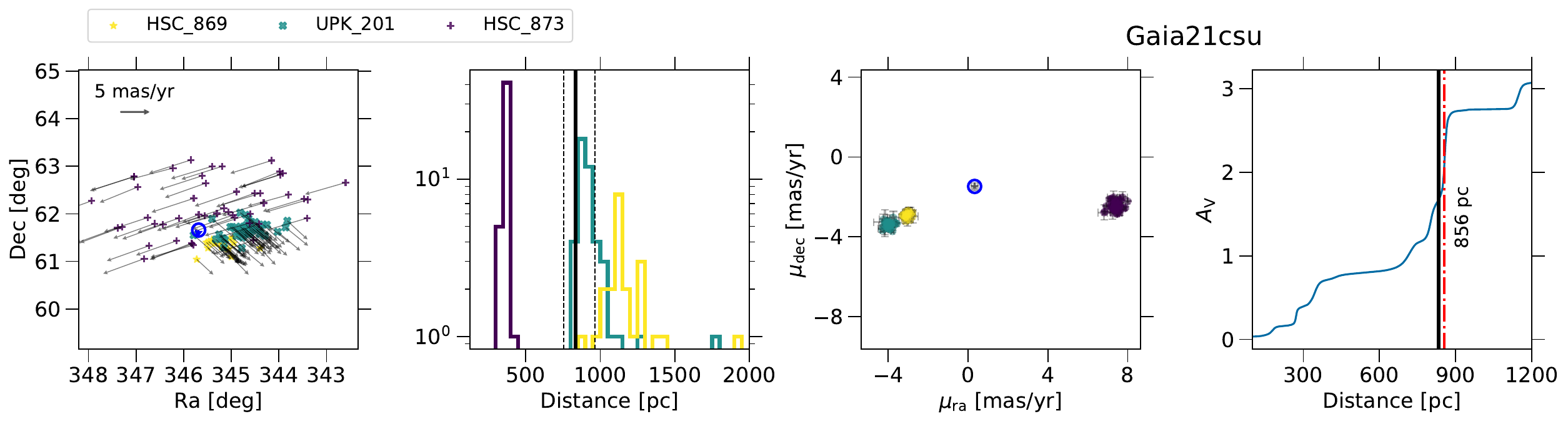}
    \caption{Continuation of Figure \ref{fig:fig1}.}
    \label{fig:fig2}
\end{figure}

\section{Observations and data reduction\label{sec:sec3}}

\begin{table*}
\small
\center
\caption{\label{tab:tab3} Journal of observations.}
\begin{tabular}{cccccccc}
\hline
\hline
ID  & Source     & \multicolumn{3}{c}{MODS}       & \multicolumn{3}{c}{LUCI}              \\
    &      &  Date       &  MJD  & T$_{int}$ (B/R)& Date     &  MJD  & T$_{int}$ (zJ/HK) \\
    &      &             &       & (s)       &               &       & (s)        \\
\hline 
 1 &Gaia21bkw & 2023 Mar 05 & 60008   &  -/1800   & 2021 Oct 15   & 59502.3 &  3600/2700 \\
   &          &    -        &   -     &  -        & 2023 Feb 28   & 60003.1 &  600/600   \\
 2 &Gaia22efa & 2023 Oct 17 & 60234.4 & 1800/1800 & 2023 Nov 23   & 60271.4 &  450/375   \\
 3 &Gaia22bvi & 2023 Oct 19 & 60236.3 & 1800/1800 & 2023 Nov 22   & 60270.5 &  750/1200  \\
 4 &Gaia22ehn & 2023 Oct 21 & 60238.3 & 1800/4500 &    -          &  -    &       -     \\
 5 &Gaia22dbd & 2024 Jan 10 & 60319.3 & 1800/1800 & 2024 Jan 09   & 60318.2 & 1800/1800  \\
 6 &Gaia21arv &    -        &   -     &      -    & 2021 Oct 15   & 59502.4 & 1800/1800  \\
 7 &Gaia23bri & 2023 Oct 19 & 60236.4 &  900/1800 & 2023 Nov 23   & 60271.4 &   2925/- \\
 & &          -             &      -  &   -       & 2024 Jan 09   & 60318.4 & 4050/3600 \\  
 8 &Gaia21ebu & 2024 Jan 10 & 60319.4 & 1800/1800 & 2024 Jan 09   & 60318.3 & 1800/1800  \\
 9 &Gaia21aul & 2023 Jun 04 & 60099.3 &  600/600  & 2023 Jun 04   & 60099.2 & 1780/160   \\
10 &Gaia23bab & 2024 Apr 11 & 60411.5 & 1800/4050 & 2024 Apr 14   & 60414.4 & 1800/1800  \\
11 &Gaia23dhi & 2024 Jun 07 & 60468.4 & 1800/900  & 2024 Jun 10   & 60471.4 &  900/900      \\
12 &Gaia24afw & 2024 Jun 07 & 60468.5 &  900/1800 &     -         &    -   &       -   \\
13 &Gaia21faq & 2023 Jun 04 & 60099.4 & 1350/-    & 2023 Jun 04   & 60099.3&  550/1800  \\
14 &Gaia24beh & 2024 Jun 08 & 60469.4 & 1800/1800 &    -          &  -     &       -   \\
15 &Gaia21fji & 2023 Oct 19 & 60236.2 & 1800/1800 &    -          &  -     &         -   \\
16 &Gaia21csu$^a$ &2023 Oct 19 & 60236.3 & 1350/1800 & 2024 Jan 09& 60318.1& 3600/3600  \\
\hline\end{tabular}	
\begin{quotation}
\textbf{Notes.}  $^a$For this source we obtained additional photometry on  2022 Dec 17 (MJD=59930).
\end{quotation}
\end{table*}

Our observations were conducted between October 2021 and June 2024 as part of a filler program at the 8.4m Large Binocular Telescope, located at Mount Graham, Arizona, USA. We acquired optical and near-IR spectra of our targets using the Multi-Object Double Spectrograph (MODS\footnote{https://scienceops.lbto.org/mods/}, Pogge et al. 2010) and the LBT Utility Camera in the Infrared (LUCI, Seifert et al. 2003), respectively. 

The journal of observations is given in Table\,\ref{tab:tab3}.
We successfully obtained simultaneous optical and near-infrared spectra for Gaia21aul and Gaia21faq. Unfortunately, simultaneous observations for the other targets were not possible due to technical issues or unfavorable weather conditions. Spectra for Gaia22dbd, Gaia21ebu, Gaia23bab, and Gaia23dhi  were obtained on subsequent or close nights. 
For Gaia21bkw, Gaia22efa, Gaia22bvi, Gaia23bri, and Gaia21csu, there is a separation of one to some months between the optical and near-infrared spectra. For the remaining targets (Gaia22ehn, Gaia21arv, Gaia24afw,  Gaia24beh, and Gaia21fji) we have only obtained either the optical or the near-infrared spectrum. Additionally, we acquired an extra near-infrared spectrum for both Gaia21bkw and Gaia23bri.

The MODS observations were conducted using the dual grating mode, which covers 
a spectral range of 350 to 950 nm across the Blue and Red channels. We used a
0$\farcs$80 slit, resulting in a spectral resolution of $\Re\,\sim$\,1500 in the Blue channel and $\Re\,\sim$\,1800 in the Red channel. To minimize any wavelength-dependent effects on the slit transmission, the slit angle was aligned with the parallactic angle. The specific observation dates and the corresponding integration times can be found in Table\,\ref{tab:tab3}.

LUCI observations were carried out with the G200 low-resolution grating coupled with the 0$\farcs$75 slit. Two datasets were acquired with the standard ABB'A'  nodding technique using the $zJ$ and $HK$ grisms. This resulted in a final spectrum spanning the 1.0$-$2.4 $\mu$m wavelength range, with a spectral resolution of $\Re\,\sim$\,1500. The integration time for each
individual source can be found in Table\,\ref{tab:tab3}.

The data reduction process was carried out using the Spectroscopic Interactive Pipeline and 
Graphical Interface  (SIPGI, Gargiulo et al. 2022), which is specifically developed for reducing long-slit MODS and LUCI spectra.

For each MODS spectral image, the data reduction involved several steps: dark and bias correction, bad-pixel mapping, flat-fielding, and the extraction of the one-dimensional spectrum by integrating along the stellar trace in the spatial direction. Wavelength calibration was achieved using arc lamp spectra.

To ensure accurate alignment between the Blue and Red spectral segments, we verified the inter-calibration by superimposing the spectral range of 5300 to 5900 \AA\, which is common to both channels. Our analysis indicated that the Blue and Red spectra were optimally aligned in all cases, without requiring any further corrections.

The raw LUCI spectral images were cleaned for bad pixels, flat-fielded, sky-subtracted, and corrected for optical distortions
in both the spatial and spectral directions. Telluric absorption features were subsequently removed using the normalized spectrum of a telluric standard star, from which its hydrogen recombination lines were also removed.
Finally, wavelength calibration was performed using arc lamp spectra.

Flux losses within the slits were accounted for by utilizing photometry from acquisition images in the $grzi$ and $JHK_s$ bands, each obtained from the combination of 5 dithered frames. 
The photometric points of the target were obtained by using as references all the visible stars in the 6$^{\prime} \times 6^{\prime}$ (4$^{\prime} \times 4^{\prime}$)  MODS (LUCI) field. For these reference stars, we accessed publicly available optical/near-IR photometry from either the Pan-STARRS catalog, or the 2MASS catalog. The resulting magnitudes are given in Appendix \ref{appendix:B}.

Given that the optical and near-infrared spectra for most sources were obtained several months apart, we chose not to re-align the MODS and LUCI spectra. This decision was made because a variability of a few tenths of magnitude could have occurred between the two observation dates. For the two sources with nearly simultaneous MODS and LUCI observations, the alignment between the optical and near-infrared spectra was within a few percent and no additional adjustments were necessary.

\section{Description of the light curves}\label{sec:sec4}

\begin{figure}[ht!]
\includegraphics{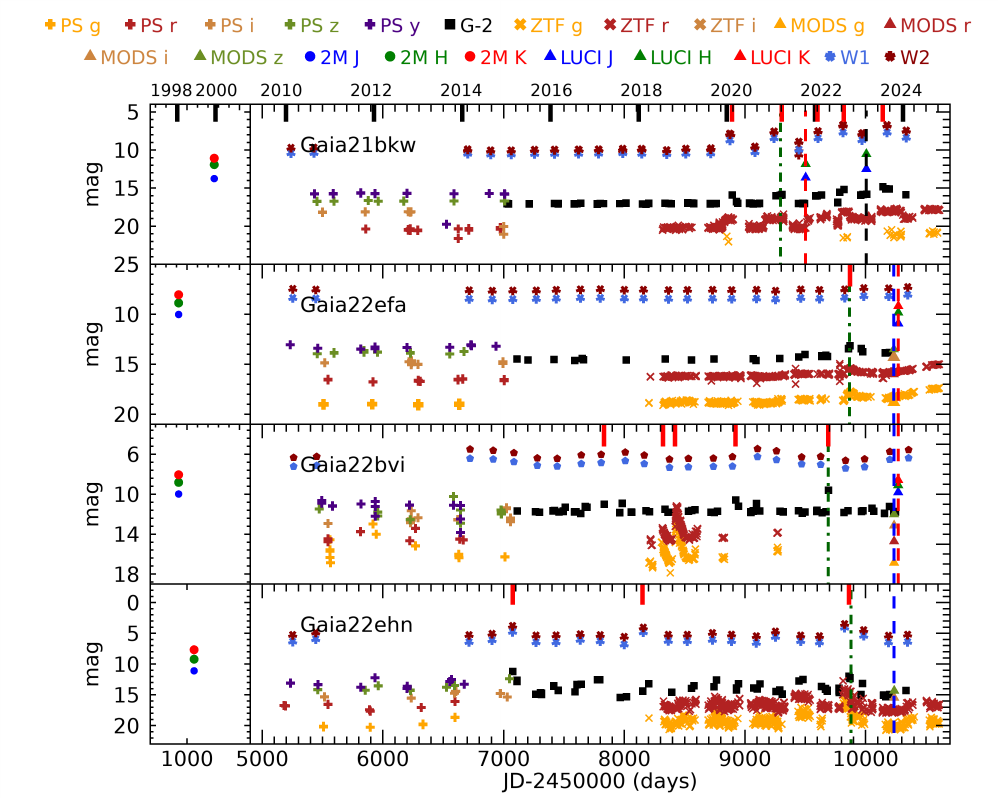}
\caption{\label{fig:fig3} Light curves of Gaia21bkw, Gaia22efa, Gaia22bvi, and Gaia22ehn. Symbols and colors identify data points from different surveys as reported in the legend.
Black ticks indicate the date of January 1st of the year displayed on top. Vertical dashed lines mark the dates of the LBT observations (blue: MODS, red: LUCI, black: MODS/LUCI spectra taken in close dates). The green dot-dashed line marks the date of the Gaia alert. The red ticks on top indicate the dates of optical bursts (see also Table\,\ref{tab:tab7}).}
\end{figure}

\begin{figure}[ht!]
\includegraphics{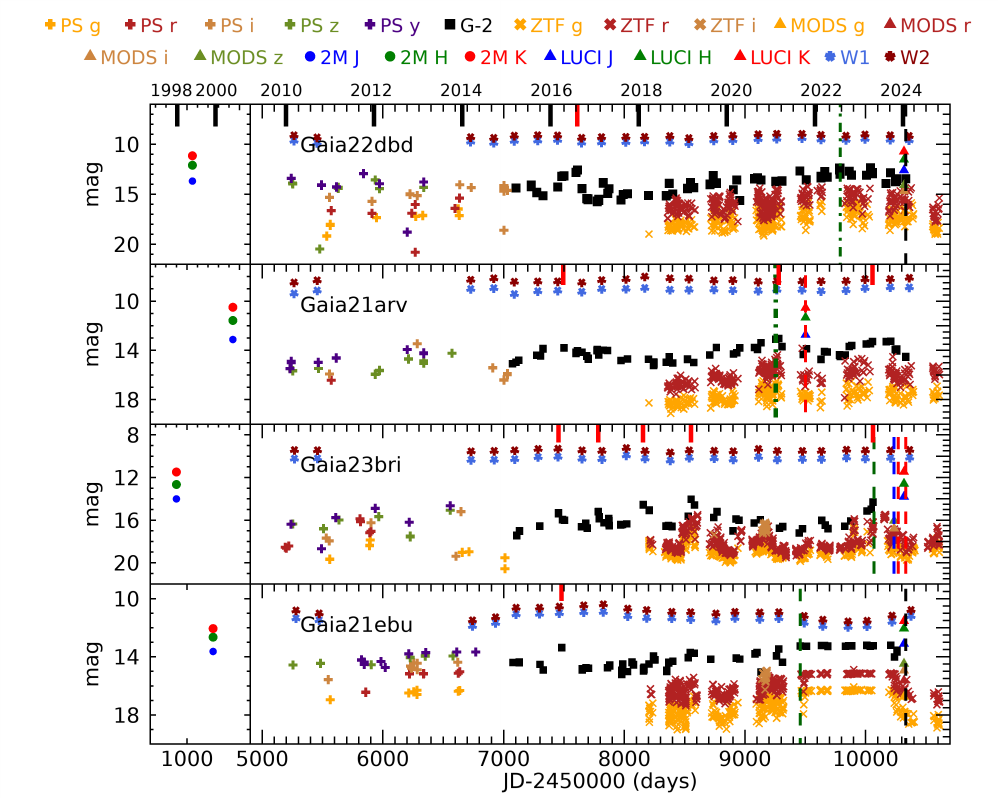}
\caption{\label{fig:fig4} Light curves of Gaia22dbd, Gaia21arv, Gaia23bri, and Gaia21ebu. Symbols and colors are the same as in Figure\,\ref{fig:fig3}.}
\end{figure}

\begin{figure}[ht!]
\includegraphics{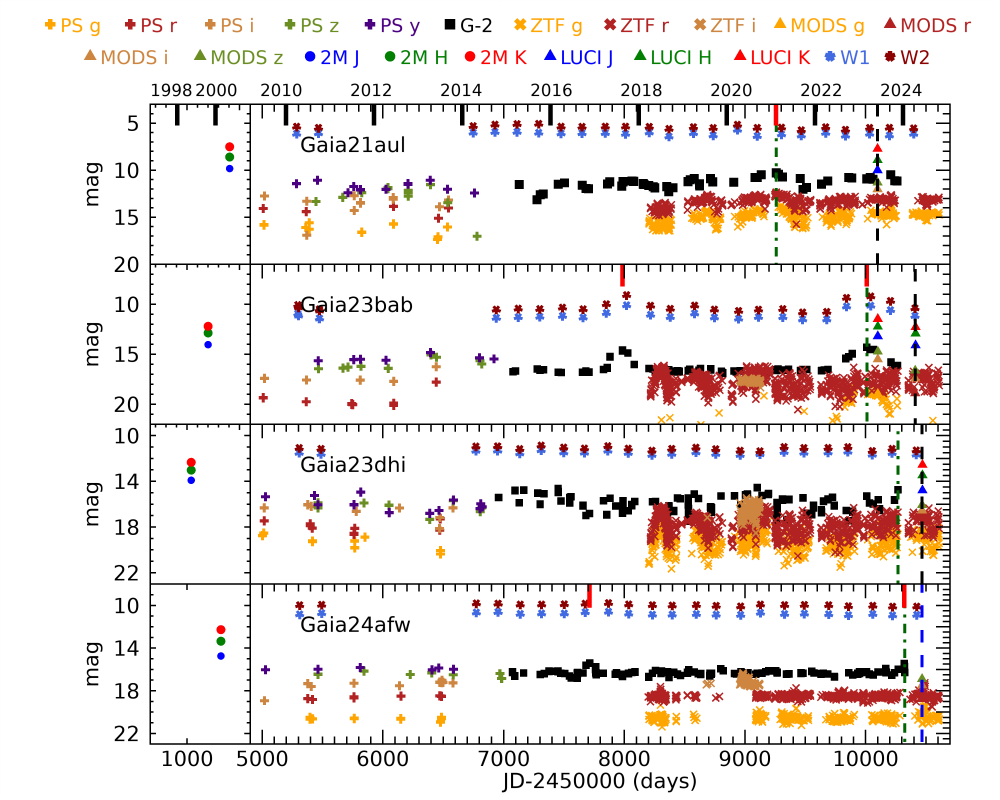}
\caption{\label{fig:fig5} Light curves of Gaia21aul, Gaia23bab, Gaia23dhi, and Gaia24afw. Symbols and colors are the same as in Figure\,\ref{fig:fig3}.}
\end{figure}

\begin{figure}[ht!]
\includegraphics{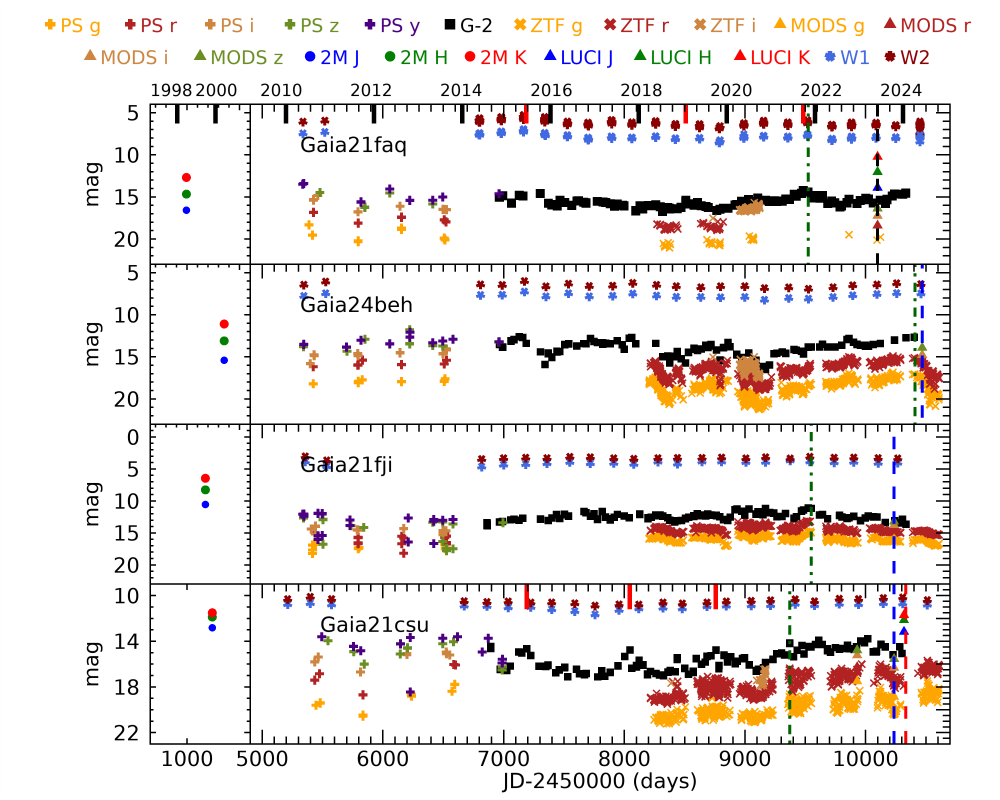}
\caption{\label{fig:fig6} Light curves of Gaia21faq, Gaia24beh, Gaia21fji, and Gaia21csu. Symbols and colors are the same as in Figure\,\ref{fig:fig3}. }
\end{figure}

The assembled light curves for the targets are presented in Figures \ref{fig:fig3}$-$\ref{fig:fig6}. These figures include photometric data gathered from public surveys (as detailed in Section\,\ref{sec:sec2.1}) along with the LBT photometry provided in Appendix\,\ref{appendix:B}. The NEOWISE light curves show a photometric point every six months, obtained by averaging the single-epoch photometry taken in each season. The single epoch data points have been checked for saturation and, in needed, corrected following the saturation corrections of  Cutri et al. (2015).  The AllWISE photometric data obtained in 2010 at 3.4 $\mu$m and 4.6 $\mu$m have also been shown for completeness. In the forthcoming Section \ref{sec:sec8.1} we will discuss these light curves from a general point of view, while here we present the light curves of the individual sources.

\begin{itemize}
    \item {\bf Gaia21bkw} was alerted by Gaia on March 19, 2021, due to a rapid brightening of 1 mag in the $G$-band. It's worth noting that a prior burst event occurred in February 2020 but no alert was issued at that time.
    Following this initial alert, the source exhibited three additional burst episodes in January 2022, August 2022, and March 2023. 
    The most recent event was the most significant, reaching a peak magnitude of 2.2 above the quiescent level of $\sim$ 19 mag in the $G$-band.
    All observed outbursts have had durations ranging from about six months to a year and have also been observed by NEOWISE, with amplitudes similar to those in the optical. The first LUCI spectrum was taken during a period of quiescence, while subsequent MODS/LUCI spectra were obtained during a rising phase of the last outburst, although not at its peak.
    \item {\bf Gaia22efa}. The Gaia alert was issued on August 6, 2022, following a brightening event of approximately one magnitude. After this episode the source returned to a quiescent state before beginning to increase in brightness again in 2024. This new rise, which started after the conclusion of the Gaia mission, has been observed by ZTF in the $g$ and $r$ bands. Our observations cover the initial part of this renewed brightening phase. 
    \item {\bf Gaia22bvi}. The Gaia alert was released on April 22, 2022. However, the source had exhibited earlier brightness variations of smaller amplitudes, which were also detected by NEOWISE and ZTF. This latter, in particular, recorded two additional bursts in 2018 that were not sampled by Gaia. Our observations were conducted while the source was in a quiescent state.
    \item {\bf Gaia22ehn}. This source displays significant variability, with peak-to-peak amplitudes reaching up to 4 magnitudes in the $G$ band and 2 magnitudes in the NEOWISE bands. The alert was issued on October 10, 2022, but our observations, conducted approximately a year later, did not capture this peak as the source had returned to a quiescent state by then. Two further outbursts have occurred: the first in early 2015 at the beginning of the Gaia mission and a second in early 2018 particularly bright in the NEOWISE bands.   
    \item {\bf Gaia22dbd}. The source exhibited a significant brightening event in 2016 that lasted for about a year, which was not alerted by Gaia. After a period of roughly one year of reduced activity, the brightness steadily rose again, peaking in 2023, when the alert was issued. Our observations cover the initial part of a declining phase in January 2024.
    \item {\bf Gaia21arv}. Gaia alerted this source on February 9, 2021. Two other peaks were observed in April 2016 and August 2023. The NEOWISE light curve shows a magnitude modulation of several tenths of magnitude in both the 3.4 and 4.6 $\mu$m filters, which is also noticeable in the Gaia and ZTF light curves. It appears to consist of a longer timescale variability component, with an amplitude of approximately 2 magnitudes and a duration of about 3 years, upon which a faster variability with a period of roughly 200 days and an amplitude of about 1 magnitude is superimposed. Our observations have been obtained during the fading phase after the 2021 burst.
    \item {\bf Gaia23bri}. The alert was issued on April 27, 2023, but similar brightening events occurred between 2016 and 2019. The light curve shows stochastic variations, which are also visible in the NEOWISE light curve, though with a lower amplitude. We observed Gaia23bri with MODS around its last peak brightness and continued to monitor its decline over the following months using LUCI.
    \item {\bf Gaia21ebu}. The Gaia light curve indicates an initial peak around April 2016, though the monitoring around this event was somewhat limited. Subsequently, in late 2018, the brightness began a gradual increase of approximately 1.7 magnitudes over three years. This rise culminated in a plateau phase, which triggered the alert. This plateau lasted for about two years, from 2021 to 2023, until the end of the Gaia monitoring. Subsequently, ZTF observed a sharp decline in brightness. Our LBT spectra were obtained at the onset of this new declining phase. Interestingly, the NEOWISE light curve exhibits a different behavior compared to the optical variations. Instead, it shows gradual fluctuations with an amplitude of about 2 magnitudes on timescales of roughly 1000 days, contrasting with the observed optical changes.
    \item {\bf Gaia21aul}. This source presents lower amplitude variability. Noticeably, $g$ and $r$ band photometries recorded by PAN-STARRS during the 2009-2014 period appear to be at a lower brightness level compared to those measured by ZTF after 2017. The Gaia alert, issued on February 13, 2021, coincided with the peak of a gradual and sustained increase in brightness. Our observations were conducted during a period of lower brightness. The NEOWISE light curve exhibits irregular variability that does not follow the same pattern as the optical light curve.  
    \item {\bf Gaia23bab}. The light curve has been commented in detail by Giannini et al. (2024) and Nagy et al. (2025). Here we only remark that the new LBT spectra have been taken in the quiescent phase after the 2023 outburst.
    \item {\bf Gaia23dhi}. This source exhibits irregular variability in both amplitude and timescale. The alert, issued on November 21st, 2023, does not show a definite peak in brightness. Our observations were conducted when the source was at an intermediate level of brightness.
    \item {\bf Gaia24afw}. The light curve for this source appears to be relatively stable, showing a typical modulation of around 0.5 magnitudes. Notably, two bright peaks, each with an amplitude of approximately 1 magnitude and lasting about a month, were recorded in November 2016 and January 2024. The alert for this source was triggered following the latter peak, and our observations were conducted immediately after this event. 
    \item {\bf Gaia21faq}. This object exhibited an erratic brightening of approximately 2 magnitudes, a variability that is also clearly evident in the NEOWISE bands. The Gaia light curve shows three distinct peaks occurring around August 2015, January 2019, and September 2021, when the alert was issued. Subsequently, the source underwent a short period of fading before its brightness started to increase again, at which point we obtained our LBT spectra.    
    \item {\bf Gaia24beh}. Between 2014 and 2021, this source underwent erratic variability with a peak-to-peak amplitude reaching up to 3 magnitudes. Following this period, the brightness began to increase, culminating in a peak on April 2, 2024, which triggered the Gaia alert. Our observations roughly coincide with this peak. Furthermore, new ZTF photometry suggests that a sudden decline in brightness is currently underway.
    \item {\bf Gaia21fji}. This source exhibits a continuous brightness modulation with a peak-to-peak amplitude exceeding 2 magnitudes. This variability is also noticeable in the NEOWISE light curve, although with a smaller amplitude. The MODS spectrum was obtained when the source was close to its minimum brightness.
    \item {\bf Gaia21csu}. Also this source presents significant brightness variations, with peak-to-peak changes of more than 3 magnitudes in $G$. Peaks were registered in June 2015, October 2017 and October 2019. Following the last peak, the source started a continuous brightening trend, which triggered an alert on June 6, 2021. Since 2022, the light curve has reached a plateau phase, during which we acquired our LBT spectra.
\end{itemize}     

\section{Extinction}\label{sec:sec5}
As anticipated in Section\,\ref{sec:sec2}, the observed variability might be at least partially due to changes in extinction along the line of sight. Therefore, precisely measuring extinction during different brightness phases is essential to distinguish extinction- from accretion- driven  variability. In this Section  we introduce the three methods used to compute \av, deferring the detailed description of each method to the subsequent relevant Sections\,\ref{sec:sec6.2},\,\ref{sec:sec7.1},\, and  \ref{sec:sec7.4}.
Specifically, we used: 1) the 'photometric' method, based on the analysis of the near-infrared colors (Section\,\ref{sec:sec6.2}), 2) the optical 'continuum-fitting' method , which relies on the slope of the optical continuum (Section\,\ref{sec:sec7.2}), and 3) the 'line-fitting' method, which considers the \av\, value providing consistency among different accretion lines (Section\,\ref{sec:sec7.4}). The resulting \av\, values, presented in Table\,\ref{tab:tab4}, generally agree within 1 magnitude.
The 'continuum-fitting' method provides \av\, with a typical uncertainty of 0.2$-$0.3 mag. The other two methods are less robust. The 'line-fitting' method's reliability depends heavily on the number of the observed emission lines and on the uncertainty in measured fluxes and adopted empirical relations. The 'photometric' method assumes that the observed sources are low-mass T Tauri stars and that the effect of scattering as due to any residual circumstellar material is negligible. We estimate the uncertainty on \av\, of these two methods to be around 0.5$-$1 mag. As a general strategy, we will adopt the \av\, derived from the 'continuum' method whenever possible, and the 'line fitting' and 'photometric' methods in the other cases.

\section{Photometric analysis}\label{sec:sec6}
\subsection{SEDs}\label{sec:sec6.1}

\begin{figure}[ht!]
\includegraphics{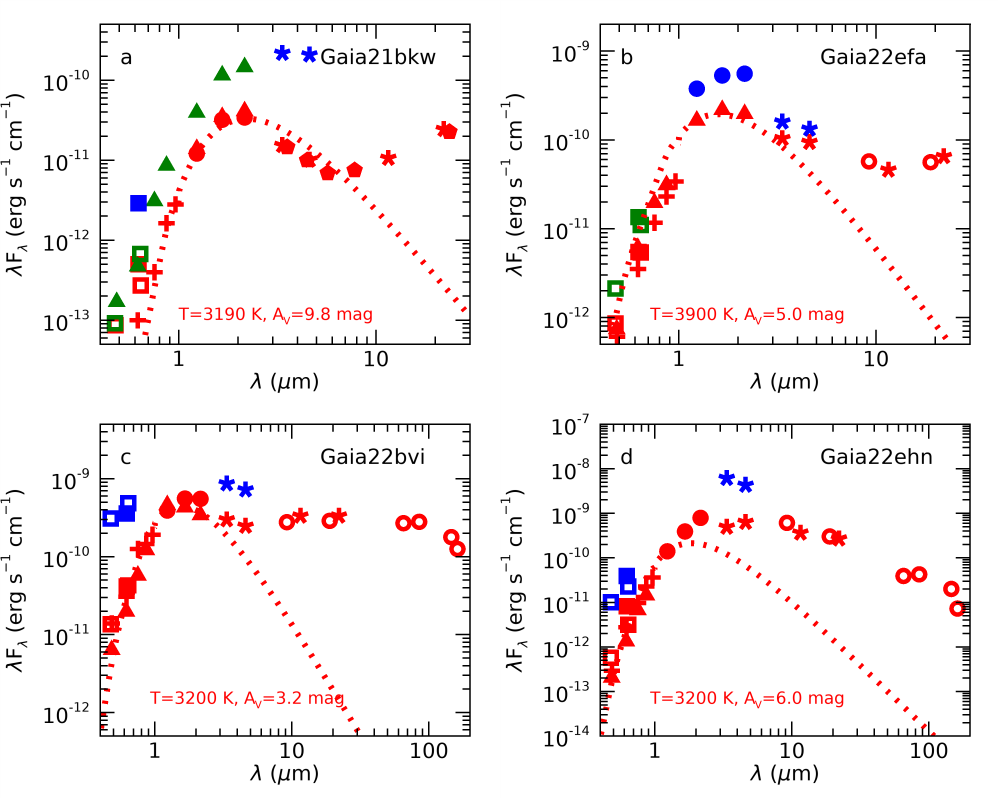}
\caption{\label{fig:fig7} Spectral Energy Distribution (SED) of Gaia21bkw (a), Gaia22efa (b), Gaia22bvi (c), and Gaia22ehn (d). Quiescence, intermediate and burst states are colored in red, green, and blue, respectively. Symbols have the following meaning. Crosses: PAN-STARRS; open squares: ZTF; filled squares: Gaia; filled triangles: LBT; filled circles: 2MASS; asterisks: AllWISE/NEOWISE; filled diamonds: Spitzer; open circles: Akari. The dotted line indicates the reddened black-body function representative of the stellar emission. The adopted temperature and extinction are reported as well.}
\end{figure}

\begin{figure}[ht!]
\includegraphics{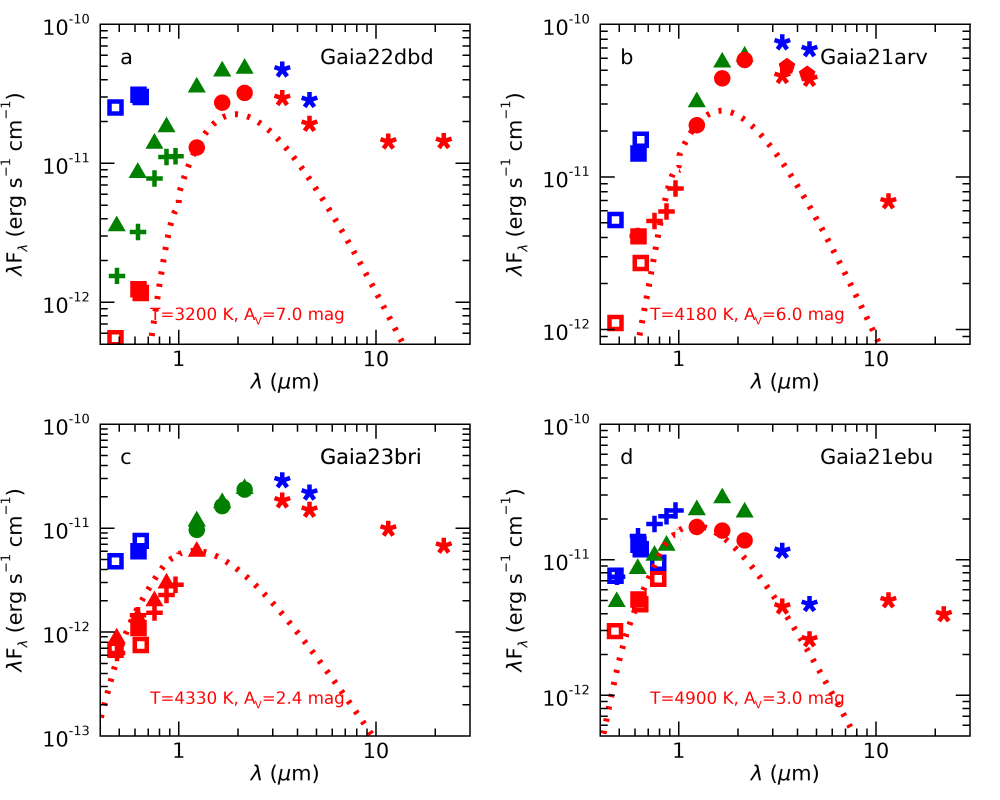}
\caption{\label{fig:fig8} As in Figure\,\ref{fig:fig7} for  Gaia22dbd (a), Gaia21arv (b), Gaia23bri (c), and Gaia21ebu (d).}
\end{figure}

\begin{figure}[ht!]
\includegraphics{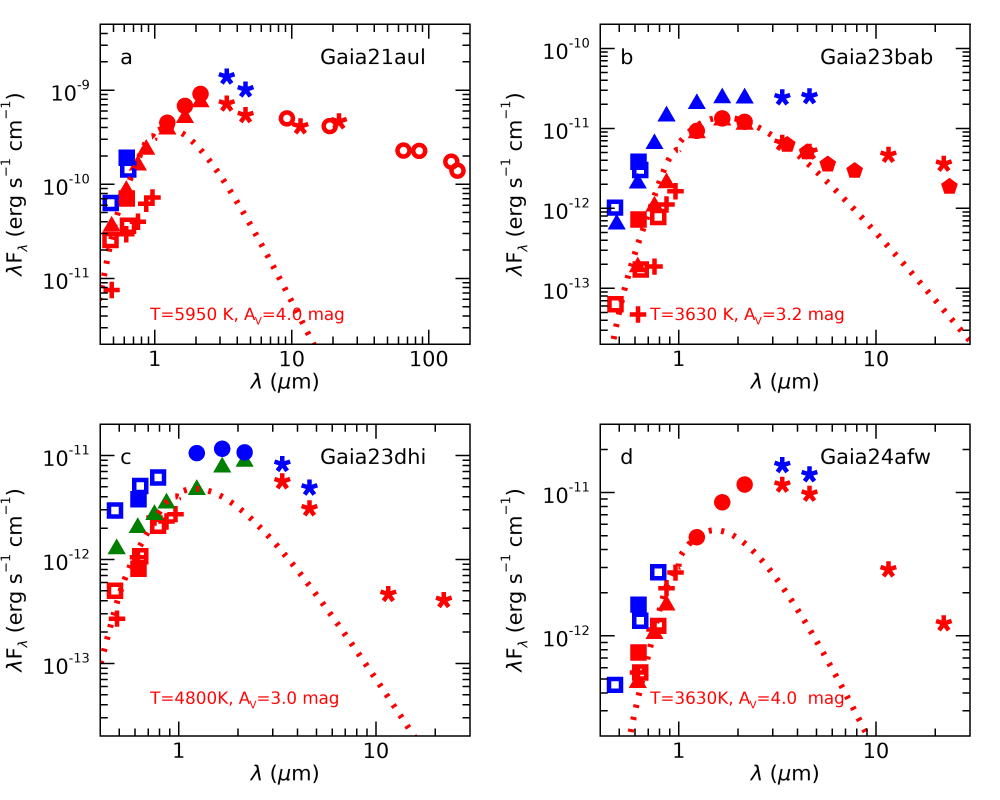}
\caption{\label{fig:fig9} As in Figure\,\ref{fig:fig7} for  Gaia21aul (a), Gaia23bab (b), Gaia23dhi (c), and Gaia24afw (d). }
\end{figure}

\begin{figure}[ht!]
\includegraphics{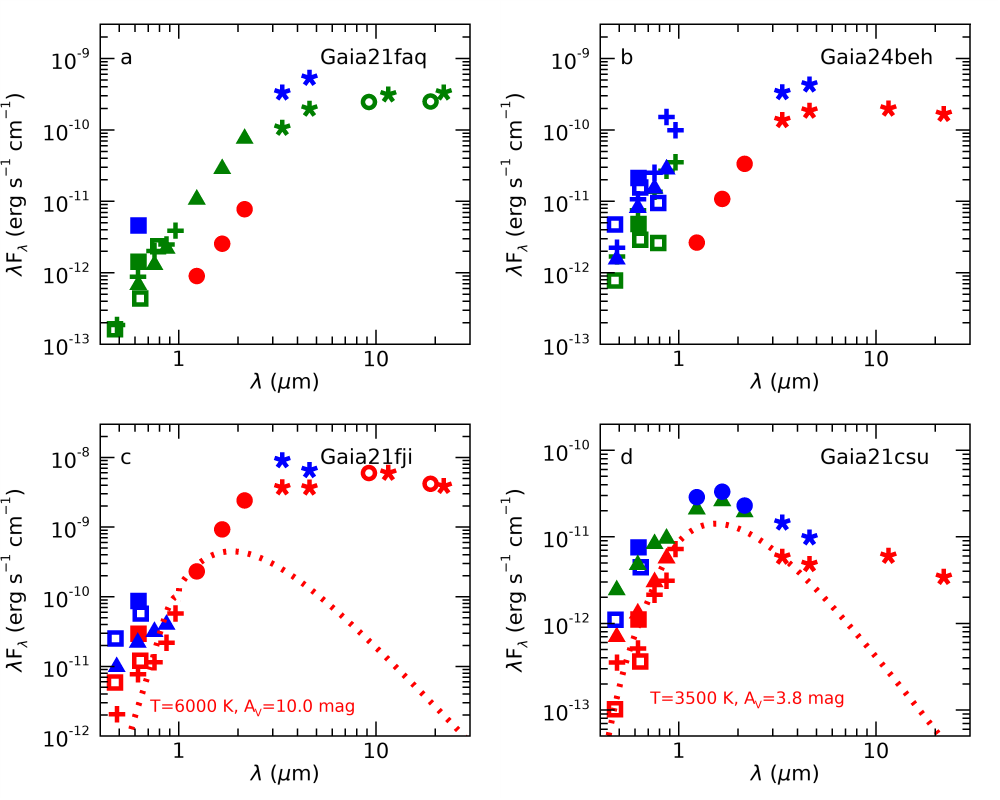}
\caption{\label{fig:fig10} As in Figure\,\ref{fig:fig7} for  Gaia21faq (a), Gaia24beh (b), Gaia21fji (c), and Gaia21csu (d).  }
\end{figure}

Figures\,\ref{fig:fig7}$-$\ref{fig:fig10} show the Spectral Energy Distribution (SED) of our sources, obtained collecting LBT photometry and public archival data. The details of the SED of each source are commented in Appendix \ref{appendix:A}, while here we give some general information. The red points represent the quiescence data, which have been taken typically in dates that may differ even by several years. In the case of multiple observations in a given band (specifically PAN-STARRS, ZTF, Gaia, and NEOWISE data) the plotted data points represent the average of the photometric measurements from the light-curve that do not exceed the historical standard deviation by more than 0.5 mag. The photometric points during burst phases are depicted in blue, while intermediate states are in green. For most of the sources, the quiescent photometric points extend up to 10-20 $\mu$m  (AllWISE or Spitzer data),  while the burst data typically extend
up to  $\sim$ 5 $\mu$m  (NEOWISE data). 

First, it's worth noting that all the sources display a Spectral Energy Distribution characteristic of a Young Stellar Object. Specifically, they show a peak in the near-infrared or at even longer wavelengths, along with a significant infrared excess. These characteristics strongly support the classification as YSOs of the objects classified by Gaia as 'YSO-candidates' (see Table\,\ref{tab:tab1}).

We plot in the same figures the black-body  function at the effective temperature (\teff) of the stellar 
photosphere (Section\,\ref{sec:sec7.2}) and reddened using the \av\, value measured during quiescence (Table\,\ref{tab:tab4}). The 
black-body was normalized using the $J$-band photometry in the quiescent phase, with the assumption that at
wavelengths around 1\,$\mu$m the emission originating from the accretion shock and the disk is minimal 
compared to the emission from the stellar photosphere.

For Gaia21bkw, Gaia22bvi,  Gaia21arv, and Gaia23bab, \teff\,  was taken from the literature (Table \ref{tab:tab2}), while the \av\, values in quiescence were estimated based on the near-infrared colors in the case of Gaia21bkw and Gaia21arv and from the continuum fitting in the case of Gaia22bvi and Gaia23bab.
For the other sources observed in the optical range during quiescence,  \teff\, and \av\, were adopted from the continuum fitting method, as will be discussed in Section\,\ref{sec:sec7.2}. This applies to Gaia22efa, Gaia22ehn, Gaia23bri, Gaia21aul,  Gaia24afw, and Gaia21csu. 
For Gaia22dbd, Gaia21ebu, and Gaia21fji,  which were observed either during a burst or in an intermediate brightness state, we estimated \av\, using the near-infrared 2MASS colors in quiescence (Section \ref{sec:sec6.2}). In the case of Gaia23dhi, which was observed with 2MASS during a burst, we estimated \av\, from the LUCI colors during an intermediate phase of brightening and used the same value also to redden the black-body function in quiescence. For these latter four sources we adopted the  \teff\, values determined as discussed in Section\,\ref{sec:sec7.2}. 
Finally, note that for Gaia21faq and Gaia24beh, we  were unable to constrain \teff\, and therefore  we did not attempt to plot the blackbody function (Section\,\ref{sec:sec7.2}). 

The integral under the black body is taken as an estimate of the stellar luminosity (\lstar) and compared with that presented in the forthcoming Section\,\ref{sec:sec7.3}. The two determinations differ in most cases by a few percent, the largest differences being in Gaia22efa, Gaia23bri,  Gaia21aul and Gaia21fji (30$-$50\%).

\subsection{Color-color diagrams}\label{sec:sec6.2}
\begin{figure}[ht!]
\includegraphics{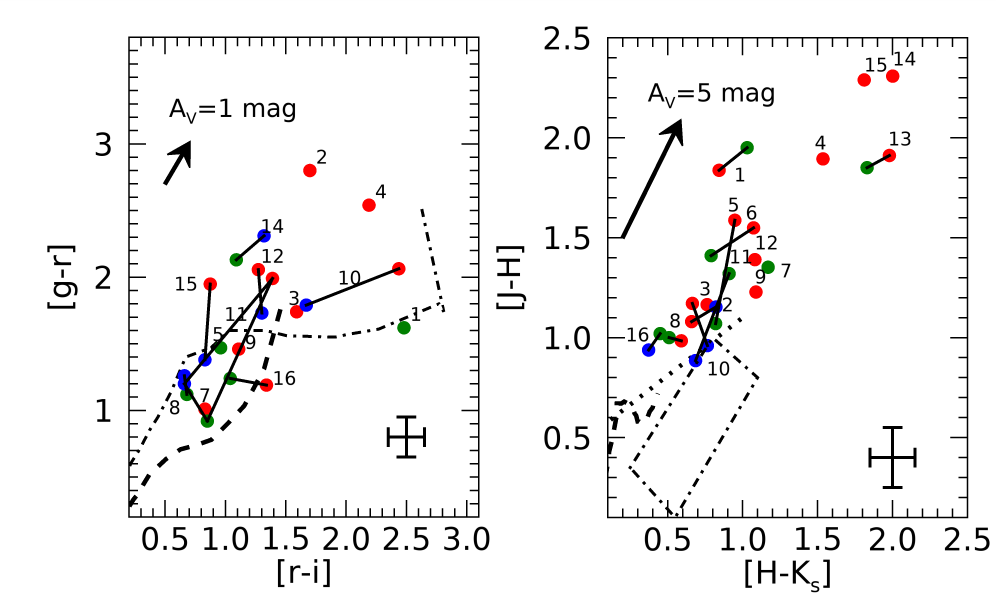}
\caption{\label{fig:fig11}  Left panel: two-color [g$-$r] vs. [r$-$i] plot. Red, green and blue dots indicate quiescent, intermediate and 
burst data points.  Sources are identified by their ID, and black segments connect data of the same source. The dashed–dotted line is the {\it locus} of young stars
with ages of 400–600 Myr (Kraus \& Hillenbrand 2007), while the  
dashed line shows the main-sequence stars. The arrow represents the 
direction of the extinction vector corresponding to \av\,=\,1 mag (reddening law of Cardelli et al. 1989). In the bottom-right corner the typical error associated to the data points is shown.
Right panel: two-color NIR [J$-$H] vs. 
[H$-$K$_s$] plot. Color code is the same as in the left panel. 
The dashed–dotted rectangle is the {\it 
locus} of un-reddened 
HAeBe stars (Hern{\'a}ndez et al. 2005) and the dotted line is the {\it 
locus} of the un-reddened T Tauri stars (Meyer et al. 1997). The arrow indicates the 
direction of the extinction vector corresponding to \av\, = 5 mag.}
\end{figure}

The [g$-$r]\,vs.\,[r$-$i] and [J$-$H]\,vs.\,[H$-$K$_s$] two-color diagrams for the targets are presented in Figure \ref{fig:fig11}. In these diagrams, data points are color-coded to represent different states: red for quiescent, green for intermediate, and blue for burst. The arrows in the plots indicate the direction of the extinction vector, following the reddening law of Cardelli et al. (1989). In the plots, the black segments connect data points of the same source observed in different brightness phases. It is noteworthy that most of these segments are not aligned with the extinction vector, which implies that changes in extinction play a relatively minor role in the observed variability. 
An \av\, estimate was obtained from the near-infrared colors by de-reddening the observed data until they overlap with the {\it locus} of the un-reddened T Tauri stars (Meyer et al. 1997), which is depicted with a dotted line in the right panel of Figure\, \ref{fig:fig11}. Based on this analysis, we conclude the following:
1) in general, our sources are subject to visual extinction ranging from 0 to 10 mag; 2) these \av\, estimates may increase if the source were an Herbig Ae/Be star, whose {\it locus} is located at lower values of the [J$-$H] color (Hern{\'a}ndez et al. 2005). As we will show in Section\,\ref{sec:sec7.3} and Figure\,\ref{fig:fig15}, this prescription applies in particular to sources \#9 and \#15, whose \av\, estimate is given as a possible lower limit in Table\,\ref{tab:tab4}; 3) the colors for sources \#13 and, to a lesser extent \#14, extend beyond the region of the observed T Tauri stars. For these sources, the \av\, estimated by ideally elongating the T Tauri straight line, should be taken with some caution.

With some exceptions, both optical and near-infrared colors 
become bluer during burst phases. The average color differences observed between quiescence and bursts are as follows:: $\Delta$([g$-$r])=0.27$\pm$0.08, $\Delta$([r$-$i])=0.37$\pm$0.12, $\Delta$([J$-$H])=0.58$\pm$0.23 and $\Delta$([H$-$K$_s$])=0.50$\pm$0.21. We interpret this blueing as evidence that the observed bursts are episodes of enhanced accretion. Indeed, this effect is commonly attributed to both the clearing of dust during the burst events (e.g., Hillenbrand et al. 2019) and the increasing contribution of accretion luminosity at UV wavelengths (e.g., Venuti et al. 2014). This interpretation is further supported by the smaller variation observed in the NEOWISE colors, $\Delta$([$W1-W2$])=0.05$\pm$0.02, indicating that typically the outer regions of the disks are less affected by the burst heating.

\section{Spectroscopic analysis}\label{sec:sec7}
\subsection{Description of the spectra}\label{sec:sec7.1}

\begin{figure}[ht!]
\includegraphics{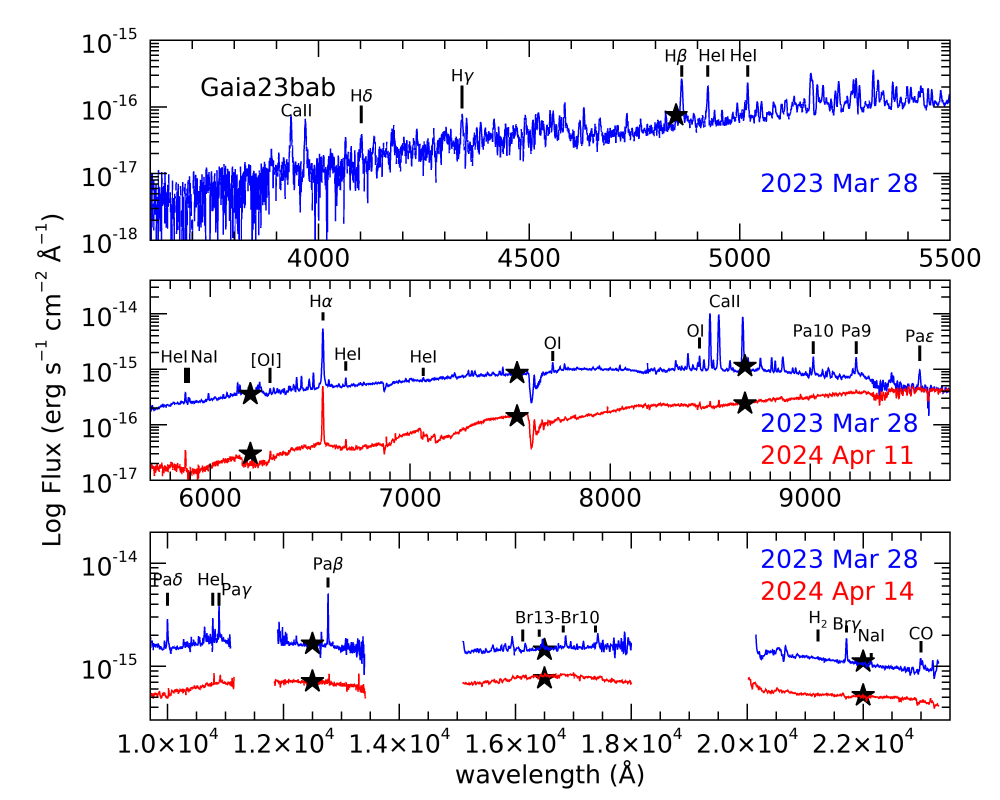}
\caption{\label{fig:fig12} MODS and LUCI spectra of Gaia23bab. Quiescence and burst  (Giannini et al. 2024) spectra are colored in red and blue, respectively. Observation dates are indicated. Black stars are the photometric points in $grizJHK_s$ filters obtained in the same night as the spectrum. The most important detected lines are labeled.}
\end{figure}
We show in Figure\,\ref{fig:fig12} the spectrum of Gaia23bab, as an example of the observed spectra. This is the only object in our sample for which we have spectra from both its burst phase (as reported in Giannini et al. 2024) and its quiescent state.
The MODS and LUCI spectra for all the remaining sources in our study can be 
found in Appendix \ref{appendix:B}, Figures\,\ref{fig:fig18}$-$\ref{fig:fig32}. 

All the observed sources share two important characteristics that are typical of accreting YSO spectra: a rising continuum from the optical to the infrared wavelengths and the presence of permitted lines in emission.
In particular, \ha\, is observed in emission in all targets, being partially absorbed in  Gaia21ebu and Gaia23dhi. The strongest  \hi\, lines are seen in burst spectra of Gaia24beh and Gaia21fji (both Class\,I sources), as well as Gaia23bab, as previously discussed in Giannini et al. (2024). Additionally, \caii\,, \oi\,, or \hei\, optical lines are detected in all spectra with the exception of Gaia22dbd, Gaia21ebu, Gaia24afw and Gaia21csu                                                  .

More than half of the analyzed spectra exhibit signatures of outflowing gas. The [\oi]\,6300\AA\, line is detected in the spectra of ten objects (Gaia22bvi, Gaia22ehn, Gaia23bri, Gaia21aul, Gaia23bab, Gaia24afw, Gaia21faq, Gaia24beh, 
Gaia21fji, and Gaia21csu), and the [\sii]\, doublet at $\lambda\lambda$\,6717, 6732 \AA\, in five objects (Gaia22ehn, Gaia21fji, Gaia21faq\, Gaia24beh, and Gaia21csu). In addition,  the [\oii]\, doublet at $\lambda\lambda$\,7320, 7330 \AA\, and the [\nii]\, 6584 \AA\, are detected in  the spectrum of Gaia22ehn, while  the \n\,1.04 $\mu$m, and [\feii]\,1.25 $\mu$m lines are detected in the spectrum of Gaia21faq.

The 
\hei\, at 1.08\,$\mu$m is observed in emission or 
absorption in all LUCI spectra, with the exceptions 
of   Gaia22dbd, Gaia21faq, and Gaia21csu. 
Additionally, we observe weak \htwo\, 2.12\,$\mu$m emission in Gaia21arv, Gaia23bab (during its burst phase), and Gaia21faq. Finally, the CO 2-0 or 3-1 bandheads and \nai\, doublet at 2.2 $\mu$m are observed in absorption in Gaia21bkw, Gaia21arv, and in Gaia21csu, in emission in Gaia21faq and within the burst spectrum of Gaia23bab.\\
Line fluxes and their corresponding 1$\sigma$ uncertainties were calculated using the SPLOT task within IRAF.  This method considers both the effective readout noise per pixel and the photon noise present in the spectral region encompassing the emission line. The fluxes of the main lines can be found in Appendix\,\ref{appendix:C}.

\subsection{Determination of the spectral type}\label{sec:sec7.2}
The first step of our study involves determining the stellar parameters of our sample sources, beginning with their spectral type (SpT).
As detailed in Section\,\ref{sec:sec2} and Table\,\ref{tab:tab2}, this is currently known for only four sources: Gaia21bkw, Gaia22bvi, Gaia21arv and Gaia23bab. 

\begin{figure}[ht!]
\includegraphics{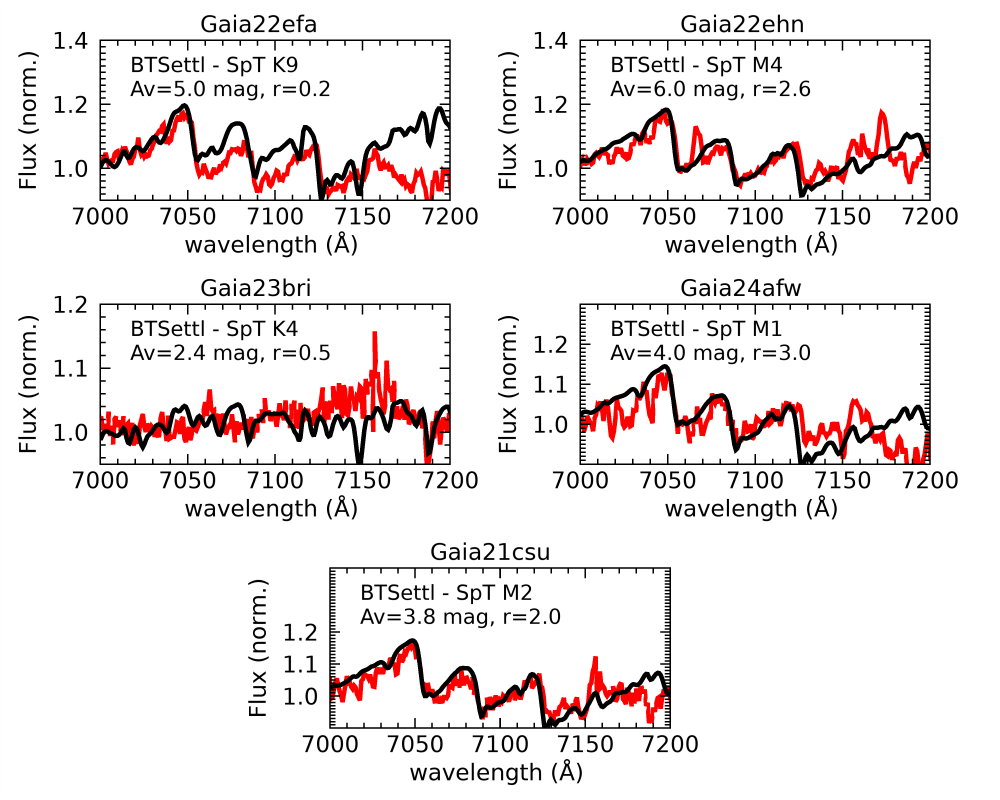}
\caption{\label{fig:fig13} Continuum fitting in the range 7000-7200 \AA\, of quiescence spectra (red). In all panels the spectral type of the best matching BTSettl model (black), and the fitted  \av\, and veiling ($r$) are labeled. }
\end{figure}

\begin{figure}[ht!]
\includegraphics{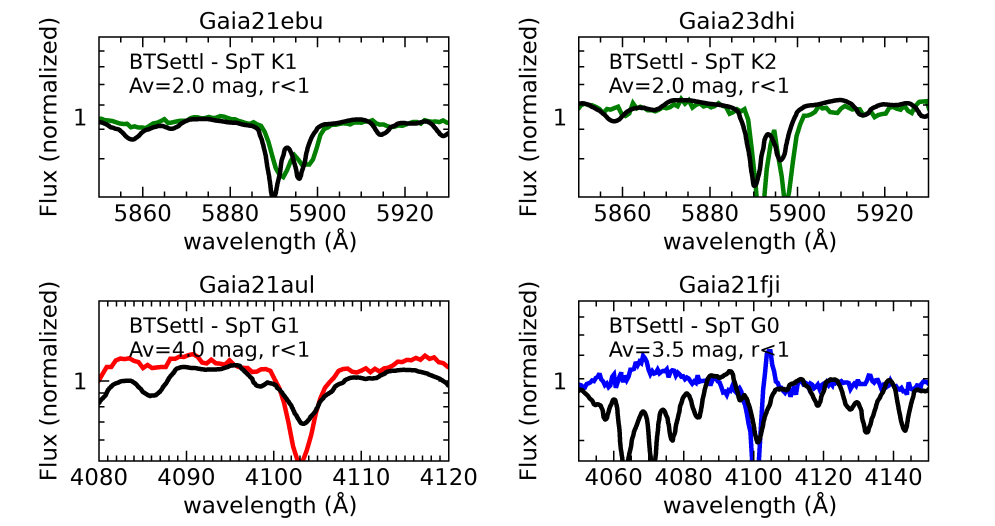}
\caption{\label{fig:fig14} Top: Fit of the spectral region around the \nai\, doublet at 5890 \AA\,  in the intermediate state spectra of Gaia21ebu and Gaia23dhi (green). Bottom: Fit of the spectral region around the  H$\gamma$ line for the quiescence spectrum of Gaia21aul (red) and the burst spectrum of Gaia21fji (blue). In all panels the spectral type of the best matching BTSettl model (black) and the fitted  \av\, and veiling ($r$) are labeled. }
\end{figure}

The standard approach for estimating SpT typically involves comparing optical photospheric lines with stellar templates. Simultaneously, extinction can be determined from the slope of the continuum. However, in cases of burst accretion, this method becomes complicated due to the presence of excess continuum emission in the UV and IR wavelengths. The UV excess is thought to originate from hot spots formed by the accretion shock on the surface of the star, while the IR excess is generated within the accretion disk itself. During periods of significant burst accretion, the UV-optical excess can become so prominent that in late-type stars the molecular bands visible in the quiescent spectrum may no longer be detectable. Consequently, a burst spectrum might be misclassified as belonging to an earlier spectral type with respect to the star's actual quiescent classification.
Examples include ASSASSN-13db, where the ratio $r = Flux(excess)/Flux(star)$ (veiling) increases from 0.4 in quiescence to over 3 during a burst (Giannini et al. 2022), and Gaia23bab, where Giannini et al. (2024) incorrectly inferred a spectral type earlier than K due to the absence of optical molecular bands in the burst spectrum.
For this reason in the present work we have focused our continuum fitting analysis only on quiescence spectra (Gaia22efa\,, Gaia22ehn\,, Gaia23bri\,, Gaia21aul\,, Gaia24afw\,, and Gaia21csu) and on spectra of other brightness phases that present prominent absorption features. These are the burst spectrum of Gaia21fji, and the intermediate state  spectra of Gaia21ebu and Gaia23dhi, as they show \hi\, lines or the \nai\, doublet at $\lambda\lambda$\,5891, 5897 \AA\, in absorption,  which are indicators of an early spectral type.
Our fitting procedure considers SpT, \av\, and $r$ as variables.  
The results of this fitting procedure are illustrated in Figures\,\ref{fig:fig13} and \ref{fig:fig14}. For late-type spectra  we fitted the continuum portion within the 7000$–$7200 \AA\, range, which encompasses prominent TiO absorption bands. For early-type spectra, we fitted either the 3700$-$4500 \AA\, range, where several Balmer lines are located (in Gaia21aul and Gaia21fji), or the spectral region around the \nai\, doublet at 5890 \AA\, (in Gaia21ebu and Gaia23dhi), following the method of Tripicchio et al. (1997) and Biazzo et al. (in preparation).

As a strategy, we have first selected a number of BTSettl templates with
temperature between 3000 and 7000 K (SpT F$-$M) and log $g$\,=\,4.5. An \av\, between 0 and 20 mag (in steps of 0.2 mag) and a  veiling between 0 and 5 (in steps of 0.1) are then applied to each template to find the values that best match the spectral features and the slope of the continuum observed in our stars. Typically, we derive SpT K$-$M, \av\, between 2 and 6 mag and $r\la$\,3.0, with the exception of Gaia21aul and Gaia21fji, for which SpT is G1 and G0, respectively. \\
A particular case is that of Gaia22dbd. The source was seen during an intermediate state and the spectrum does not present lines or bands in absorption. We have provisionally estimated the effective temperature by fitting the SED through the optical and $J$ photometry in quiescence and assuming the \av\, derived from the near-infrared colors. Then, this is converted to spectral type using the value tabulated in Table 6 of Pecaut \& Mamajek (2013) for  5$-$30 Myr old stars. The estimated uncertainty is about two sub-classes.  Finally, for Gaia21faq and Gaia24beh, we were unable to derive the SpT because their intermediate and burst spectra do not present absorption features and there are no optical points available during quiescence to fit the SED with a black-body function.


\subsection{Other stellar parameters}\label{sec:sec7.3}
\begin{table*}
\center
\caption{\label{tab:tab4} Extinction determinations}
\begin{tabular}{cc|ccc|ccc|ccc}
\hline
\hline
ID  &  Source  & \multicolumn{3}{c|}{Quiescence} & \multicolumn{3}{c|}{Intermediate}                               & \multicolumn{3}{c}{Outburst} \\
    &               &  cont. &   lines      &   col-col      &  cont.   &  lines       &  col-col  &  cont. &   lines  & col-col\\
\hline
\hline
1   & Gaia21bkw     &  -       &     10.0        &    9.8        & 10.0        &   10.0       &   10    &   -  &  -        &   -    \\
2   & Gaia22efa     &  5.0      &    6.1         &     2        &   -         &    -         &    -      &   -  &  -        &  2  \\
3   & Gaia22bvi     &  3.2      &    2.8         &    3          &   -         &    -         &    -      &   -  &  -        &  -    \\ 
4   & Gaia22ehn     &  6.0      &    5.0         &    5          &   -         &    -         &    -      &   -  &  -        &  -     \\
5   & Gaia22dbd     &   -       &     -          &    7          & 2.2         &    -         &    -      &   -  &  -        & 1     \\
6   & Gaia21arv     &   -       &    -           &    6          &6.6$^a$      &    -         &    6      &   -  & -         &  -     \\
7   & Gaia23bri     &  2.4      &    2.3         &    -          &  -          &    -         &   2.5     &   -  &  -        &  -     \\
8   & Gaia21ebu     &  -        &     -          &    3           & 2.0         &    -         &   3.5       &   -  &  -        &   -   \\ 
9   & Gaia21aul     &  4.0        &    4.0         &   $\ga$3          &  -          &    -         &    -      &   -  &  -        &   -    \\
10  & Gaia23bab     & 3.2       &    4.3         &    3.5         &  -          &    -         &    -      & 5.5$^1$&  -      &   -    \\
11  & Gaia23dhi     &     -     &    -           &     -         &  2.0        &    -         &   3      &   -  &  -        &  0     \\
12  & Gaia24afw     & 4.0       &     5.0          &     4         &  -          &    -         &    -      &    -  &    -       &  -     \\
13  & Gaia21faq     &  -        &  -             &     5         &  6.0        & 5.0          &   5       &   -  &   -       &  -    \\
14  & Gaia24beh     & -         &  -             &    $\ga$10        &  -          &    -         &    -      & 5.2  &    5.2    &  -    \\
15  & Gaia21fji & -         &  -             &    10         &  -          &    -         &    -      & 3.5  &    4.4    &     -   \\
16  & Gaia21csu     &3.8        &  3.0           &               &  1.4          &    -       &    3      &  -   &    -      &  2.5   \\ 
\hline
\hline
\end{tabular}
\begin{quotation}
\textbf{Note.} $^a$In this source we fitted the shape of the near-infrared continuum. 
\textbf{Reference.}$^1$Giannini et al. 2024.
\end{quotation}
\end{table*} 
 
Based on the spectral type of our sources, we derived other stellar parameters, including stellar luminosity, radius, and mass (Table\,\ref{tab:tab5}). This derivation was performed using near-infrared photometry and considering extinction during the quiescent phase.
Assuming that the $J$-band emission is minimally affected by accretion spots and emission of the circumstellar disk, we calculated the bolometric magnitude (\Mbol\,) using the formula:
\begin{equation}
    M_{bol} = m(J) + 5 - 5 \log_{10}[d(pc)] + BC_J
\end{equation}

where $m(J)$ is the extinction-corrected $J$ magnitude, $d(pc)$ is the distance in parsecs, and $BC_J$ is the bolometric correction in the $J$ band. For $BC_J$, we adopted the values provided in Table 6 of Pecaut \& Mamajek (2013) for sources aged between 5 and 30 Myr. For our objects, which have spectral types ranging from G0 to M4, the corresponding $BC_J$ values are between 0.98 and 1.91. For Gaia21faq and Gaia24beh we were only able to determine a lower limit for \Mbol\,. This is because we could not add a BC$_J$ correction in applying equation 1, as their spectral types are unknown.\\
An estimate of the stellar luminosity, \lstar\,, can be calculated as:

\begin{equation}
    log_{10}L_\mathrm{*} = 0.4 [M_\mathrm{bol} - M_{\mathrm{\odot}}]
\end{equation}

where $M_{\mathrm{\odot}}$ represents the bolometric luminosity of the Sun, which is equal to 4.74 mag (Mamajek et al. 2015).

Assuming black-body emission, the stellar radius (\rstar) can be calculated using the formula: 

\begin{equation}
   R_\mathrm{*} =\,1/2\,T_\mathrm{eff}^2/\sqrt{L_{\mathrm{*}}/\pi\sigma}
\end{equation}

where $\sigma$ is the Stefan-Boltzmann constant and \teff\, is the effective temperature. This latter was derived based on the correspondence with the spectral type (SpT) from Pecaut \& Mamajek (2013). With the exception of Gaia22efa, Gaia21aul and Gaia21fji,  stellar luminosities do not exceed a few solar luminosities, which aligns with the K$-$M spectral types determined from both the optical continua and SED fitting.

\begin{figure}[ht!]
\includegraphics{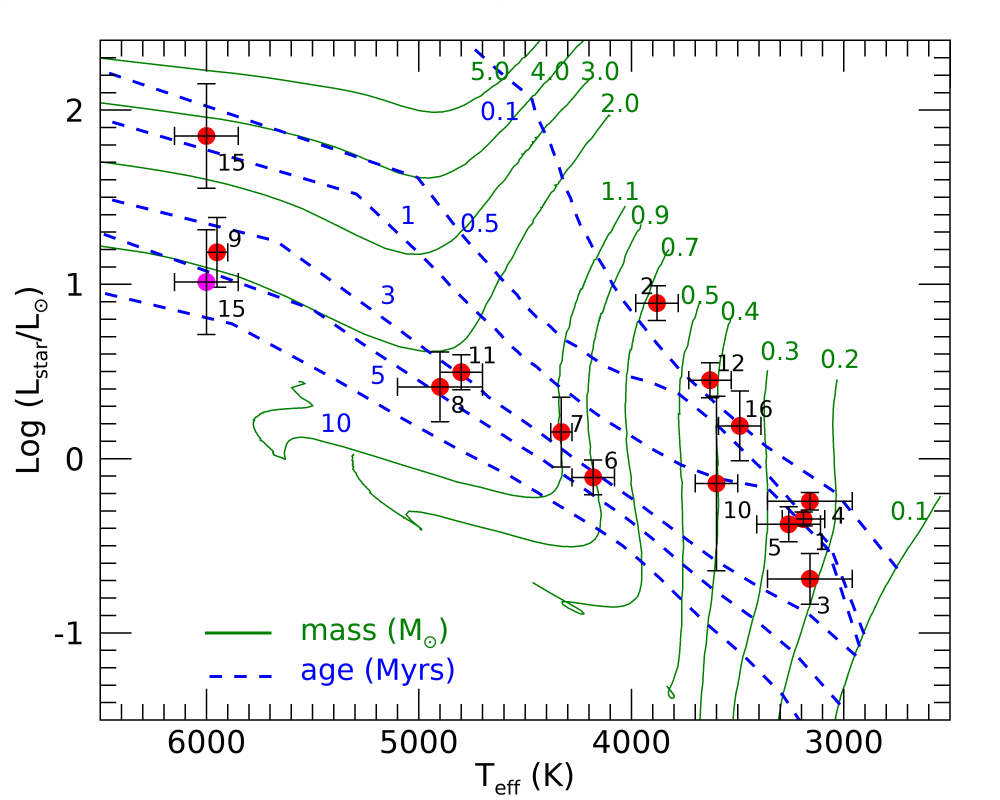}
\caption{\label{fig:fig15} Evolutionary tracks of Siess (2000) in the range 0.1$-$5.0 \msun\, (green filled lines) and for ages 0.1$-$10 Myrs (dashed blue lines). The data of the sources of our sample are plotted in red and labeled. In magenta is the point corresponding to the distance estimate of 225 pc for source \#15 (Gaia2fji). }
\end{figure}

The stellar masses have been determined by comparing \lstar\, and \teff\, with the evolutionary tracks of Siess et al. (2000). As shown in Figure\,\ref{fig:fig15}, this comparison typically yields a mass range of 0.25$-$2.5 \msun\, and ages $\la$ 5 Myrs. 
In the case of Gaia21fji (\#15), the data corresponding to the two different distance determinations have been plotted in magenta and red (magenta: $d$=225 pc; red: $d$=600 pc). Both sets indicate  this is an intermediate-mass object (\mstar\,=\,2$-$3.5 \msun\,) with age $\sim$ 1$-$5 Myr.\\

\subsection{Accretion luminosity and mass accretion rate}\label{sec:sec7.4}
\begin{table*}
\center
\caption{\label{tab:tab5} Stellar parameters}
\begin{tabular}{ccCccccccc}
\hline
\hline
ID  &  Source      & $\alpha_{2-24}$     & Class &   SpT$^a$   &  \Mbol     &  \lstar   &  \rstar    &  \mstar  & age  \\
\hline
    &              &                     &       &                 &  (\msun)   &  (\lsun)  &  (\rsun)   &  (\msun) & (Myr)  \\
\hline
\hline
1   & Gaia21bkw         &  -0.17         & flat  & M4$^1$    &  5.6           &  0.45$^1$       &   2.2             &  0.25$^1$  & 0.5-1\\ 
2   & Gaia22efa         &  -0.47         & II    & K9        &  2.5           &  7.8            &   6.1             &  0.65      & 0.1  \\
3   & Gaia22bvi         &-0.01 (0.05$^2$)& flat  & M4$^2$    &  4.4           &  1.4            &   2.27$^2$        &  0.35$^2$    & 2    \\
4   & Gaia22ehn         & -0.47          & II    & M4        &  5.3           &  0.6            &   2.5             &  0.25      & 0.5-1\\ 
5   & Gaia22dbd         & -0.34          & II    & M3-M4$^a$ &  5.9           &  0.35           &   1.9             &  0.25      & 1-2\\
6   & Gaia21arv         &  -             & -     & K3$^3$    &  5.0           &  0.8            &   1.7             &  0.9       & 3    \\
7   & Gaia23bri         & -0.53          & II    & K4        &  4.3           &  1.4            &   2.1             &  1.1       & 3\\
8   & Gaia21ebu         & -0.54          & II    & K1        &  3.4           &  3.4            &   2.5             &  1.8      & 5 \\
9   & Gaia21aul         & -0.47          & II    & G1        &  1.8           &  15.3           &   3.6             &  2.5       & 5  \\
10  & Gaia23bab         & -0.57          & II    & M1$^4$    &  5.1           &  0.72$^4$       &   2.3$^4$         &  0.4$^4$   & 1    \\
11  & Gaia23dhi         & -0.61          & II    & K2        &  3.1           &  4.4            &   3.0             &  1.8       & 3 \\
12  & Gaia24afw         & -0.81          & II    & M1        &  3.4 &         3.3               &   4.6   &  0.4       & 0.1-0.5   \\
13  & Gaia21faq         & 1.6            & I     & -         &  $>$4.8        &      $<$0.9     &   -               &   -        & -    \\
14  & Gaia24beh         & 1.1            & I     & -         &  $>$3.2        &      $<$4.2     &    -              &   -        & -    \\
15  & Gaia21fji         & 0.54           & I     &  G0       & 1.9$-$ -0.23   &    6.4$-$46     &   2.3$-$6.2       &  2$-$3.5   & 1-5  \\
16  & Gaia21csu         & -              & -     &  M2       &  4.3           &  1.5            &   3.4             &  0.35      & 0.1\\
\hline
\hline
\end{tabular}
\begin{quotation}
\textbf{References.}$^1$Fiorellino et al. 2021; $^2$Dahm \& Hillenbrand 2020; $^3$Kounkel et al. 2019; $^4$Nagy et al. 2024.
\textbf{Notes.} $^a$SpT estimated by fitting the SED. 
\end{quotation}
\end{table*} 
 
\begin{table*}
\center
\caption{\label{tab:tab6} Accretion parameters}
\begin{tabular}{cc|cc|cc|cc}
\hline
\hline
ID  &  Source       & \multicolumn{2}{c|}{Quiescence}   & \multicolumn{2}{c|}{Intermediate} & \multicolumn{2}{c}{Outburst} \\
    &               & log (\lacc)  &   log(\macc)    & log (\lacc) &   log(\macc)     & log (\lacc) &   log(\macc)\\
\hline
    &                 &  (\lsun)     &   (\msunyr)        &   (\lsun)   &   (\msunyr)  )&(\lsun)   &   (\msunyr)\\
\hline
\hline
1   & Gaia21bkw  &   $-$2.10 (S) &    $-$8.55 (S)   & $-$0.85 (S)&  $-$7.31 (S) & $-$0.08 (P) & $-$6.54 (P)     \\
2   & Gaia22efa  &$-$0.36 (S)    &    $-$6.79 (S)   &    -       &    -         &  -          &   -             \\
3   & Gaia22bvi  &   $-$1.13 (S) &    $-$7.71 (S)   &    -       &    -         &  $-$0.22 (P)   &  $-$6.80 (P)    \\ 
4   & Gaia22ehn  &   $-$1.33 (S) &    $-$7.74 (S)   &    -       &    -         & 0.03 (P) &  $-$6.35  (P)   \\
5   & Gaia22dbd  &   $-$1.60 (P) &    $-$8.02 (P)   & $-$2.42 (S)& $-$8.84 (S)  & $-$2.28 (P) &  $-$8.70 (P)    \\
6   & Gaia21arv  &    -          &                  & $-$1.28 (S)&  $-$8.45 (S) & -           &  -              \\
7   & Gaia23bri  &   $-$0.80 (S) &   $-$7.92  (S)   & $-$0.34 (S)&  $-$7.20 (S) & 0.00 (P)    &  $-$7.11 (P)    \\
8   & Gaia21ebu  &     -         &     -            & $>$\,$-$1.7 (S)&  $>$$-$8.9  (S)  &  -       &   -        \\ 
9   & Gaia21aul  &   $-$0.31 (S) &   $-$7.51  (S)   &    -       &    -          & 0.41 (P)   &  $-$6.79 (P)    \\
10  & Gaia23bab  &   $-$1.66 (S) &   $-$8.30  (S)   &    -       &    -          & 0.64  (S)  &  $-$5.99 (S)    \\
11  & Gaia23dhi &    -       &      -  & $>$\,$-$1.8 (S)&  $>$$-$9.0 (S)    &  -       &   -      \\
12  & Gaia24afw  &  $-$0.60 (S) &  $-$6.98  (S) &    -       &    -          & $-$0.05 (P) &  $-$6.43 (P)   \\
13  & Gaia21faq      &  -           &      -       & 0.07 (S)    &               &0.22  (P)    &        -       \\
14  & Gaia24beh      &  -           &      -       &    -       &    -           & 2.00  (S)   &  -  \\
15  & Gaia21fji$^g$  &  -           &      -       &    -       &    -           & 0.17$^a$-1.07$^b$ (S) &  $-$7.26$^a$-$-$6.13$^b$ (S)    \\
16  & Gaia21csu      &  $-$1.7 (S)  &  $-$8.1 (S)  &  $-$1.62 (S)  &  $-$8.0 (S) & $-$1.30(P)  &  $-$7.71 (P)      \\ 
\hline
\hline
\end{tabular}
\begin{quotation}
\textbf{Note.}{In parenthesis is indicated the method used to derive the \av\,: S:spectroscopy, P:photometry. $^a$Computed assuming d=225 pc. $^b$Computed assuming d=600 pc.  } 
\end{quotation}
\end{table*} 
To determine if a source is a genuine EY, it is crucial to quantify its key accretion parameters, namely the accretion luminosity (\lacc) and the mass accretion rate (\macc), during both quiescent and burst phases.
A robust method for calculating \lacc\, and  \macc\, involves observing lines excited within the accretion columns.
For sources where we have optical and/or near-infrared spectra, we have estimated \lacc\, based on its correlation with the luminosities (\lumi\,) of specific accretion lines. 

This correlation is defined by the empirical relations (\lacci\, vs. \lumi\,) established by Alcal{\'a} et al. (2014, 2017) for various \hi\, recombination lines from the Balmer, Paschen, and Brackett series, as well as for some  \hei\,, \oi\, and \caii\, lines. To determine \lacc\,, we have developed a grid of \av\, values ranging from 0 to 20 magnitudes (in increments of 0.2 magnitudes).  For each \av\, value, we calculated the accretion luminosity derived from each line ($L_{acc(i)}$). The \av\, and \lacc\, (computed as the average of the individual \lacci\,)  pair that yields the minimum dispersion among the $L_{acc (i)}$ values is considered the most accurate estimate for these two quantities. We estimate a typical error on \av\, within 1.0 mag  and on \lacc\, of 0.2$-$0.3 dex. 

With reference to Table\,\ref{tab:tab6}, we applied the method described above to 
all the spectra exhibiting accretion lines in emission. The accretion luminosity and 
mass accretion rate estimates for Gaia21ebu and Gaia23dhi should be considered lower limits due to 
partial absorption of the \hi\, lines in their spectra. In the case of Gaia21bkw we get a value for 
\lacc\, about a factor of 1.5 higher than that estimated by Fiorellino et al. (2021).  This result is consistent with the source being in a fainter state during the Fiorellino et al. observation period. Specifically, their $JHK_s$ photometries were approximately 0.2 magnitudes fainter than the LUCI values. 

As outlined in Section \ref{sec:sec2}, the majority of our sources have only been observed during a single brightness phase, predominantly during quiescence. Since it is fundamental to measure the variation of \lacc\, across quiescent and burst states, we have estimated \lacc\,  using photometric data when spectral data for a particular phase is lacking. Specifically, we computed,  as representative of \laccb/\laccq\, the de-reddened photometric ratio  of   \lburst/\lquiesc\, between the luminosity at $\lambda$ $\sim$ 6000 \AA\, at the burst peak and during quiescence. The photometric bands used are the Gaia $G$-band and the ZTF, PAN-STARRS and MODS $r$-band. 
To validate this procedure we have applied the same method to sources for which we have line fitting results, finding  agreement within 20$-$50\%. 

In cases where a value for \av\, was not available for a specific phase, we have used the average of determinations from other phases. For these cases, our estimate of \lacc\, relies on the assumption that \av\, remains constant over time, which may not be accurate. For these sources, obtaining spectra for the missing phase(s) in the future would be beneficial to strengthen the results presented.

Once \lacc\, is derived, the mass accretion rate \macc\, was estimated as 
\begin{equation}
\dot{M}_{\mathrm{acc}}=  (1 -R_\mathrm{*}/R_{\rm in})^{-1} L_\mathrm{acc}R_\mathrm{*}/G M_\mathrm{*}
\end{equation}
 (Gullbring et al. 1998), where R$_{\rm in}$ is the inner-disk radius, assumed  $\sim$ 5\rstar\, (Hartmann et al. 1998), and G is the gravitational constant.
 
We were able to determine \lacc\, and \macc\, both in quiescence and in outburst for nine sources. The two sources with the most remarkable variability are Gaia21bkw and Gaia23bab, in which \lacc\, and \macc\, increase by more than 2 orders of magnitude. In four sources (Gaia22bvi, Gaia22ehn, Gaia23bri, and Gaia21aul), \lacc\, and \macc\, vary of factors between 5 and 30, while in further two objects, Gaia24afw, and Gaia21csu, the variation is lower than a factor of 3.  In Gaia22dbd, the observed variability appears to be due to extinction rather than accretion variability ($\Delta$\av\,$\sim$ 6 mag). Indeed, the intrinsic accretion luminosity during burst is even lower than what is measured during the quiescent phase. Similarly, variable extinction may also play a role in the variability observed in Gaia21csu, since \av\, decreases by approximately one magnitude between the quiescent and burst phases.

\section{Discussion}\label{sec:sec8}
\subsection{Light curve analysis}\label{sec:sec8.1}
\begin{table*}
\small
\center
\caption{\label{tab:tab7} Light curve analysis}
\begin{tabular}{ccccccc}
\hline
\hline
Source    &  $\Delta$G/$\Delta$W1 & Maximum  & Gaia & Average &  Ris./Decl.   & Group\\
          &            & amplitude &         peaks & duration & speed  &  \\
          &  (mag)     &  (mag)           &            &    (days)         &  (mmag/d)  &     \\
\hline
Gaia21bkw & 2.2/2.9        &  2.3             & 5          &   284              & 14/10        &  A \\
Gaia22efa & 1.1/0.5        &  1.5             & 1          &   530              & 20/9.5       &  A \\
Gaia22bvi & 2.1/1.2        &  2.3             & 5$^a$      &   273              & 13/12        &  A \\
Gaia22ehn & 2.9/2.7        &  4.2             & 3$^b$      &   180              & 27/22        &  B \\
Gaia22dbd & 1.3/0.5        &  3.4             & 1          &   362              & 9/40         &  B \\
Gaia21arv & 1.3/0.5        &  2.2             & 3          &   637              & 2.5/4        &  B \\
Gaia23bri & 2.1/0.5        &  3.4             & 5          &   375              & 11/10        &  B \\  
Gaia21ebu & 1.2/0.9        &  2.0             & 1          &   324              & 8/7          &  B \\
Gaia21aul & 1.0/0.7        &  2.9             & 1          &   551              & 3/9          &  B \\
Gaia23bab & 2.3/1.4        &  2.6             & 2$^c$          &   589              & 11/8         &  A \\
Gaia23dhi & 1.6/0.8        &  3.0             &  -         &    -               & -            &  B \\
Gaia24afw & 0.9/0.3        &  1.2             & 2          &   106              & 16/18        &  A \\   
Gaia21faq & 1.4/1.2        &  2.5             & 3          &   336              &  6/3         &  B \\
Gaia24beh & 2.0/1.0        &  3.9             &  -         &  -                 & -            &  B \\
Gaia21fji & 2.0/1.0        &  2.8             &   -        &   -                & -            &  B \\
Gaia21csu & 2.1/1.0        &  3.3             &  3         &  372               & 10/11        &  B \\
\hline
\end{tabular}	
\begin{quotation}
\textbf{Notes}. $\Delta$G : $G_{peak}$-$G_{median}$; $\Delta$W1: $W1_{peak}$-$W1_{median}$; Maximum amplitude: $G_{min}$-$G_{max}$; Average duration: time elapsed between two subsequent quiescent levels; Rising/declining speed: time elapsed from quiescence to peak and viceversa. $^a$Two peaks detected by ZTF; $^b$One peak detected in W1/W2.$^c$A further peak was detected by PAN-STARSS in 2013, just before the beginning of the Gaia mission.
\end{quotation}
\end{table*} 
In Section \ref{sec:sec4}, we presented the light curves for each source in our sample, as shown in Figures\, \ref{fig:fig3}$-$\ref{fig:fig6}. We now broaden our discussion on these light curves, emphasizing the similarities and differences we observed throughout the ten-year Gaia dataset. One particularly interesting feature is the significant diversity in light curves, even among sources with comparable mass, age, and luminosity.  This likely indicates that the mechanisms responsible for the observed variability are inherently varied and influenced by multiple factors.

From a visual inspection of Figures\, \ref{fig:fig3}$-$\ref{fig:fig6}, and we can note that our sources can be broadly categorized into two main groups (Table\,\ref{tab:tab7}). The first group ('A'), includes light curves displaying distinct peaks above a relatively stable baseline, such as the EXor source Gaia23bab. The second group ('B') consists of sources displaying significant variability throughout the observation period, with some never achieving a stable state. This variability manifests in several ways, including irregular brightness fluctuations, dips, low-frequency modulation, or a continuous increase in brightness starting from a specific date with superposed higher-frequency variability. 
Alongside accretion variability, we observe low-amplitude variations with timescales of days, which likely originate from stellar activity and rotation, as well as regular or abrupt changes in extinction, potentially caused by warps, clumps within the disk, or geometrical effects (Venuti et al. 2015, Bouvier et al. 2007). 

It is also noteworthy that the mid-IR light curve does not always consistently reflect the trends seen in the Gaia data.
For example, the W1/W2 light curve of Gaia22bvi shows a low-frequency modulation that is not present in the $G$-band data. Similarly, for Gaia22ehn, a burst detected in W1/W2 during 2018 has no counterpart in the $G$ light curve. In the cases of Gaia21arv and in Gaia23dhi, brightness increases in $G$-band  align with minima in W1/W2, while a burst detected in April 2016 in Gaia21ebu corresponds to a low-frequency rising and decreasing in the W1/W2 light curve. 

As a general observation, we note that the wide variety we see in the Gaia light curves is not reflected in  the optical and infrared spectra, that all display the same emission lines.  Similarly, the spectral energy distributions across our sample are uniformly characteristic of YSOs.

To identify common features and properties within the light curves, we conducted a more quantitative analysis, the results of which are presented in Table~\ref{tab:tab7}. To account for the fluctuations seen in the optical ($G$) and mid-infrared (W1) light curves across different sources, we applied our analysis method to both the Gaia and NEOWISE datasets. First,  to minimize the influence of outliers and burst events, we have taken as Gaia brightness baseline the median magnitude
 (G$_{median}$) during the ten-year of the  operations. 
Then, we identified a 'Gaia peak’ (G$_{peak}$) as any event where the brightness, sampled by at least two photometric measurements, increased by at least one magnitude compared to the baseline level ($\Delta$G\,=\,G$_{peak}$ - G$_{median} \ga$ 1).  Analogously, we have defined the peaks in the W1 band. Finally, we note that a similar analysis performed on the $r$-band ZTF data reveals the same peaks found with the Gaia data. The only exception is the presence of two additional peaks detected in the Gaia22bvi light curve between 2018 and 2019, during which no Gaia observations were performed.\\
The identified peaks are visually represented in Figures\,\ref{fig:fig3}$-$\ref{fig:fig6} with a thick red line.
We underline that our definition excludes episodes of strong irregular variability, as well as phenomena characterized by a gradual increase in brightness or a sustained period of high brightness, as observed for example in Gaia22efa and Gaia21ebu. 
We classify as ’very active’, objects exhibiting at least an episode of brightness increase with $\Delta$G\,$\ga$\,2 mag or $\Delta$W1\,$\ga$\,1 mag, assuming that the amplitude of the variation decreases with the wavelength (e.g. Lorenzetti et al. 2007). As shown in the second column of Table\,\ref{tab:tab7}, these are Gaia21bkw, Gaia22bvi, Gaia22ehn, Gaia23bri, Gaia23bab, Gaia21faq, Gaia24beh, Gaia21fji, and Gaia21csu.  
According to the $\alpha$ values and ages presented in Table\,\ref{tab:tab5}, this group of ’very active’ objects includes the three Class I and the two flat-spectrum sources, along with three Class II sources, almost all with ages $\la$ 2 Myr. Conversely, the ’less active’ group is composed of six Class II sources, along with one source that has an unknown spectral index. These sources have ages ranging from 2 to 5 Myr, with the exception of Gaia22efa and Gaia24afw, which have ages less than 0.5 Myr. Gaia22efa, however, might be more active than one can think on the base of its ten-year light curve, as demonstrated by the steep rising in brightness occurring since last year. 

We also report in Table\,\ref{tab:tab7} the maximum amplitude ($G_{min}$-$G_{max}$) in the Gaia light curve. When this value does not exceed $\Delta$G by more than 10$-$30\,\%, the light curve has a relatively constant baseline. Conversely, the light curve may exhibit irregular variability, dips, or a non-flat slope. As said, this latter case involves the majority of our sources.
Another interesting aspect to consider is the frequency of enhanced brightness events. 
Notably, the objects categorized as ’very active’ also experience peak of brightness more frequently, typically with three to five peaks in 10 years. Conversely, with a couple of exceptions, the light curves of less active objects display just one peak.\\
Our analysis indicates that the burst duration generally falls between 200 and 600 days, and that both the rising and the declining speed are from a few to a dozen milli-magnitudes per day, with no significant difference observed between the more and less active source groups.  This result is confirmed from the analysis of the ZTF light curves, whose optimal $\sim$ 2 days cadence is more suited for accurately capturing events with durations comparable to the sampling frequency. 

\subsection{Accretion vs. Stellar parameters}\label{sec:sec8.2}
\begin{figure}[ht!]
\includegraphics{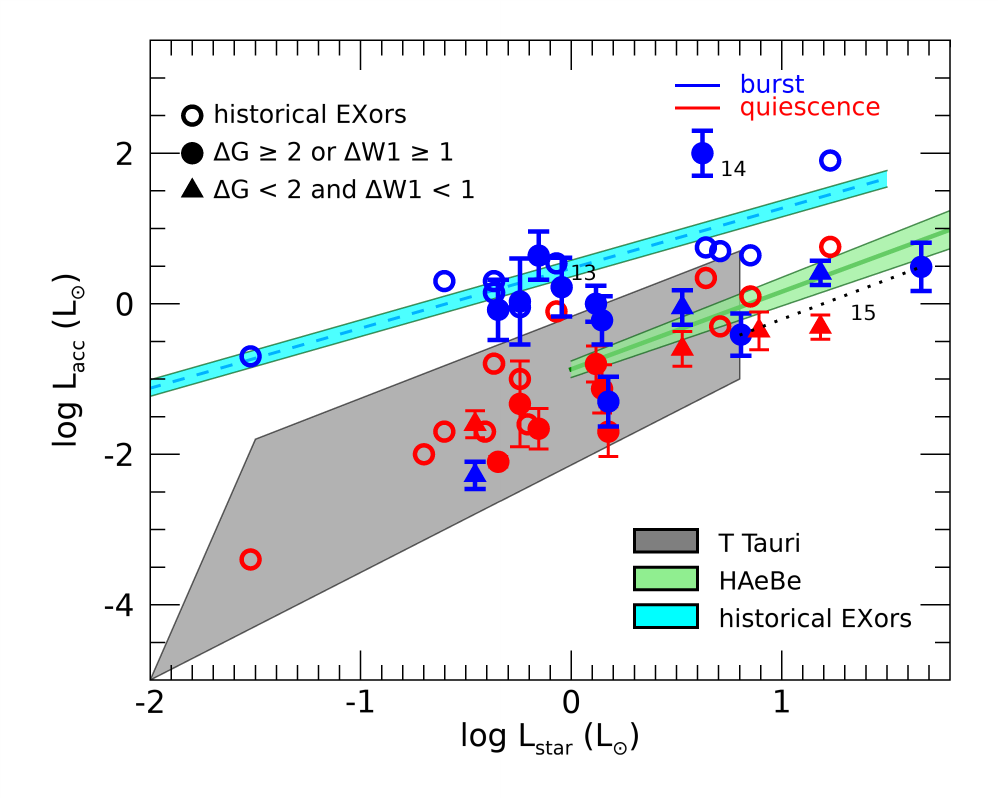}
\caption{\label{fig:fig16} \lacc\, vs. \lstar\,. Sources in quiescence and in burst are colored in red and blue, respectively. Filled circles are sources with $\Delta$\,G\,$\ga$ 2 or $\Delta$\,W1\,$\ga$\,1, while triangles are sources with $\Delta$\,G\,$<$\,2 and $\Delta$\,W1\,$<$\,1. Gray and green shaded areas represent the {\it loci} of T Tauri (Manara et al. 2024) and HAeBe stars 
(Wichittanakom et al. 2020). The blue  dotted line and the shaded area is the fit through a sample of historical EXor bursts (Giannini et al. 2024), plotted with blue (burst) and red (quiescence) open circles. The three Class I sources are indicated with their identification number, and the dotted line connects the data of source \#15 for the two distance determinations.}
\end{figure}

In this section, we will discuss the parameters derived from the emission lines, specifically the accretion luminosity and the mass accretion rate.
Figure \,\ref{fig:fig16} shows \lacc\, versus \lstar\, for the quiescence (red) and burst (blue) phases of the sources in our sample. The most active sources are represented by filled circles, while others are shown with filled triangles. For comparison, the {\it loci} of accreting T Tauri stars in several star-forming regions (Manara et al. 2021) and HAeBe stars (Wichittanakom et al. 2020) are displayed as gray and green shaded areas, respectively. The dashed cyan line and shaded area represent the fit through a sample of nine historical EXors during burst (Giannini et al. 2024), shown with blue open circles. Additionally, quiescence data for 12 known EXors are plotted with red open circles.

From this plot, we note several features:
1. When in quiescence, all our sources, as well as known EXors, occupy the same parameter space as T Tauri or HAeBe stars.
2. The variability observed in ’less active’ sources’ is consistent with that seen in T Tauri stars. 
3. During burst, four of the nine ’very active’ sources lie within the {\it locus} of EXor bursts, and two are at the upper end of the {\it locus} of T Tauri sources.
4. The position of the Class I Gaia24beh (\#14) is noteworthy, as its \lacc\, value is approximately one order of magnitude above the location of the EXor bursts. The Class I object Gaia21faq (\#13) is within the {\it locus} of the EXors, while the \lacc\, of the other Class I source Gaia21fji (\#15) appears consistent with an HAeBe star.

\begin{figure}[ht!]
\includegraphics{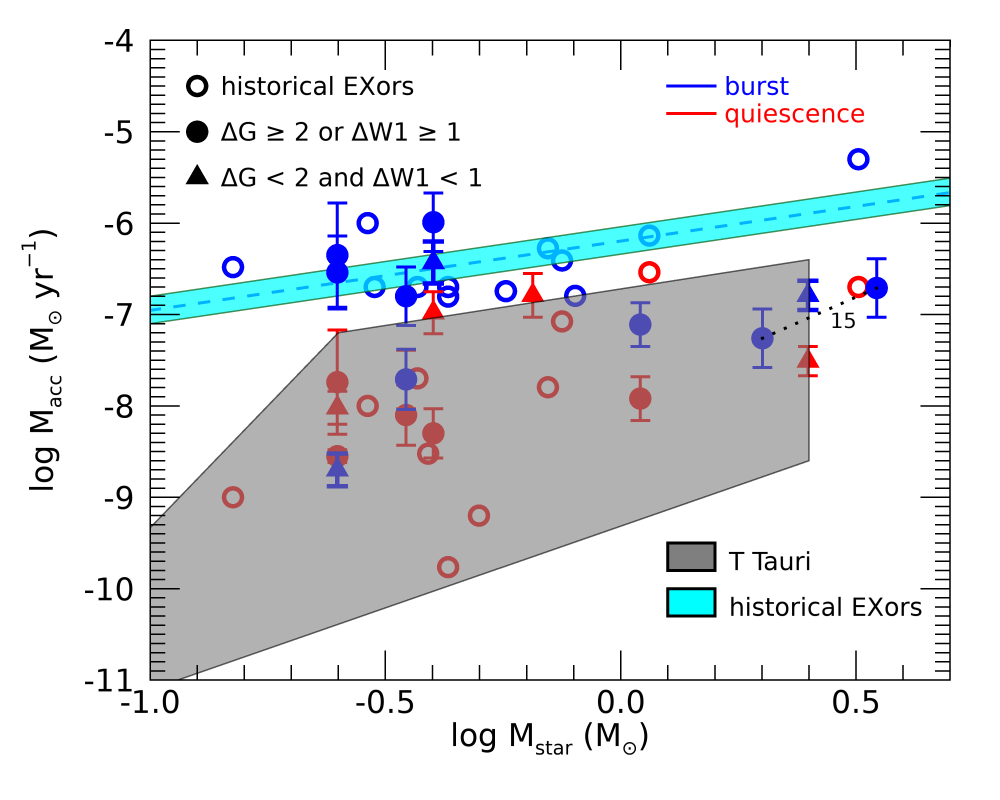}
\caption{\label{fig:fig17}  \macc\, vs. \mstar\,. Colors and symbols have the same meaning as in Figure\,\ref{fig:fig16}.}
\end{figure}

The plot of \macc\, vs. \mstar\, shown in Figure\,\ref{fig:fig17} supports conclusions similar to those drawn from the \lacc\, vs. \lstar\, plot. In particular, all sources in their quiescent state are located within the parameter space of the T Tauri sources. During bursts, the ’very active’ objects are found to be on or slightly below the {\it locus} of EXor bursts, although with some exceptions.
Unfortunately, we were  unable  to compute \mstar\, for the two Class\,I Gaia24beh and Gaia21faq, due to a lack of data needed to derive their spectral type and related parameters. For Gaia21fji, our estimates suggest values characteristic of intermediate-mass objects.

From our analysis, we can draw two main conclusions. First, sources showing photometric variations of approximately 2 magnitudes in the optical are strong EY candidates. Second, when these objects are in quiescence, they do not show significant differences from classical T Tauri or HAeBe stars.
Although our statistical sample is small, this finding supports the idea that all low- and intermediate-mass objects {\it may potentially} experience significant episodes of burst accretion.
This variability could therefore be a key process contributing to the considerable spread observed in the relations between accretion and stellar parameters.

\section{Summary and Conclusions}\label{sec:sec9}
This paper presents the ten-year light curves and optical/near-infrared spectroscopic LBT follow-up for a sample of 16 sources alerted by Gaia between 2021 and 2024 (8 known YSOs and 8 YSO candidates). While our study was based on a small statistical sample and observations were limited by unfavorable atmospheric conditions and technical issues, we were able to achieve several key results, which are summarized below.

\begin{itemize}
    \item[1.] In terms of spectral index, the sample is composed of three Class I sources, two flat spectrum sources, nine Class II sources, and two sources with unknown class.
    \item[2.] The light curves of our sample can be classified into two distinct groups: those with clear peaks above a stable baseline (Gaia21bkw, Gaia22efa, Gaia22bvi, Gaia23bab, and Gaia24afw) and those exhibiting significant variability across the entire observation period (Gaia22ehn, Gaia22dbd, Gaia21arv, Gaia23bri, Gaia21ebu, Gaia21aul, Gaia23dhi, Gaia21faq, Gaia24beh, Gaia21fji, Gaia21csu). This complex nature of the light curves suggests that multiple processes likely contribute to the observed variability.   
    \item[3.] All sources display a SED characteristic of a YSO. Specifically, they show a peak in the near-infrared or even at longer wavelengths, along with a significant infrared excess. These characteristics strongly support the classification as genuine YSOs of the 8 objects classified by Gaia as ’YSO-candidates’. 
    \item[4.] Color-color diagrams indicate that the observed variability in most of the sources is primarily due to episodes of enhanced accretion, with changes in extinction playing a minor role.
    \item[5.] The spectra show characteristics typical of YSOs, specifically a rising continuum in the optical range and the presence of emission lines. It is particularly notable that over half of the spectra display atomic forbidden lines. These lines are key indicators of gas in ejection, which is tightly linked to ongoing accretion.
    \item[6.] We have derived the stellar parameters for most of our sources through the analysis of the continuum, emission lines, and SED. With the exception of Gaia21aul and Gaia21fji, which are intermediate-mass stars, our sample primarily consists of low-mass objects with Spectral Type K-M, and age  from 0.1 to 5 Myr.   
    \item[7.] From the analysis of the light curves, we have identified as ’very active’ objects those exhibiting at least an episode of brightness increase with $\Delta$G$\ga$\,2 mag, or $\Delta$W1$\ga$\,1 mag. These sources not only show outbursts with the largest amplitudes but also experience brightness peaks more frequently, typically from three to five peaks within a ten-year period. This group includes the three Class I sources and the two flat-spectrum sources.      
    \item[8.] We have derived the accretion luminosity and the mass accretion rate in various phases of brightness from the luminosities of the accretion lines. For nine sources, we were able to determine these quantities during both quiescence and burst phases.  Specifically, for Gaia21bkw and Gaia23bab, we observed an increase in accretion luminosity and mass accretion rate of more than two orders of magnitude during burst phases. In two sources (Gaia22dbd and Gaia21csu), the observed variability is likely mainly attributable to extinction variations. For the remaining sources, the increase in accretion parameters typically ranges from a factor of 5 to 30. 
    \item[9.] The plot \lacc\, vs.\,\lstar\, reveals that all our sources, when in quiescence, occupy the same parameter space as T Tauri or HAeBe stars. This result supports the idea that all low- and intermediate-mass objects may experience significant episodes of burst accretion. 
   \item[10.]  When in burst, the ’very active’ sources fall in the parameter space of EXor bursts or close to the upper end of the T Tauri  {\it locus}, therefore being  strong EY candidates. Notably, the Class I Gaia24beh exhibits an exceptionally high accretion luminosity, approximately an order of magnitude greater than EXors of comparable mass. Furthermore, the \macc\, vs.\,\mstar\, relation supports conclusions consistent with those drawn from \lacc\, vs.\,\lstar\, relation.   
\end{itemize}
 A significant advancement in our understanding of YSO variability is expected with the upcoming Legacy Survey on Space and Time (LSST\footnote{https://rubinobservatory.org/}), conducted by the Vera C. Rubin Observatory.
The Rubin LSST's rapid cadence and high sensitivity will allow for systematic coverage and monitoring of most known star-forming regions. When combined with spectroscopic monitoring
from instruments like X-Shooter and SoXS at ESO, these observations will offer an unprecedented opportunity to determine the percentage of eruptive variables, especially in regions of moderate extinction.

\begin{appendix}
\section{Notes on the individual sources}\label{appendix:A}
In this Appendix we give some notes on the individual sources. Specifically, 
we report on : (a) distance determination with reference to Figures\,\ref{fig:fig1}-\ref{fig:fig2}; (b) brightness level at the epoch of the LBT observation; (c) notes to the individual SED, with reference to Figures \ref{fig:fig7}$-$\ref{fig:fig10}.
\begin{itemize}
\item[\bf{Gaia21bkw}]
\begin{itemize}
    \item[(a)] Distance taken from the literature.
    \item[(b)] Observed with LBT twice, once during quiescence and once in an intermediate state.
    \item[(c)] SED with excess at wavelengths $\ga$ 8 $\mu$m. 
The LBT photometric data in its quiescent state aligns well with the 2MASS , AllWISE, and Spitzer datasets. Furthermore, the data from the intermediate state show good agreement with the ZTF data. We also observe a blue excess. Notably, the Gaia photometric point during the burst phase does not match any other data points.
A blackbody function (BB) with  \teff\, =\,3190 K (as reported by Fiorellino et al. 2021), when reddened by $A_V$ = 9.8 mag, provides a good fit to the quiescence photometric points across the optical to the mid-infrared range (Figure\,\ref{fig:fig7}, panel a). 
    \end{itemize}
\item[\bf{Gaia22efa}]
    \begin{itemize}
        \item [(a)] This star is in the Auriga-California Molecular cloud at distance between 450 and 510 pc. The source has a large distance of 922$_{-187}^{+619}$pc from Bailer-Jones et al. (2021, BJ21), and has a high RUWE of 2.29. The cluster NGC\_1579 from Hunt \& Reffert (2024) has a median distance of 521 pc, while the extinction rises at 532 pc. We adopt this latter value as distance to Gaia22efa.
        \item [(b)] Object observed during quiescence. 
        \item [(c)] The SED shows a notable excess at wavelengths greater than 3.4 $\mu$m (Figure\,\ref{fig:fig7}, panel b). The LUCI photometric data points are lower than the 2MASS data, which suggests a previous high brightness level that has not been observed since 2014. The BB function 
        with \teff\,=\,3900 K, reddened by \av\,=\,5.0 mag (as determined from the continuum fitting), provides a reasonable fit to the quiescent photometric data points across the optical to near-infrared spectrum.
    \end{itemize}
\item[\bf{Gaia22bvi}]
    \begin{itemize}
        \item[(a)] Distance taken from the literature.
        \item[(b)] Object observed in quiescence.
        \item[(c)] The SED extends up to 160 $\mu$m based on Akari data and exhibits an excess at wavelengths longer than approximately 2.2 $\mu$m (as shown in Figure \ref{fig:fig7}, panel c). Our continuum fit indicates \av\,= 3.2 mag and \teff\,= 3200 K, which aligns with the findings of Dahm \& Hillenbrand (2020). A BB function accurately fits the quiescent data from the optical range up to the $K$ band photometry. Additionally, burst data are available from the Gaia, ZTF, and NEOWISE surveys.
        \end{itemize}
\item[\bf{Gaia22ehn}] 
    \begin{itemize}
        \item[(a)] The source is located in LDN 1536 in Taurus. We adopt the value estimated by Nagy et al. (in preparation) of 152 pc.
       \item[(b)] Object observed in quiescence.
       \item[(c)] The SED extends up to 160 $\mu$m (Akari data), showing excess at $\lambda$ $\ga$ 1.6 $\mu$m (Figure\,\ref{fig:fig7}, panel d). The BB fits nicely the optical and near-infrared photometries, \teff\,= 3200 K  and \av\,= 6.0 mag being derived from continuum fitting. 
       The SED extends up to 160 $\mu$m based on Akari data and exhibits an excess at wavelengths longer than approximately 1.6 $\mu$m (as shown in Figure\,\ref{fig:fig7}, panel d). 
      \end{itemize}
\item[\bf{Gaia22dbd}] 
    \begin{itemize}
        \item[(a)] We adopt the BJ21 distance of 345$_{-5}^{+7}$ pc which matches that of Orion, where the source is located.
        \item[(b)] This object was observed during an intermediate state of brightness.
        \item[(c)] The photometric data in the SED (Figure\,\ref{fig:fig8}, panel a) correspond to four distinct brightness levels of the source. As can be seen, both blue and red excesses are present. In particular, the 2MASS data were acquired when the source was at its faintest state, and were specifically used by means of the [J-H] and [H-K$_s$] colors to determine \av\, = 7.0 mag  during quiescence. \teff\, has been estimated from the SED fit with a significant degree of uncertainty.
       \end{itemize}
\item[\bf{Gaia21arv}]
    \begin{itemize}
        \item[(a)] The photogeometric distance is in agreement with that of Orion, where the object is located. The distance of 390$\pm$12 pc is in agreement with the cluster NGC\_1980.        
        \item[(b)] This object was observed solely with LUCI during a period of intermediate brightness. However, this brightness level does not appear to significantly exceed the 2MASS data points, which characterize its quiescent state.
        \item[(c)] The SED shows a moderate excess that sharply decreases at mid-infrared wavelengths  (Figure\,\ref{fig:fig8}, panel b). Our continuum fit yields \teff\,= 4180 K, which aligns with the findings of Kounkel et al. (2019). The optical and $J$-band data are well-fitted by a BB function with a reddening of \av\, = 6.0 mag, consistent with the near-infrared colors. Additionally, the ZTF $g$-band photometry indicates a blue excess.
    \end{itemize}
\item[\bf{Gaia23bri}]
    \begin{itemize}
        \item[(a)] Its BJ21 distance is 1974$_{-396}^{+671}$ pc and its RUWE is 1.04. The proper motions in Fig.\ref{fig:fig1} suggest it can be a member of HSC\_1421. Its median distance is 1514 pc. The spatial distribution and lower distance uncertainty also strengths this idea. The extinction rises at 1586 pc, which we adopt as distance.
        \item[(b)] For this source we have an optical/near-infrared spectrum in quiescence and a near-infrared spectrum obtained when the source was in an intermediate brightness state, with  magnitudes similar to those observed by 2MASS.
        \item[(c)] This SED (Figure\,\ref{fig:fig8}, panel c) displays a remarkable infrared excess. A \teff\,= 4330 K and \av\,=2.4 mag, as derived from the continuum fitting in quiescence, are used to plot the BB function, normalized to the LUCI datum in obtained in quiescence. This function is slightly above MODS and PAN-STARRS data points, maybe because the source was brighter at the time of LUCI observations. 
    \end{itemize}
\item[\bf{Gaia21ebu}] 
    \begin{itemize}
        \item[(a)] The BJ21 distance is 1383$_{-164}^{+199}$ pc. It is inside the Rosette Nebula (d\,$\sim$\,1300 pc) so the distance is in agreement with its location. The distance histogram agrees with the adopted distance as well.
        \item[(b)] The source was observed in an intermediate state that shows little difference from both its quiescent and burst phases.
        \item[(c)] The source exhibits an excess primarily in the AllWISE data at $\lambda$ $\ga$ 10 $\mu$m. A BB function, reddened by \av\,=3.0 mag as determined from the near-infrared colors, provides the best fit to the quiescence points for a temperature \teff\,=\,4900 K (as shown in \ref{fig:fig8}, panel d).
    \end{itemize}
\item[\bf{Gaia21aul}]
    \begin{itemize}
        \item[(a)] The distance of 379$_{-11}^{+12}$ pc from BJ21 aligns with the distance distribution of UPK\_39 members.
        \item[(b)] This object was observed during quiescence.
        \item[(c)] The SED shows an excess extending up to 160 $\mu$m, based on Akari data. Our continuum fitting indicates \teff\,=\,5950 K and \av\,=\,4.0 mag. The reddened BB function effectively fits the optical (Gaia, ZTF and MODS) and near-IR data. However, the PAN-STARRS data points are notably below the other observations, a trend also visible in the light curve. These data points can be accurately fitted using the same black body function, provided it is reddened by an \av\, of approximately 5 mag (Figure\,\ref{fig:fig9}, panel a).
    \end{itemize}
\item[\bf{Gaia23bab}]
    \begin{itemize}
        \item[(a)] Distance taken from the literature.
        \item[(b)] This source was observed both in quiescence and in burst. 
        \item[(c)] The SED in burst is already described in Giannini et al. (2024), while in this paper we present data obtained in quiescence (Figure\,\ref{fig:fig9}, panel b). The continuum fitting of the quiescence data provides \teff\,=\,3630 K and \av\,=3.2 mag. The reddened blackbody function fits the Gaia, MODS, LUCI, and 2MASS data. ZTF and PAN-STARRS photometries are below the fit, possibly due to source variability across the different surveys. 
    \end{itemize}
\item[\bf{Gaia23dhi}] 
    \begin{itemize}
        \item[(a)] The photogeometric distance and the position angle of the proper motion vector suggests it is a member of HSC\_432. We adopt the BJ21 distance of 2735$_{-451}^{+638}$ pc.      
        \item[(b)] This object was observed during an intermediate state of brightness.
        \item[(c)] The photometric data probe three distinct brightness levels, with the 2MASS data specifically representing the highest (Figure\,\ref{fig:fig9}, panel c). Through continuum fitting, we have estimated \teff\,=\,4800 K. The BB at this temperature, reddened by 
        \av\,=\,3.0 mag (as determined from near-infrared colors) provides a good fit to the optical Gaia, PAN-STARRS, and ZTF data. We have normalized this BB function using the photometric point in the $J$-band obtained with LUCI.
    \end{itemize}
\item[\bf{Gaia24afw}] 
    \begin{itemize}
        \item[(a)] Although the proper motion slightly differs to that of the cluster NGC\_6823, distance suggests it is a member of it. We use the BJ21 distance of 2150$_{-570}^{+973}$ pc.
        \item[(b)] This object was observed during quiescence. 
        \item[(c)] The SED shows a peak around 3 $\mu$m, followed by a decline in the mid-infrared range, where a noticeable excess is present (Figure\,\ref{fig:fig9}, panel d). Our fit to the optical continuum indicates \teff\,=\,3630 K and \av\,= 4.0 mag.  We have used these values to plot the reddened BB function, which provides a good fit through
        all the optical data.
    \end{itemize}
\item[\bf{Gaia21faq}]
    \begin{itemize}
        \item[(a)] The star is in the Cygnus-X SFR at 1400 pc, which falls in the range of values from BJ21 (1190$_{-258}^{+474}$ pc).
        \item[(b)] This object has been observed during an intermediate state of brightness.
        \item[(c)]  The SED shows a steep increase with wavelengths up to approximately 3 $\mu$m, followed by a plateau at longer wavelengths (Figure\,\ref{fig:fig10}, panel a). The spectral index suggests that this is a Class I object. Analysis of 2MASS photometry indicates that Gaia21faq was in a fainter state about 30 years ago compared to its current brightness. Data from PAN-STARRS, ZTF, LBT, and AllWISE are consistent with each other, showing an intermediate brightness level relative to the peak of the Gaia light curve. The 2MASS colors indicate \av\,= 5.0 mag.  A similar value is obtained from continuum fitting, suggesting that the extinction did not significantly change during the different brightness phases observed. No BB fit was performed.
        \end{itemize}
\item[\bf{Gaia24beh}]
    \begin{itemize}
        \item[(a)] The spatial distribution shows that this source can be located on the edge of the cluster UPK\_127. We use the distance 741$_{-15}^{+17}$ pc from BJ21.
        \item[(b)] Source  observed during burst.
        \item[(c)] The 2MASS data reveal a considerably fainter brightness phase for Gaia24beh compared to what has been observed over the following three decades, similar to the case of Gaia21faq. As illustrated in Figure\,\ref{fig:fig10}, (panel b), the SED exhibits a steep increase towards longer wavelengths, classifying Gaia24beh as a Class I source. The near-infrared color-color diagram in quiescence indicates an extinction  of 10 magnitudes, which is higher than the value of 5.2 magnitudes estimated from continuum fitting in burst. This discrepancy suggests that a variation in extinction plays a part in the observed brightening.  No BB fit was performed. 
    \end{itemize}
\item[\bf{Gaia21fji}]
    \begin{itemize}
        \item[(a)] The star is located toward Cygnus-L988, a cloud that supposedly has a distance of 600 pc, which aligns with both the peak of the histogram of LDN\_988e and the extinction increases at 626 pc. The BJ21 distance is 225$_{-18}^{+22}$ so it is separated from the other two estimates. We will adopt both the distances (Nagy et al., in preparation)
        \item[(b)] This object was observed during burst. 
         \item[(c)] The photometric data of Gaia21fji  suggest it is a Class I source  (Figure\,\ref{fig:fig10}, panel c). The quiescent data are fitted by a BB at $T$=6000 K reddened by \av\,= 10.0 mag, in agreement with the SpT derived from the 370$-$450 nm fitting and the \av\, obtained from the near-infrared color-color diagram.        
    \end{itemize}
\item[\bf{Gaia21csu}] 
    \begin{itemize}
        \item[(a)] This star is in LDN1218. The extinction rises at 856 pc, which aligns with the BJ21 distance of 834$_{-78}^{+93}$.
        \item[(b)] This source was observed during quiescence with MODS and in an intermediate state with MODS and LUCI. 
        \item[(c)] In the SED (Figure\,\ref{fig:fig10}, panel d) we observe an excess at $\lambda$ $\ga$ 4.4 $\mu$m. Our continuum fitting indicates \teff\,=3500 K and \av\,=3.8 mag, which fits most of the optical data.
    \end{itemize}

\end{itemize}

\section{LBT photometry}\label{appendix:B}
In this Appendix we show the tables of the optical ($griz$) and near-infrared ($JHK_s$) magnitudes obtained with MODS and LUCI, respectively.
\begin{table*}
\small
\center
\caption{\label{tab:tabB1} Optical photometry.}
\begin{tabular}{ccccccc}
\hline
\hline
ID  & Source     & Date          & g      & r      & i      & z     \\
\hline
1 & Gaia21bkw  &  2023 Mar 05  & $>$20.4$^a$ & 18.78  &  16.3$^a$     & 14.9$^a$     \\   
2 & Gaia22efa  &  2023 Oct 17  & 18.8    &  16.0  & 14.3   &  13.0 \\
3 & Gaia22bvi  &  2023 Oct 19  & 16.46  & 14.72  & 13.13  &  12.02 \\
4 & Gaia22ehn  &  2023 Oct 19  & 20.19 & 17.65  &  15.46 &  14.33 \\
5 & Gaia22dbd  &  2024 Jan 10  & 17.10  & 15.63  &  14.67 &  14.08 \\
7 & Gaia23bri  &  2023 Oct 19  & 18.62  & 17.61  &  16.78 &  16.06 \\
8 & Gaia21ebu  &  2024 Jan 10  & 16.75  & 15.63  &  14.95 &  14.47 \\
9 & Gaia21aul  &  2023 Jun 04  & 14.59   & 13.13 & 12.02 & 11.31\\
10 & Gaia23bab  &  2024 Apr 11  & -      & 19.80  & 17.46  & 16.44  \\
11 & Gaia23dhi  &  2024 Jun 07  & 18.22  & 17.30  & 16.45  & 15.88\\
12 & Gaia24afw  &  2024 Jun 07  & 20.41  & 18.78  & 17.49  & 16.70\\
13 & Gaia21faq  &  2023 Jun 04  &  -     & 18.39  & 17.24  &  16.38 \\
14 & Gaia24beh  &  2024 Jun 08  & 18.00  & 15.69  & 14.37  & 13.51\\
15 & Gaia21fji  &  2023 Oct 19  & 15.99  & 14.61  & 13.78  &  13.13 \\
16 & Gaia21csu  &  2022 Dec 17  & 17.51  & 16.27  & 15.23  &  14.78 \\
      &         &  2023 Oct 19  & 18.86  & 17.77  & 16.33  &  15.35\\
\hline\end{tabular}	
\begin{quotation}
\textbf{Notes.} Errors are within 0.1 mag. $^a$Taken from the ZTF catalog.
\end{quotation}
\end{table*}  
\begin{table*}
\small
\center
\caption{\label{tab:tabB2} Infrared photometry.}
\begin{tabular}{ccccccc}
\hline
\hline
ID & Source     & Date      &   J    &  H     & K$_s$      \\
\hline
1 & Gaia21bkw   &  2021 Oct 15      &  13.61  & 11.83  & 10.88  \\
  &             &  2023 Feb 28      &  12.48  & 10.53  &  9.50  \\
2 & Gaia22efa   &  2023 Nov 23      &  10.92  & 9.84   &  9.18  \\
3 & Gaia22bvi   &  2023 Nov 22      &   9.81  &  9.10  & 8.58   \\
5 & Gaia22dbd   &  2024 Jan 09      & 12.60   &  11.53 &10.71 \\
6 & Gaia21arv   &  2021 Oct 15      &  12.74  &  11.31 & 10.44    \\
7 & Gaia23bri   &  2023 Nov 23      &  14.53  &  -     &  -        \\
  &             &  2024 Jan 09      &  13.80  & 12.58 & 11.47       \\
8 & Gaia21ebu   &  2024 Jan 09      &  13.05  & 12.05  &11.54      \\ 
9 & Gaia21aul   &  2023 Jun 04      &  10.00  &  8.82  &  7.73     \\
10& Gaia23bab  &  2024 Apr 11       &  14.11  & 12.93 & 12.29      \\
11& Gaia23dhi  &  2024 Jun 10       &  14.79  &  13.47  &  12.56  \\
13& Gaia21faq  &  2023 Jun 04      &  13.90  & 12.05  & 10.22     \\
16& Gaia21csu  &  2024 Jan 09      &  13.17  & 12.15  & 11.70     \\
\hline\end{tabular}	
\begin{quotation}
\textbf{Note.} Errors are within  0.2 mag. 
\end{quotation}
\end{table*}

\section{LBT spectra}\label{appendix:C}
Here we show the optical and near-infrared spectra of the sources of our sample. Quiescence, intermediate, and burst spectra are colored in red, green, and blue, respectively. Observation dates are indicated. Black stars are the photometric points obtained in the same night as the spectrum. The most important lines detected are labeled.

\begin{figure}
\epsscale{0.8}
\includegraphics{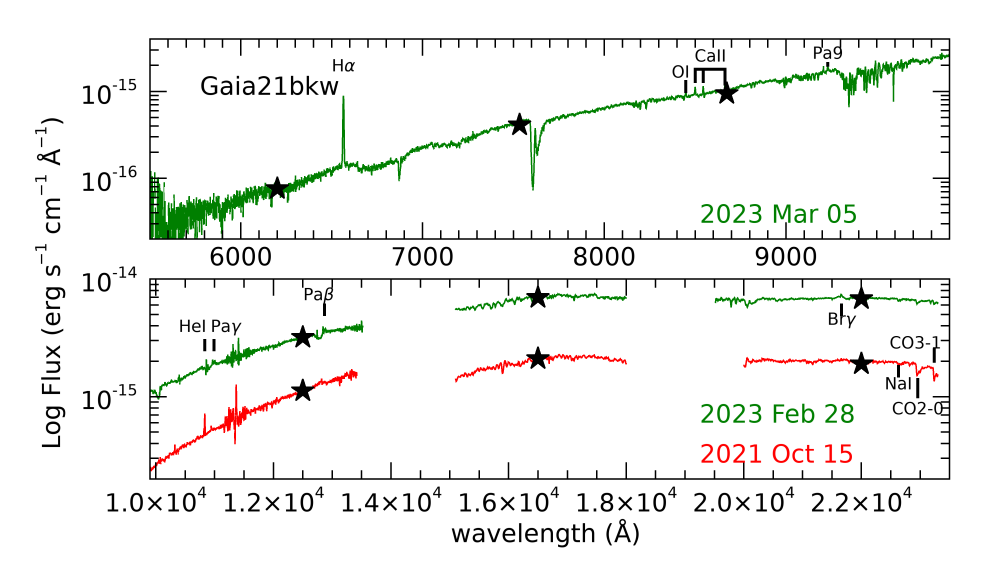}
\caption{\label{fig:fig18} Quiescence and intermediate state spectrum of Gaia21bkw.  }
\end{figure}

\begin{figure}
\epsscale{0.8}
\includegraphics{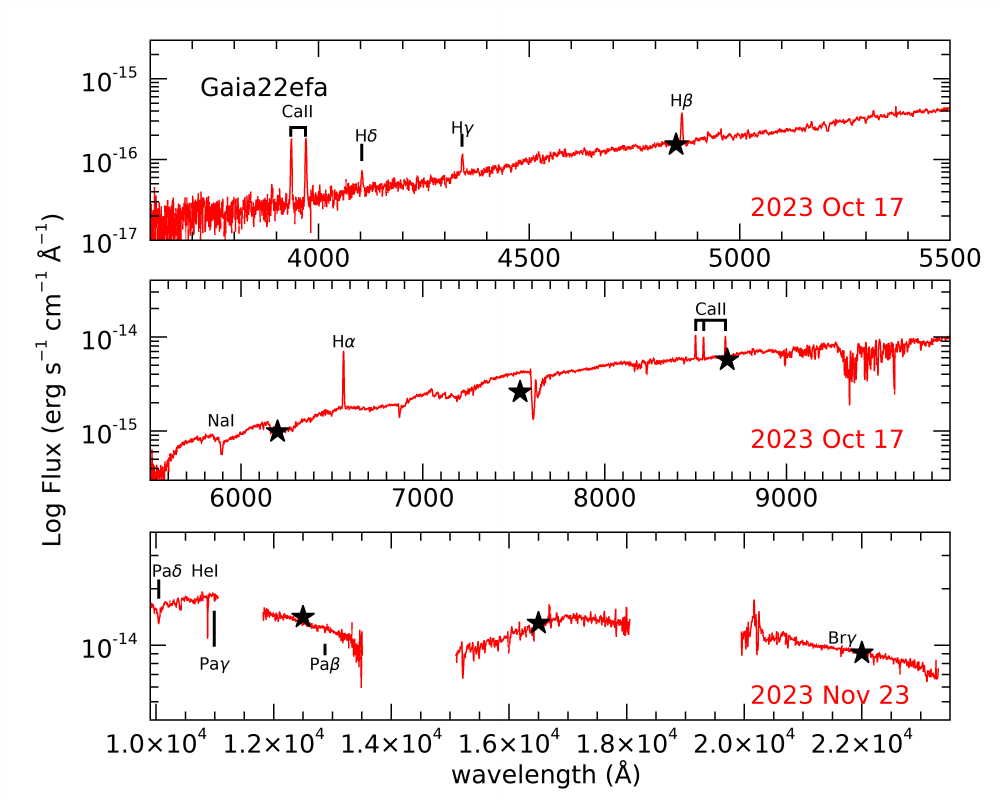}
\caption{\label{fig:fig19} Quiescence spectrum of Gaia22efa.}
\end{figure}

\begin{figure}
\epsscale{0.8}
\includegraphics{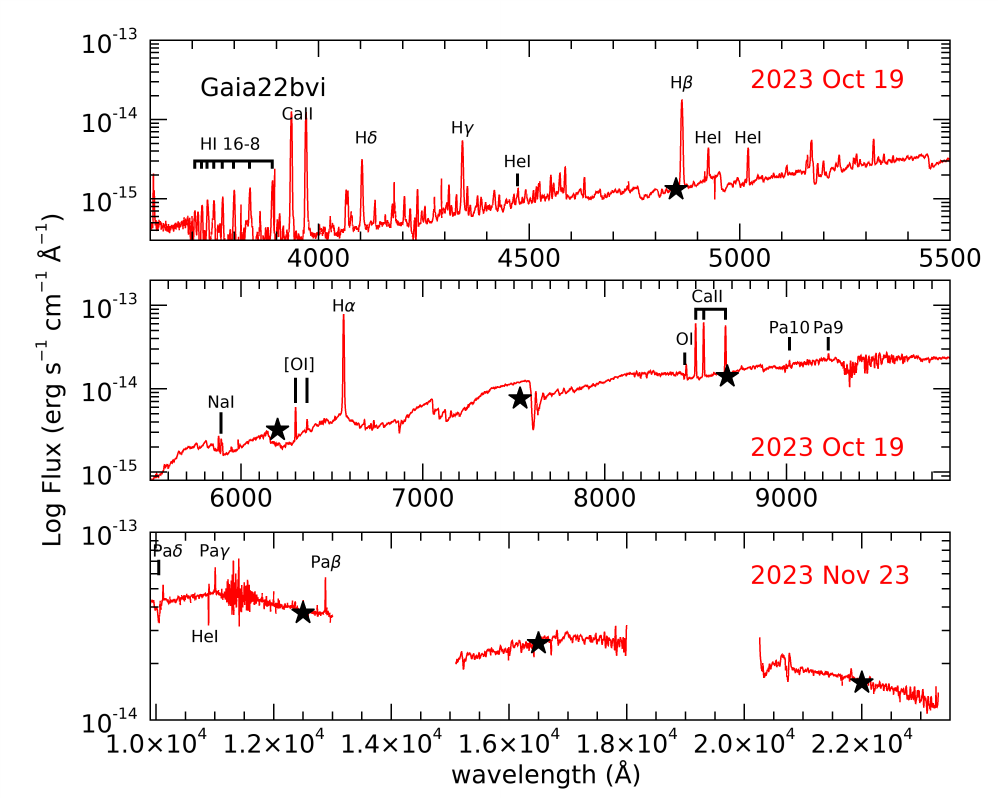}
\caption{\label{fig:fig20} Quiescence spectrum of Gaia22bvi. }
\end{figure}

\begin{figure}
\epsscale{0.8}
\includegraphics{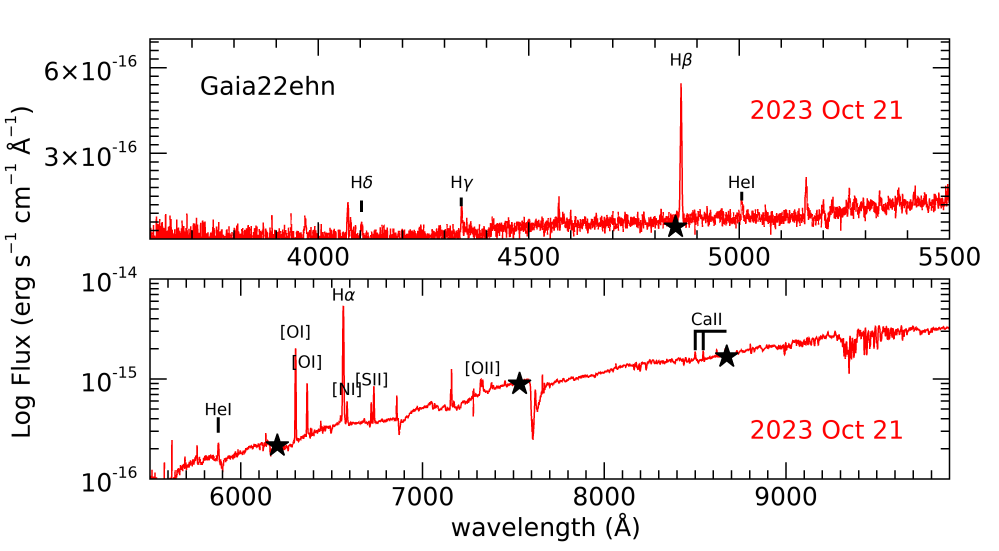}
\caption{\label{fig:fig21} Quiescence spectrum of Gaia22ehn.}
\end{figure}

\begin{figure}
\epsscale{0.8}
\includegraphics{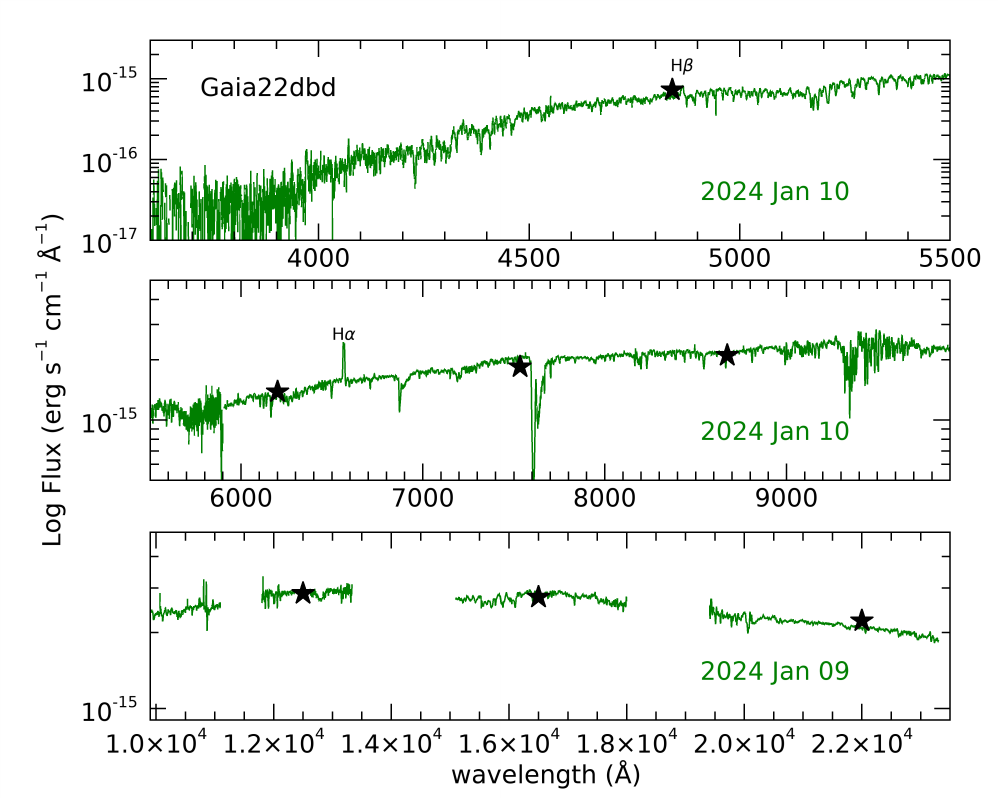}
\caption{\label{fig:fig22} Intermediate state spectrum of Gaia22dbd.}
\end{figure}

\begin{figure}
\epsscale{0.8}
\includegraphics{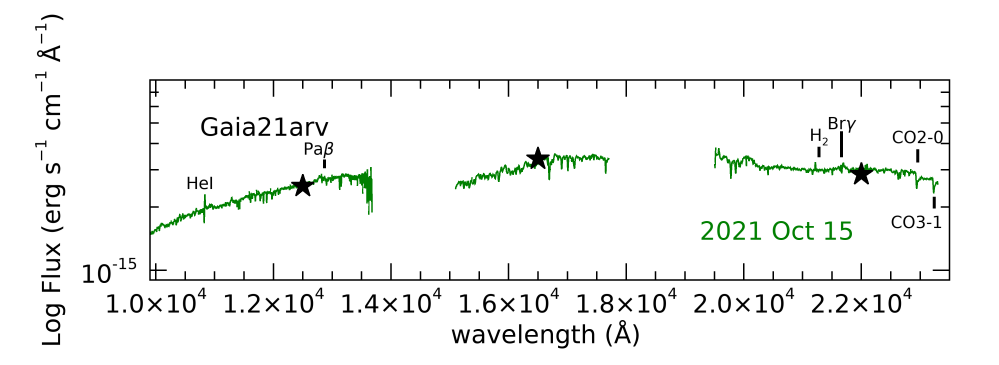}
\caption{\label{fig:fig23} Intermediate state spectrum of Gaia21arv.}
\end{figure}

\begin{figure}
\epsscale{0.8}
\includegraphics{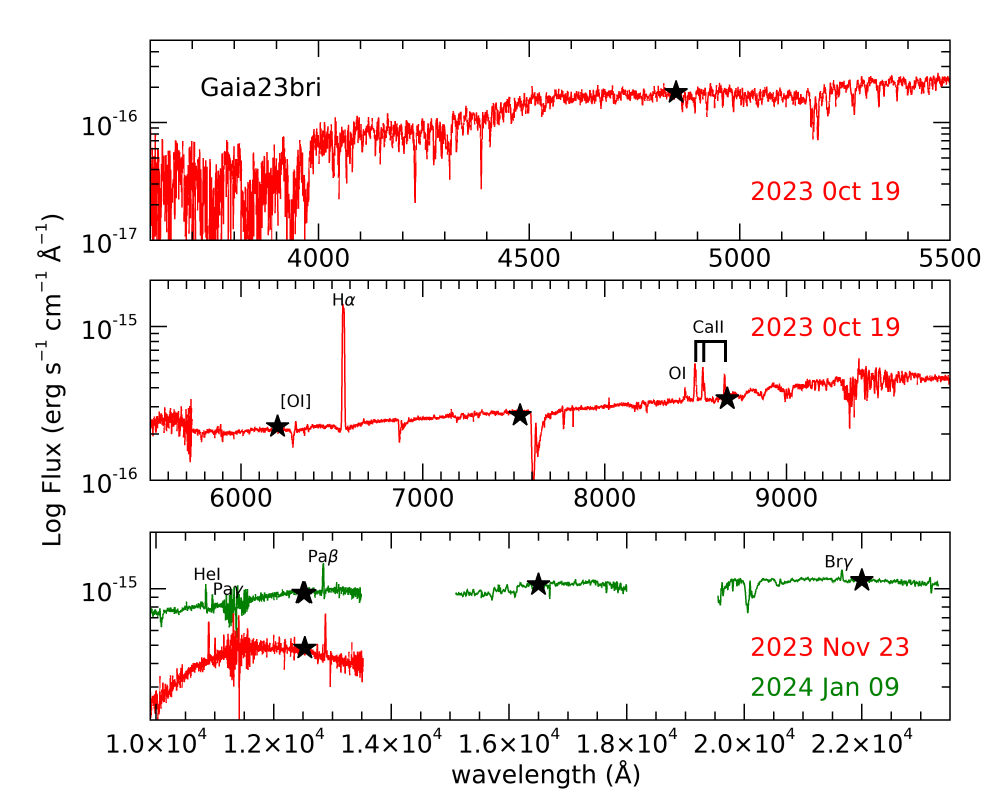}
\caption{\label{fig:fig24} Quiescence and intermediate state spectrum of Gaia23bri.}
\end{figure}

\begin{figure}
\epsscale{0.8}
\includegraphics{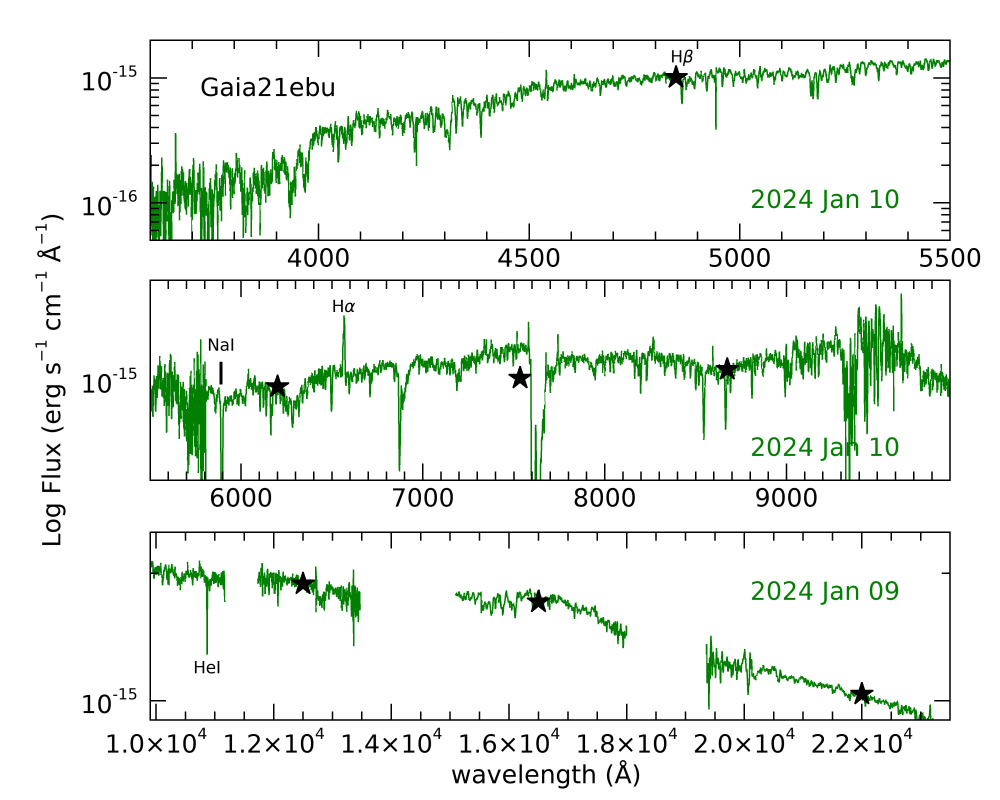}
\caption{\label{fig:fig25} Intermediate state spectrum of Gaia21ebu.}
\end{figure}

\begin{figure}
\epsscale{0.8}
\includegraphics{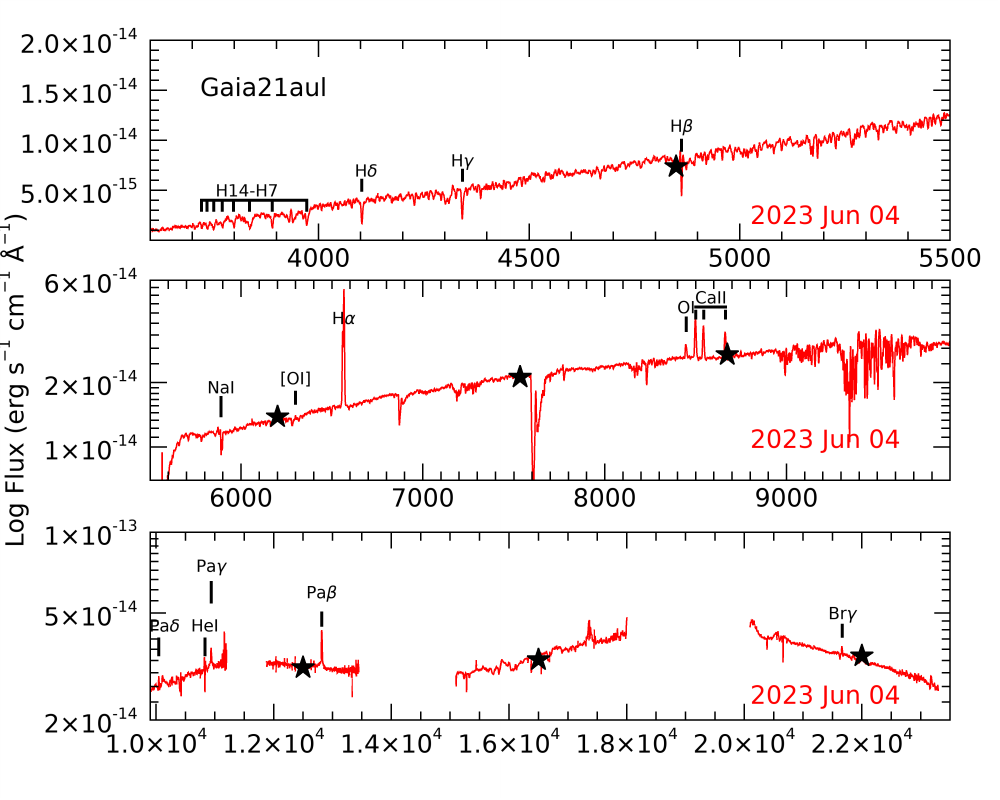}
\caption{\label{fig:fig26} Quiescence spectrum of Gaia21aul.}
\end{figure}

\begin{figure}
\epsscale{0.8}
\includegraphics{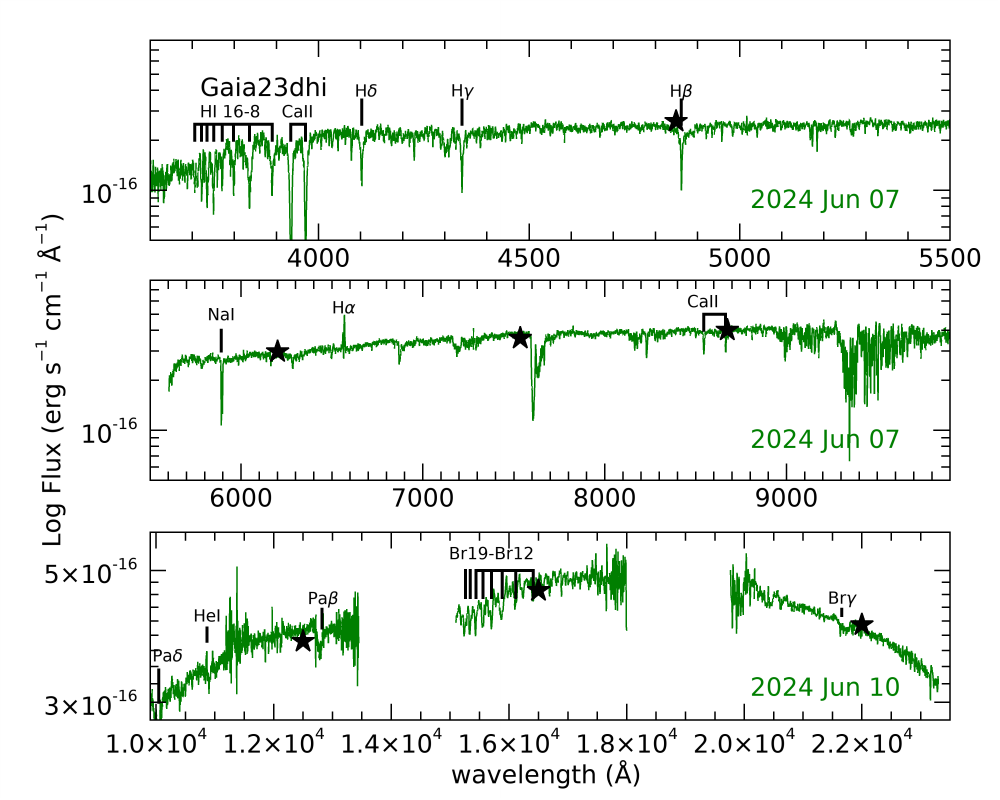}
\caption{\label{fig:fig27} Intermediate state spectrum of Gaia23dhi.}
\end{figure}

\begin{figure}
\epsscale{0.8}
\includegraphics{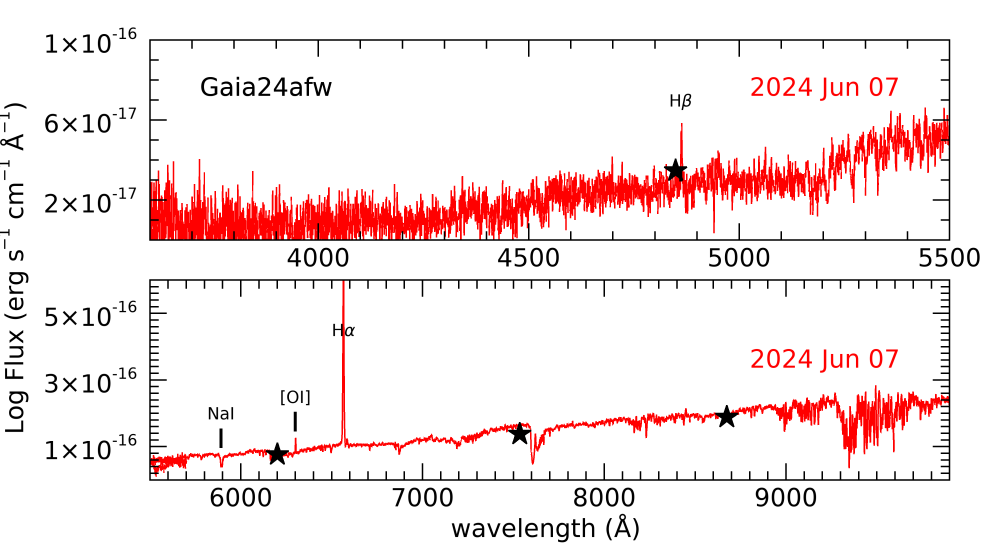}
\caption{\label{fig:fig28} Quiescence spectrum of Gaia24afw.}
\end{figure}

\begin{figure}
\epsscale{0.8}
\includegraphics{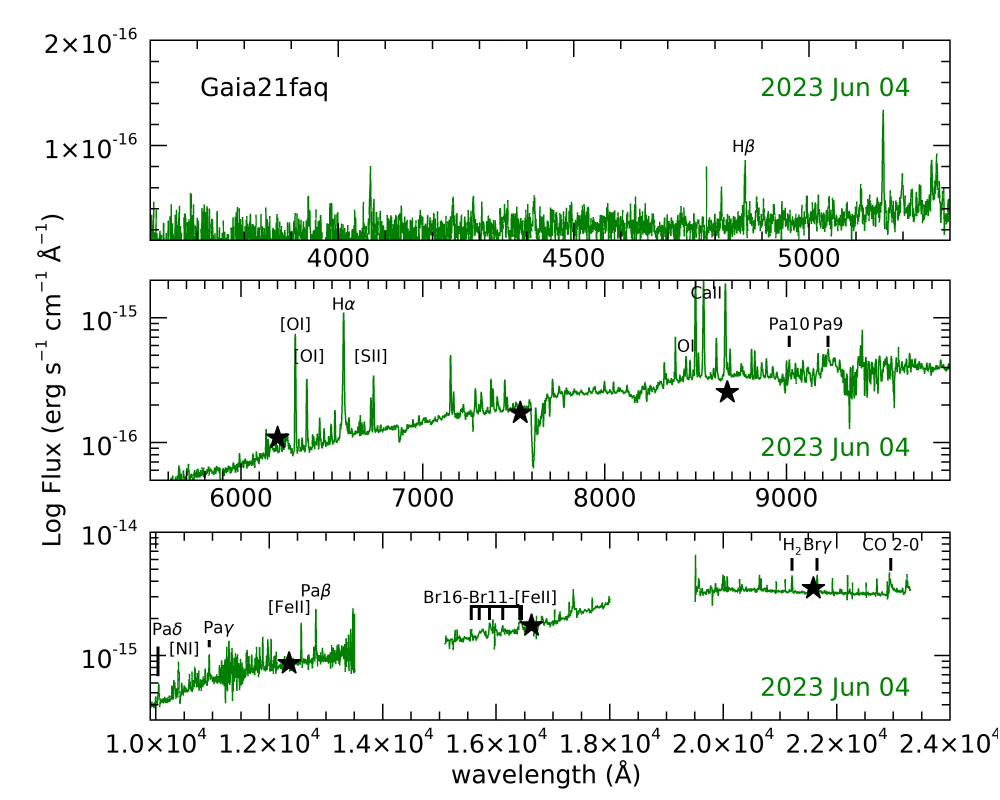}
\caption{\label{fig:fig29} Intermediate state spectrum of Gaia21faq. }
\end{figure}

\begin{figure}
\epsscale{0.8}
\includegraphics{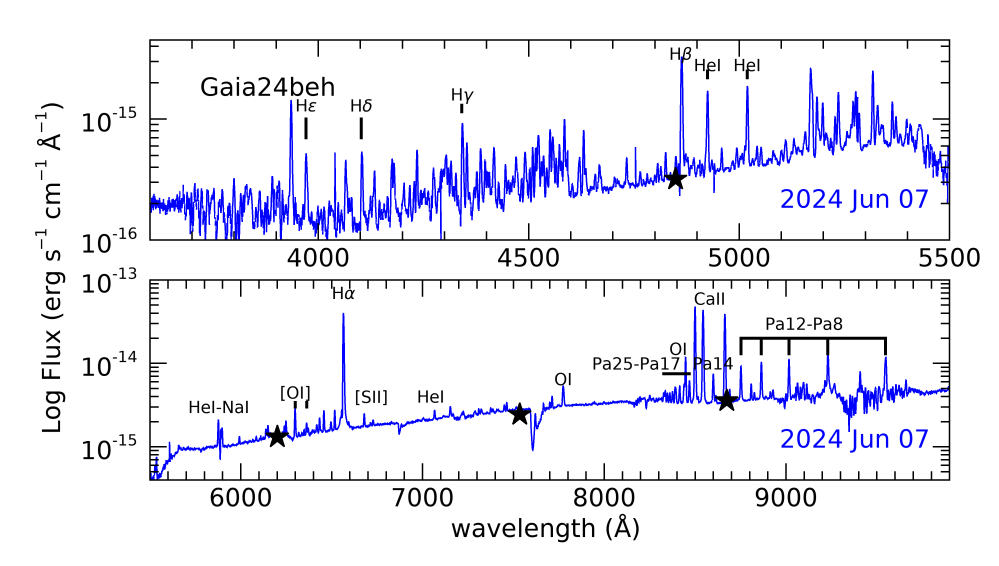}
\caption{\label{fig:fig30} Burst spectrum of Gaia24beh.}
\end{figure}

\begin{figure}
\epsscale{0.8}
\includegraphics{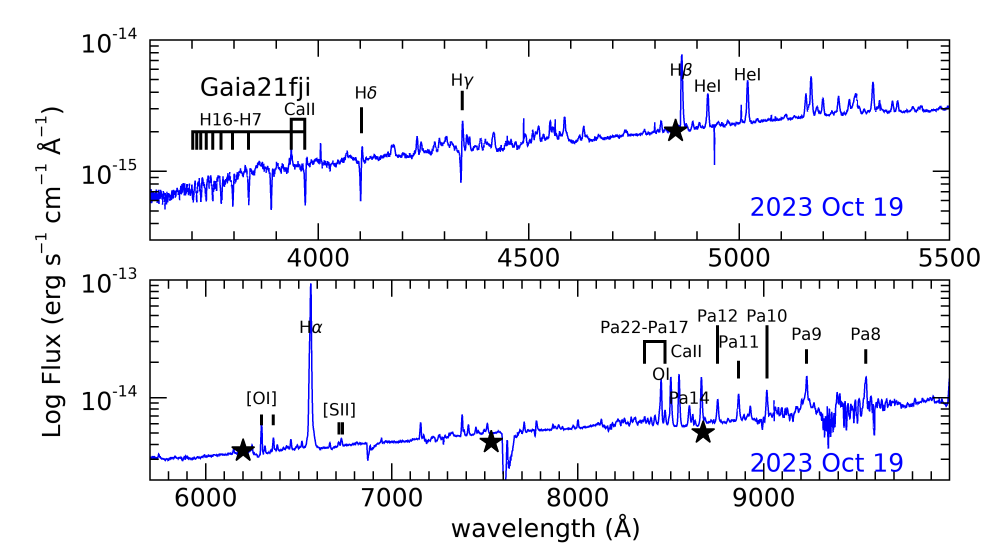}
\caption{\label{fig:fig31} Burst spectrum of Gaia21fji.}
\end{figure}

\begin{figure}
\epsscale{0.8}
\includegraphics{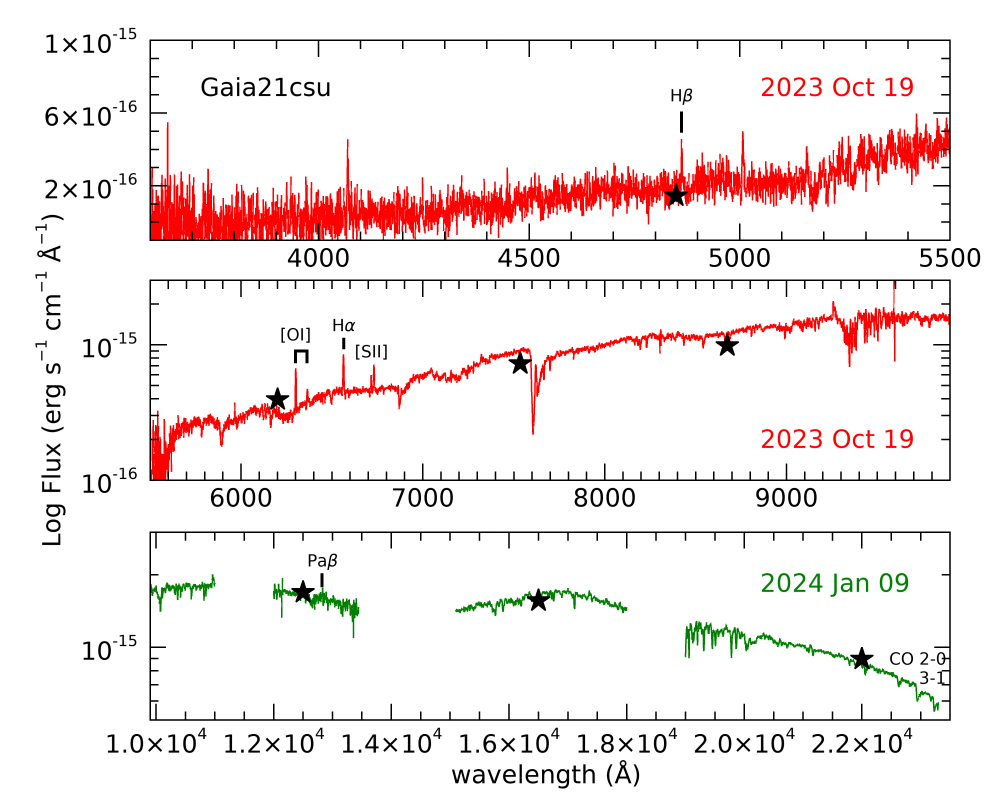}
\caption{\label{fig:fig32} Quiescence and intermediate spectrum of Gaia21csu.}
\end{figure}

\section{Line fluxes}\label{appendix:D}
In this Appendix we report the fluxes of Hydrogen
(Table\,\ref{tab:tabC1}) and other accretion lines (Table\,\ref{tab:tabC2}), as well as fluxes of ejection lines (Table\,\ref{tab:tabC3}).

\begin{table}
\scriptsize
\caption{\label{tab:tabC1} Fluxes of main Hydrogen lines.}
\begin{tabular}{cccccccccc}
\hline
Source    &  Date        &   H$\delta$  & H$\gamma$    & H$\beta$ &  H$\alpha$   & Pa$\delta$ & Pa$\gamma$ & Pa$\beta$ &Br$\gamma$ \\ 
\hline
          &              & \multicolumn{7}{c}{F$\pm\Delta$F (10$^{-16}$ erg s$^{-1}$ cm$^{-2}$)}\\
\hline
\hline
Gaia21bkw & 2021 Oct 15  &              &              &               &            &              & 5$\pm$2     &  11.3$\pm$3.2 &          \\
          & 2023 Feb 28  &              &              &               &            &              &17.2$\pm$1.9 &  65.3$\pm$3.2 &86.8$\pm$4.1\\
          & 2023 Mar 05  &              &              &               & 46.7$\pm$0.4&             &             &             &          \\
Gaia22efa & 2023 Oct 17  & 1.7$\pm$0.14 & 2.2$\pm$0.14 &9.4$\pm$0.1    & 318$\pm$0.2 &             &             &             &          \\
Gaia22efa & 2023 Nov 23  &              &              &               &             &$-$940$\pm$49&$-$150$\pm$49&$-$105$\pm$38&$-$102$\pm$47 \\
Gaia22bvi & 2023 Oct 19  &105.6$\pm$1.1 &189.1$\pm$1.4 &713.9$\pm$1.1  &4736$\pm$1.8 &             &             &             &          \\
          & 2023 Nov 23  &              &              &               &             &$-$2320$\pm$160&488.6$\pm$130&1729$\pm$80&          \\
Gaia22ehn & 2023 Oct 19  &2.6$\pm$0.4   &4.1$\pm$0.3   &19.6$\pm$0.2   &288.0$\pm$0.5&             &             &             &          \\
Gaia22dbd & 2024 Jan 10  &              &              &13.5$\pm$1.2   & 96.8$\pm$1.2&             &             &             &          \\
Gaia21arv & 2021 Oct 15  &              &              &               &             &             &             &17.2$\pm$4.6 & 66.3$\pm$11\\
Gaia23bri & 2023 Oct 19  &              &              &               &140.1$\pm$0.9&             &             &             &            \\
          & 2023 Nov 23  &              &              &               &             &             &14.1$\pm$1.7 &44.6$\pm$1.7 &            \\
          & 2024 Jan 09  &              &              &               &             &             &23.6$\pm$1.7 &78.3$\pm$2.0 & 40.5$\pm$3\\
Gaia21ebu & 2024 Jan 10  &              &              &$-$12.5$\pm$1.0&31.8$\pm$0.5 &             &             &             &            \\
Gaia21aul & 2023 Jun 04  &$-$108$\pm$10 &$-$104$\pm$7.4&$-$95.5$\pm$8.8&2429$\pm$22  &$-$852$\pm$70& 907$\pm$43   & 2565$\pm$88            & ($-$224) 442$\pm$110\\
Gaia23bab & 2024 Apr 11  &              &              &                &25.7$\pm$0.1 & 7.8$\pm$1.1& 13.0$\pm$1.0 & 13.1$\pm$1.0 & 9.1$\pm$2.3\\
Gaia23dhi & 2024 Jun 07  &$-$6.1$\pm$0.1&$-$5.2$\pm$0.1&$-$7.5$\pm$0.1  &3.1$\pm$0.1 &$-$33.4$\pm$0.1&           &$-$27 $\pm$0.1&$-$4.0$\pm$0.1\\
Gaia24afw & 2024 Jun 07  &              &              &  1.1$\pm$0.4   &36.9$\pm$0.1 &               &           &              &              \\
Gaia21faq & 2023 Jun 04  &              &              &  1.6$\pm$0.4   &65.8$\pm$0.6 & 34.1$\pm$1.9  & 47.7$\pm$1.2  &   172.6$\pm$1.4     &  282.0$\pm$2.3 \\
Gaia24beh & 2024 Jun 07  & 17.8$\pm$0.4 & 33.6$\pm$0.4 &132.7$\pm$0.4   &2631$\pm$1.1 &               &              &                     &    \\
Gaia21fji & 2023 Oct 19  &($-$18) 9.6$\pm$1.7&($-$44) 28.5$\pm$1.6 &236$\pm$1.6   &7047$\pm$3.3 &     &              &                     &    \\
Gaia21csu & 2023 Oct 19  &              &              &6.4$\pm$0.5     &26.9$\pm$1.0 &               &              &                     &    \\
          & 2024 Jan 09  &              &              &                &             &               &              &    16.2$\pm$2.4     &    \\

\hline
\hline
\end{tabular}	
\begin{quotation}
\textbf{Note.} Negative numbers in parenthesis represent fluxes of absorption components.
\end{quotation}
\end{table} 

\begin{table}
\scriptsize
\caption{\label{tab:tabC2} Fluxes of other accretion lines.}
\begin{tabular}{ccccccccc}
\hline
Source    &  Date        &   OI 8448  & HeI 4925    & HeI 5020  &  HeI 5876  &  CaII 8500     & CaII 8544    & CaII 8664  \\ 
\hline
          &              & \multicolumn{6}{c}{F$\pm\Delta$F (10$^{-16}$ erg s$^{-1}$ cm$^{-2}$)}\\
\hline
\hline
Gaia21bkw & 2023 Mar 05  &  6.7$\pm$0.4 &            &            &             &  11.5$\pm$0.4 & 9.9$\pm$0.4 & 8.8$\pm$0.5   \\
Gaia22efa & 2023 Oct 17  &              &            &            &             & 237$\pm$0.2   & 219$\pm$0.2 & 203$\pm$0.2 \\
Gaia22bvi & 2023 Oct 19  & 257.3$\pm$1.7&95.6$\pm$1.1&96.6$\pm$1.0&46.2$\pm$1.8 & 2576$\pm$1.6 & 2767$\pm$2.0 &2415$\pm$1.8 \\
Gaia22ehn & 2023 Oct 19  &              &            &            & 4.6$\pm$0.7 & 22.1$\pm$0.7 & 15.5$\pm$0.6 & 15.8$\pm$0.5 \\
Gaia23bri & 2023 Oct 19  &  6.1$\pm$0.4 &            &            &             & 24.5$\pm$0.6 &  19.6$\pm$0.7& 12.9$\pm$0.6 \\
Gaia21aul & 2023 Jun 04  &  402$\pm$19  &            &            &             &1230$\pm$18   &  978$\pm$22  & 771$\pm$19   \\
Gaia23bab & 2024 Apr 11  &  2.9$\pm$0.1 &            &            & 1.1$\pm$0.1 &2.5$\pm$0.1   &  2.4$\pm$0.1 & 2.2$\pm$0.1   \\
Gaia23dhi & 2024 Jun 10  &              &            &            &             &              &$-$6.8$\pm$0.1&$-$3.9$\pm$0.1 \\
Gaia21faq & 2023 Jun 04  & 10.7$\pm$0.6 &            &            &             & 107$\pm$0.6  & 104$\pm$0.5  &  94$\pm$0.5    \\
Gaia24beh & 2024 Jun 07  &519.8$\pm$0.6 &55.8$\pm$0.4&57.7$\pm$0.4&65.9$\pm$0.5 &2886$\pm$0.7  & 2947$\pm$0.7  &  2538$\pm$0.7    \\
Gaia21fji & 2023 Oct 19  &744.2$\pm$3.3 &84.6$\pm$1.8&112.3$\pm$1.9&            &76.7$\pm$2.4  & 84.2$\pm$3.0  &  81.4$\pm$3.2    \\
\hline
\hline
\end{tabular}	
\begin{quotation}
\textbf{Note.} Negative numbers in parenthesis represent fluxes of absorption components.
\end{quotation}
\end{table} 
\begin{table}
\scriptsize
\caption{\label{tab:tabC3} Fluxes of main ejection lines.}
\begin{tabular}{ccccccccc}
\hline
Source    &  Date          &   HeI 1.08         & [OI]6300        & [SII]6732     & [NI]1.04     &[NII]6584    &[FeII]1.25 & H$_2$ 2.12 \\ 
\hline
          &                & \multicolumn{7}{c}{F$\pm\Delta$F (10$^{-16}$ erg s$^{-1}$ cm$^{-2}$)}\\
\hline
\hline
Gaia21bkw & 2021 Oct 15  &    24.0$\pm$1.8       &                  &             &              &            &             &            \\
          & 2023 Feb 28  &($-$8) 26.0$\pm$1.6    &                  &             &              &            &             &            \\
Gaia22efa & 2023 Nov 23  &$-$721$\pm$39          &                  &             &              &            &             &            \\    
Gaia22bvi & 2023 Oct 19  &                       &190.5$\pm$1.7     &             &              &            &             &            \\   
          & 2023 Nov 23  &$-$933$\pm$67          &                  &             &              &            &             &            \\     
Gaia22ehn & 2023 Oct 19  &                       & 83.9$\pm$0.6     & 20.4$\pm$0.6&              & 11.0$\pm$0.6&            &            \\     
Gaia21arv & 2021 Oct 15  &($-$24.3) 33.5$\pm$4.1 &                  &             &              &             &            &35.1$\pm$0.6\\   
Gaia23bri & 2023 Oct 19  &                       &  2.2$\pm$0.4     &             &              &             &            &            \\     
          & 2023 Nov 23  &37.2$\pm$1.3           &                  &             &              &             &            &            \\       
          & 2024 Jan 09  &37.9$\pm$1.7           &                  &             &              &             &            &            \\       
Gaia21ebu & 2024 Jan 09  & $-$51.2$\pm$1.1       &                  &             &              &             &            &            \\       
Gaia21aul & 2023 Jun 04  &($-$53.9) 16.7$\pm$43  &24.7$\pm$18       &             &              &             &            &            \\       
Gaia23bab & 2024 Apr 11  & 11.4$\pm$1.0          &0.5$\pm$0.1       &             &              &             &            &            \\       
Gaia23dhi & 2024 Jun 07  & $-$33.4$\pm$9.1       &                  &             &              &             &            &            \\       
Gaia24afw & 2024 Jun 07  &                       & 1.5$\pm$0.1      &             &              &             &            &            \\       
Gaia21faq & 2023 Jun 04  &                       &35.6$\pm$0.5      &11.7$\pm$0.5 & 71.2$\pm$1.8 &             &96.5$\pm$1.2&184.5$\pm$2.5\\  
Gaia24beh & 2024 Jun 07  &                       &94.7$\pm$0.6      &17.8$\pm$0.7 &              &             &             &           \\       
Gaia21fji & 2023 Oct 19  &                       &198.2$\pm$2.8     &56.0$\pm$3.3 &              &             &             &           \\       
Gaia21csu & 2023 Oct 19  &                       &19.4$\pm$1.1      &13.8$\pm$0.9 &              &            &              &           \\       
\hline
\hline
\end{tabular}
\begin{quotation}
\textbf{Note.} Negative numbers in parenthesis represent fluxes of absorption components.
\end{quotation}
\end{table} 
\end{appendix}

\begin{acknowledgements}
We acknowledge ESA Gaia, DPAC and the Photometric Science Alerts Team (http://gsaweb.ast.cam.ac.uk/alerts). This work has made use of data from the European Space Agency (ESA) mission
{\it Gaia} (\url{https://www.cosmos.esa.int/gaia}), processed by the {\it Gaia}
Data Processing and Analysis Consortium (DPAC,
\url{https://www.cosmos.esa.int/web/gaia/dpac/consortium}). Funding for the DPAC
has been provided by national institutions, in particular the institutions
participating in the {\it Gaia} Multilateral Agreement.
We acknowledge support from the Large Grant INAF 2022 “YSOs Outflows, Disks and Accretion: towards a global framework for the evolution of planet forming systems (YODA)” and from PRIN-MUR 2022 20228JPA3A “The path to star and planet formation in the JWST era (PATH)”.
This project has received funding from the European Research Council (ERC) via the ERC Synergy Grant ECOGAL (grant 855130). Views and opinions expressed are however those of the author(s) only and do not necessarily reflect those of the European Union or the European Research Council Executive Agency. Neither the European Union nor the granting authority can be held responsible for them. 
We acknowledge the Hungarian National Research, Development and Innovation Office grant OTKA FK 146023. We acknowledge support from the ESA PRODEX contract nr. 4000132054. Zs. N. was supported by the János Bolyai Research Scholarship of the Hungarian Academy of Sciences.
This research has made use of the Spanish Virtual Observatory (https://svo.cab.inta-csic.es) project funded by MCIN/AEI/10.13039/501100011033/ through grant PID2020-112949GB-I00.
\end{acknowledgements}

%
%

\end{document}